\pdfoutput=1

\documentclass[10pt]{article}

\usepackage[left=.75in,right=.75in,top=.75in,bottom=.75in]{geometry}

\usepackage[utf8]{inputenc}

\usepackage{amssymb,amsmath,latexsym,amsthm,amsfonts,mathtools}
\DeclareMathOperator*{\argmax}{arg\,max}
\DeclareMathOperator*{\argmin}{arg\,min}

\usepackage{mathrsfs}

\usepackage{multirow}
\usepackage{graphicx}
\usepackage{textcomp,siunitx}
\usepackage{float}
\usepackage{subcaption}

\usepackage{booktabs}
\usepackage{enumitem}

\usepackage[hyperref]{xcolor}
\definecolor{ao(english)}{rgb}{0.0, 0.5, 0.0}

\usepackage{hyperref}
\hypersetup{colorlinks, breaklinks, citecolor=blue, linkcolor=ao(english), urlcolor=blue}

\usepackage[nameinlink,noabbrev,sort&compress]{cleveref}

\crefname{assumption}{Assumption}{Assumptions}

\usepackage{setspace}

\usepackage{cite}

\theoremstyle{plain}
\newtheorem{theorem}{Theorem}
\newtheorem{lemma}{Lemma}
\newtheorem{corollary}{Corollary}

\newtheorem{assumption}{Assumption}

\theoremstyle{remark}

\theoremstyle{definition}
\newtheorem{definition}{Definition}

\newtheorem{problem}{Problem}

\newcommand{\sign}{\mathrm{sign}}

\usepackage{newtxmath}

\usepackage{accents}
\newcommand\munderbar[1]{%
	\underaccent{\bar}{#1}}

\usepackage{scalerel}
\usepackage{tikz}
\usetikzlibrary{automata, shapes, arrows, calc, arrows.meta, fit, positioning}
\usetikzlibrary{svg.path}

\definecolor{orcidlogocol}{HTML}{A6CE39}
\tikzset{
	orcidlogo/.pic={
		\fill[orcidlogocol] svg{M256,128c0,70.7-57.3,128-128,128C57.3,256,0,198.7,0,128C0,57.3,57.3,0,128,0C198.7,0,256,57.3,256,128z};
		\fill[white] svg{M86.3,186.2H70.9V79.1h15.4v48.4V186.2z}
		svg{M108.9,79.1h41.6c39.6,0,57,28.3,57,53.6c0,27.5-21.5,53.6-56.8,53.6h-41.8V79.1z M124.3,172.4h24.5c34.9,0,42.9-26.5,42.9-39.7c0-21.5-13.7-39.7-43.7-39.7h-23.7V172.4z}
		svg{M88.7,56.8c0,5.5-4.5,10.1-10.1,10.1c-5.6,0-10.1-4.6-10.1-10.1c0-5.6,4.5-10.1,10.1-10.1C84.2,46.7,88.7,51.3,88.7,56.8z};
	}
}

\newcommand\orcidicon[1]{\href{https://orcid.org/#1}{\mbox{\scalerel*{
				\begin{tikzpicture}[yscale=-1,transform shape]
					\pic{orcidlogo};
				\end{tikzpicture}
			}{|}}}}

\title{Cooperative Target Capture using Predefined-time Consensus over Fixed and Switching Networks}

\author{Abhinav~Sinha\textsuperscript{\orcidicon{0000-0001-6419-2353}}
	\thanks{Corresponding author.\newline The authors are with the Intelligent Systems \& Control Lab, Department of Aerospace Engineering, Indian Institute of Technology Bombay, Powai- 400076, Mumbai, India. e-mails: abhinavsinha@aero.iitb.ac.in, srk@aero.iitb.ac.in.}	
	\and Shashi~Ranjan~Kumar\textsuperscript{\orcidicon{0000-0001-6446-7281}}
}

\date{}

\begin{document}

\maketitle
\doublespacing

\begin{abstract}
	We propose predefined-time consensus-based cooperative guidance laws for a swarm of interceptors to simultaneously capture a target capable of executing various kinds of motions. Unlike leader-follower cooperative guidance techniques, the swarm of interceptors has no leader and each interceptor executes its own guidance command in a distributive fashion, thus obviating the residency of the mission over a single interceptor. We first design cooperative guidance commands subject to target's mobility, followed by the same when the target is stationary, and also account for interceptors' changing topologies. We further show through rigorous analysis that the proposed cooperative guidance laws guarantee consensus in the interceptors' time-to-go values within a predefined-time. The proposed design allows a feasible time of consensus in time-to-go to be set arbitrarily at will during the design regardless of the interceptors' initial time-to-go values, thereby ensuring a simultaneous target interception. We also demonstrate the efficacy of the proposed design via simulations for interceptors connected over static and dynamic networks.\medskip
	
\end{abstract}


\section{Introduction}\label{sec:introduction}
Recently, terminal time-constrained guidance strategies \cite{1597196,ZHU2019818,HOU2019105142,SINHA2021106824,SINHA2021106776} have drawn a lot of attention, especially in scenarios where a swarm of interceptors are required to capture a target simultaneously \cite{LI2018243,SONG2017193,SONG201531,YANG2021106958,CHEN2021106523,ZHANG2020105641,LYU2019100,ZHAO2019308,JIANG2018426,SU2019556,SU2017147,WANG20151,SU201891}. While advanced electronic countermeasures may make it difficult for a single interceptor to penetrate the adversarial zone, merits of information processing and communication technology can be harnessed to establish group coordination in the swarm for simultaneous target interception \cite{9274339,doi:10.2514/1.G005367,9000526}. It is also possible to have benefits, such as spatial diversity, fault tolerance, and increased likelihood of mission success by introducing cooperation in the swarm.

Guidance strategies for a single interceptor with constraints on the time of target capture \cite{1597196,ZHU2019818,HOU2019105142,SINHA2021106824,SINHA2021106776} have paved way for advanced guidance strategies by augmenting a feedback control term with the baseline guidance command. For example, the work in \cite{Jeon2010,Zhao2015,ZHANG20151438,CHEN20195692} and references therein, have used this concept to design guidance commands for multi-interceptor swarm against a single target. Authors in \cite{SHIYU2010103} transformed the guidance problem into a nonlinear tracking problem by using the concept of a virtual leader. Authors in \cite{doi:10.2514/1.G001609} designed tangential acceleration to improve the effectiveness of the control inputs. A distributed, adaptive and optimal approach to capture a target using group cooperation was presented in \cite{doi:10.1080/00207179.2018.1533251} for a swarm of interceptors. Against stationary target, authors in \cite{SONG201531} designed a consensus scheme of impact time by accounting for the communication noise over the multi-interceptor interaction network. Some studies have also extended the two-dimensional cooperative target capture paradigm for three-dimensional engagements \cite{SU2019556,CHEN2021106523,YANG2021106958,ZHAO20151104,doi:10.2514/1.G005367,doi:10.1177/09544100211010524,SONG2017193,LYU20191294}.

Since engagements only last for a short duration of time, having interceptors agree on certain quantities of interest early in the engagement is of utmost importance, especially in scenarios demanding a simultaneous target capture. This essentially means that an accurate control over engagement duration, and the time required by the interceptors to agree upon those relevant quantities are needed. Usually, controlling an interceptor's time-to-go, a quantity representative of its engagement duration, can steer the interceptor on a path requisite for time-constrained interception. Consequently, the time-to-go can be regarded as the quantity of interest upon which interceptors may agree to simultaneously capture the target.

To this aim, several studies have been proposed that have made attempts to ensure consensus in interceptors' time-to-go. For example, authors in \cite{ZHAO2019308,9147677,9123223,LYU2019100} have presented asymptotic consensus in interceptors' time-to-go to realize a simultaneous target interception. Cooperative guidance strategies based on finite-time consensus in interceptors' time-to-go have been put forth in \cite{9447834,9266104,7122426,9000526,SONG2017193,LYU20191294,YANG2021106958}. Although finite-time cooperative guidance laws lead to finite-time consensus in interceptors' time-to-go values and a satisfactory guidance precision, the performance of their respective designs are largely dictated by their initial engagement parameters. In scenarios where initial parameters are unavailable or poor, the time of consensus may be difficult to estimate before homing. In worst case, the time of consensus may even become larger than the total engagement duration, thereby degrading the guidance precision and eventually, causing mission failure. To cope up with such scenarios, fixed-time convergent guidance commands have been proposed in cooperative guidance applications \cite{CHEN2021106523,LI2018243,doi:10.1080/00207179.2019.1662947,9274339,sinha2021threeagent}, wherein the time of convergence/consensus was made independent of the initial engagement parameters, thus ensuring satisfactory guidance precision. However, the controller gains required to ensure a particular time of consensus might be overestimated, and the design parameters may need to be re-tuned if a different time of consensus is required.

It is also worth noting that a majority of the existing literature on cooperative guidance (for example, see \cite{9123223,SONG2017193,9447834,9000526,9147677,SU201891,WANG20151,LYU2019100,CHEN2021106523,YANG2021106958,doi:10.1080/00207179.2019.1662947,LI2018243,LYU20191294,doi:10.2514/1.G005367,9266104,9274339} and references therein) were proposed for the case when interceptors' interaction network is fixed. That is throughout the engagement, an interceptor, say $I_1$, exchanges the time-to-go information with the same set of other interceptors in the swarm (referred to as the neighbors of $I_1$). However, due to practical limitations on sensing and communication, it is more pragmatic to account for the interceptors' changing interactions. This means that it is not guaranteed for a subset of interceptors that are initially neighbors to remain so throughout the engagement. For a cooperative guidance law to be effective under such limitations, a proposed design should provision for the interceptors' switching topologies. On the other hand, cooperative guidance designs considering interceptors' switching topologies are few and far between (for example, see \cite{ZHANG2020105641,Zhao2017NoDy,Zhao2019NoDy,SUN20141202,ZHAO20171570}). The work in \cite{ZHANG2020105641,Zhao2017NoDy,SUN20141202,ZHAO20171570} only designed cooperative guidance commands against a stationary target, while authors in \cite{Zhao2019NoDy} addressed the problem of intercepting a maneuvering target. Note that simultaneously capturing a mobile adversary is more challenging. Authors in \cite{ZHANG2020105641} designed a two-stage guidance scheme which included switching in the guidance commands at the onset of the second phase. A similar concept of two-stage design was also reported in \cite{Zhao2017NoDy} wherein the authors designed cooperative techniques among different subgroups of interceptors. In \cite{Zhao2019NoDy,SUN20141202}, a leader-follower cooperative design was adopted for target capture. It should be noted that the failure of leader interceptor in such scenarios may lead to the failure of the mission. The study in \cite{Zhao2017NoDy,ZHAO20171570} used consensus in interceptors' range-to-go and their look angles while using the baseline proportional-navigation command. In spite of the various cooperative guidance designs that considered interceptors' switching topologies, stricter control over the time of consensus in coordination variable was overlooked.

Motivated by the aforementioned works and the need to achieve an enhanced guidance precision in a time-constrained scenario, we aim to provide a systematic and unified distributed cooperative design. In this paper, we propose a novel predefined-time consensus-based design for cooperative target capture. In light of the existing literature, we summarize the merits of our work as follows:
\begin{itemize}
	\item First of all, our study addresses the problem of simultaneous target capture when the cooperating interceptors exchange the information of time-to-go over fixed and switching topologies. Owing to practical limitations on sensing and communication, our consideration of interceptors' switching topologies is closer to practice.
	\item Our design does not include a swarm leader, and hence the residency of the cooperative mission over a single interceptor may be irrelevant, unlike the leader-follower cooperation discussed in \cite{Zhao2019NoDy,SUN20141202,9000526,9266104}.
	\item Unlike the existing works \cite{ZHANG2020105641,Zhao2017NoDy,SUN20141202,ZHAO20171570}, we also take the motion of the target into account since target's mobility adds further challenges in a cooperative guidance design. Thus, we design cooperative guidance commands against a target that may execute maneuvers, move with a constant speed, or remain stationary.
	\item Unlike asymptotic \cite{ZHAO2019308,9147677,9123223,LYU2019100}, finite-time \cite{9447834,9266104,7122426,9000526,SONG2017193,LYU20191294,YANG2021106958} and fixed-time \cite{CHEN2021106523,LI2018243,doi:10.1080/00207179.2019.1662947,9274339,sinha2021threeagent} time-to-go error convergent guidance schemes, we present a predefined-time consensus-based cooperative guidance scheme for simultaneous target capture. Through the cooperative guidance commands in the proposed scheme, a feasible time at which the interceptors' time-to-go values achieve consensus can be arbitrarily preassigned at will. Having a strict control over the time of consensus is beneficial since engagements last for a short duration only, and letting interceptors align themselves on their requisite collision courses within a predefined-time may improve the cooperative guidance precision. This predefined-time can be directly specified as a parameter in the guidance commands, thereby making the design independent of the initial engagement parameters and preventing overestimation of the controller gains. To the best of our knowledge, predefined-time consensus has not been applied previously in the cooperative guidance literature. 
	\item We treat the cooperative guidance problem intact in the nonlinear form, thereby preventing the errors caused due to linearization of the engagement kinematics and interceptors' small heading angle assumptions.
	\item We validate our design over a dual-controlled interceptor dynamic model \cite{SINHA2021106776} to show the superior performance of the proposed scheme.
\end{itemize}
The remainder of this paper is organized as follows. After introduction in \Cref{sec:introduction}, necessary background and problem formulation are presented in \Cref{sec:problemformulation}. Derivations of the cooperative guidance commands are presented in \Cref{sec:fixed,sec:switching} while simulations are shown in \Cref{sec:simulations}. Finally, \Cref{sec:conclusions} concludes the paper and discusses some outlook towards future investigations.

\section{Background and Problem Formulation}\label{sec:problemformulation}
We now provide a brief background of the elements necessary for the design of proposed cooperative guidance commands.

\subsection{Kinematics of Interceptor-target Cooperative Engagement}
\begin{figure}[h!]
	\centering
	\includegraphics[width=.6\linewidth]{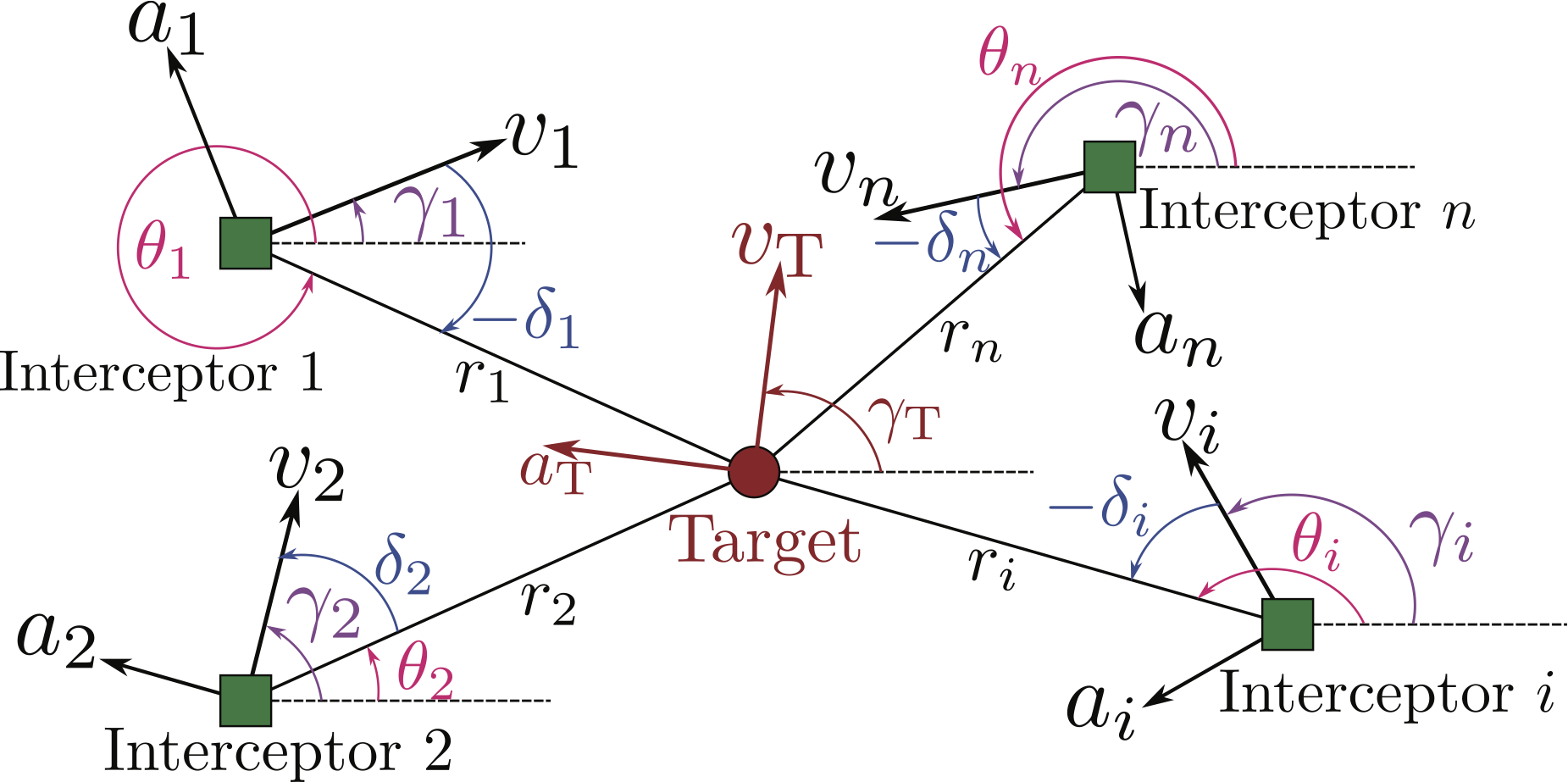}
	\caption{Engagement geometry for cooperative target pursuit.}
	\label{fig:enggeo}
\end{figure}

We consider a planar cooperative engagement scenario shown in \Cref{fig:enggeo} where multiple interceptors aim to simultaneously capture a single target capable of executing various kinds of motions. The speed of the $i$\textsuperscript{th} interceptor, $\forall\,i=1,2,\ldots,n$, is denoted by $v_i$ while that of the target is $v_\mathrm{T}$, such that $v_i>v_\mathrm{T}~\forall\,i$. For $i$\textsuperscript{th} interceptor, the relative separation and the line-of-sight (LOS) angle with respect to the target are denoted by $r_i$ and $\theta_i$, respectively. The $i$\textsuperscript{th} interceptor has a heading angle, $\gamma_i$, whereas its look angle is $\delta_i$. In the interest of our analysis, we assume that the speeds of the vehicles remain invariant throughout the engagement, $\delta_i\in(-\pi/2,\pi/2)$, and that each vehicle can alter its orientation using lateral acceleration applied perpendicular to its velocity. This essentially implies that $a_i$ is steering control for the $i$\textsuperscript{th} interceptor while that of the target is $a_\mathrm{T}$.
The kinematics of cooperative engagement is governed by the equations,
\begin{subequations}\label{eq:enggeo}
	\begin{align}
		v_{r_i} =\dot{r}_i =& v_\mathrm{T}\cos(\gamma_\mathrm{T}-\theta_i) - v_{i}\cos\delta_i,\label{eq:rdot}\\
		v_{\theta_i} =r_i\dot{\theta}_i =& v_\mathrm{T}\sin(\gamma_\mathrm{T}-\theta_i) - v_{i}\sin\delta_i, \label{eq:thetadot}\\
		\dot{\gamma}_{i} =& \dfrac{a_{i}}{v_{i}},\quad  \dot{\gamma}_\mathrm{T} = \dfrac{a_\mathrm{T}}{v_\mathrm{T}}, \label{eq:aMaT} \\
		\delta_i =&\gamma_{i}-\theta_i,\label{eq:sigma}	
	\end{align}
\end{subequations}
where $v_{r_i}$ and $v_{\theta_i}$ represent the relative velocity components along and perpendicular to the corresponding LOS, respectively, such that
\begin{align}
	\dot{v}_{r_i} =& \dfrac{v_{\theta_i}^2}{r_i}+a_i \sin\delta_i - a_\mathrm{T} \sin\left(\gamma_\mathrm{T}-\theta_i\right), \label{eq:vr}\\
	\dot{v}_{\theta_i}=&-\dfrac{v_{r_i}v_{\theta_i}}{r_i}-a_i\cos\delta_i+a_\mathrm{T} \cos\left(\gamma_\mathrm{T}-\theta_i\right).\label{eq:vtheta}
\end{align}

\subsection{Graph Theory for Cooperative Guidance}
The interaction among multiple cooperating interceptors can be conveniently modeled via graphs. A graph is an ordered pair $\mathcal{G} = (\mathcal{V,E})$ consisting of a finite vertex set $\mathcal{V} = \{1, 2, \ldots, n\}$ representing $n$ interceptors, and an edge set $\mathcal{E} \subset \big\{\{i,j\} | i,j \in \mathcal{V} \;\text{and}\; i\neq j\big\}$, corresponding to the communication links among the interceptors. Cardinality of $\mathcal{E}$ is usually considered to be $E$, unless otherwise specified. The neighborhood of a vertex $i$ is the set of vertices adjacent to it, and is denoted by $\mathcal{N}_i$, that is, $\mathcal{N}_i = \{j\in\mathcal{V}|\{i,j\}\in\mathcal{E}\}$. For an undirected graph with the vertex set $\mathcal{V}$, we have the Laplacian matrix, $\mathcal{L}(\mathcal{G})$, with entries $l_{ij}$,  as $-1$ if $\{i,j\}\in\mathcal{E}$, $|\mathcal{N}_i|$ if $i=j$, and $0$ otherwise. The eigenvalues of $\mathcal{L}(\mathcal{G})$, denoted by $\lambda(\mathcal{L}(\mathcal{G}))$, are non-negative and the null space of the Laplacian is spanned by $\mathbf{1}_n$, i.e., the vector of all ones. The Laplacian matrix can also be represented in terms of the graph incidence matrix, $\mathcal{F}(\mathcal{G})$, such that $\mathcal{L}(\mathcal{G}) = \mathcal{F}(\mathcal{G})\mathcal{F}^\top(\mathcal{G})$.
\begin{lemma}\cite{MesbahiEgerstedt}\label{lem:ac}
	The smallest non-zero eigenvalue of $\mathcal{L}(\mathcal{G})$ (denoted by $\lambda_2(\mathcal{L}(\mathcal{G}))$ and referred to as the algebraic connectivity of the graph, $\mathcal{G}$ satisfies $x^\top\mathcal{L}(\mathcal{G})x\geq \lambda_2(\mathcal{L}(\mathcal{G}))\|x\|_2^2>0$ for some $x\in\mathbb{R}^n$.
\end{lemma}
\begin{definition}
	A switched dynamic network is an ordered pair,  ${\mathcal{G}}_{\sigma(t)}=({\mathcal{G}},\sigma(t))$ such that ${\mathcal{G}}=\{\mathcal{G}_1,\mathcal{G}_2,\ldots,\mathcal{G}_p\}$ is a collection of graphs having the same vertex set and $\sigma(t):[t_0,\infty]\to[1,2,\ldots,p]$ is a piecewise continuous switching signal that determines the interceptors' communication topology at each instant of time.
\end{definition}

\subsection{Notions of Predefined-time Stability}	
Consider a system whose dynamics is given by
\begin{equation}\label{eq:toysystem}
	\dot{z} = -\dfrac{1}{T_s}f(z),~\forall\,t\geq t_0,~f(0)=0,~z(t_0)=z_0,
\end{equation}
where $z\in\mathbb{R}^n$ is the state variable, $T_s>0$ is a parameter and $f:\mathbb{R}^n\to\mathbb{R}^n$ is nonlinear and continuous on $z$ everywhere, except at the origin.
\begin{assumption}\label{asm:psi}
	Let $\psi(\ell) = \phi(|\ell|)^{-1}\sign(\ell)$ such that $\ell\in\mathbb{R}$ and the function $\phi:\mathbb{R}_+\to{\mathbb{R}}\setminus\{0\}$ satisfies $\phi(0)=\infty$, $\phi(\ell)<\infty\;\forall\,\ell\in\mathbb{R}_+\setminus\{0\}$, and $\int_{0}^{\infty}\phi(\ell)d\ell=1$.
\end{assumption}
\begin{lemma}\cite{ALDANALOPEZ202011880}\label{lem:predeftime}
	If there exists a continuous positive definite radially unbounded function, $\mathscr{V}:\mathbb{R}^n\to\mathbb{R}$, whose time derivative along the trajectories of \eqref{eq:toysystem} satisfies $\dot{\mathscr{V}}(z)\leq-\dfrac{1}{T_s}\psi(\mathscr{V}(z))$ for all $z\in\mathbb{R}^n\setminus\{0\}$, and $\psi(z)$ satisfies \Cref{asm:psi}, then the system \eqref{eq:toysystem} is fixed-time stable with a predefined upper bound on the settling time, $T_s$.
\end{lemma}
\begin{lemma}\cite{ALDANALOPEZ202011880}\label{lem:omega}
	A predefined-time consensus function $\Omega:\mathbb{R}\to\mathbb{R}$ is a monotonically increasing function, if there exists a function $\hat{\Omega}:\mathbb{R}_+\to\mathbb{R}_+$, a non-increasing function $\beta:\mathbb{N}\to\mathbb{R}_+$, and a parameter $d\geq 1$, such that for all $z=(z_1,z_2,\ldots,z_n)^\top,~z_i\in\mathbb{R}_+$, the inequality
	\begin{equation}
		\hat{\Omega}\left(\beta(n)\|z\|\right) \leq \beta(n)^d\sum_{i=1}^{n}\Omega(z_i)
	\end{equation}
	holds, and $\psi(\ell) = \ell^{-1}\hat{\Omega}(|\ell|)$ satisfies \Cref{asm:psi}.
\end{lemma}
\begin{lemma}\label{lem:pdt2}\cite{https://doi.org/10.1002/rnc.4715}
	For a continuous positive definite radially unbounded function, $\mathscr{V}:\mathbb{R}^n\to\mathbb{R}$, such that $\mathscr{V}(0)=0$ and $\mathscr{V}(z)>0$ for all $z\in\mathbb{R}^n\setminus\{0\}$, if the relation
	\begin{equation}
		\dot{\mathscr{V}}(z) \leq - \dfrac{\Gamma\left(\dfrac{1-k\mathfrak{m}}{\mathfrak{n}-\mathfrak{m}}\right)\Gamma\left(\dfrac{k\mathfrak{n}-1}{\mathfrak{n}-\mathfrak{m}}\right)}{T_s\mathscr{M}^k \Gamma(k)\left(\mathfrak{n}-\mathfrak{m}\right)}\left(\dfrac{\mathscr{M}}{\mathscr{N}}\right)^{\frac{1-k\mathfrak{m}}{\mathfrak{n}-\mathfrak{m}}}\left[\mathscr{M}\mathscr{V}(z)^\mathfrak{m} +\mathscr{N}\mathscr{V}(z)^\mathfrak{n} \right]^{k},
	\end{equation}
	where $\Gamma(\cdot)$ is the \emph{gamma} function such that $\Gamma(\ell)=\int_{0}^{\infty}e^{-t}t^{\ell-1}dt$, holds for some $\mathscr{M},\mathscr{N},\mathfrak{m},\mathfrak{n},k>0$ satisfying the constraints $k\mathfrak{m}<1$ and $k\mathfrak{n}>1$, then the origin of the system $\dot{z}=f(z),$ is predefined-time stable with a predefined-time, $T_s$.
\end{lemma}

\subsection{Other Useful Results}
\begin{lemma}\label{lem:norm}\cite{hardy1988cambridge}
	For some $z=[z_1,z_2,\ldots,z_n]^\top \in\mathbb{R}^n$ and $p\in\mathbb{N}$, the following relation holds:
	\begin{equation}
		\|z\|_p = \sqrt[p]{\sum_{i=1}^{n}|z_i|^p},
	\end{equation} 
	such that $\|z\|_s\leq\|z\|_t$ for any $s\geq t$.
\end{lemma}
\begin{lemma}\cite{https://doi.org/10.1002/rnc.4715}\label{lem:jenson2}
	Let $s=(s_1,s_2,\ldots,s_n) \in\mathbb{R}_+$ be a sequence of positive numbers. For some $\mathscr{M},\mathscr{N},\mathfrak{m},\mathfrak{n},k>0$ satisfying the constraints $k\mathfrak{m}<1$ and $k\mathfrak{n}>1$, the following relations holds:
	\begin{equation}
		\left(\dfrac{1}{n}\sum_{i=1}^{n} s_i\right) \left[\mathscr{M}\left(\dfrac{1}{n}\sum_{i=1}^{n} s_i\right)^\mathfrak{m} + \mathscr{N}\left(\dfrac{1}{n}\sum_{i=1}^{n} s_i\right)^\mathfrak{n}\right]^k \leq \dfrac{1}{n}\sum_{i=1}^{n} s_i\left(\mathscr{M}s_i^\mathfrak{m}+\mathscr{N}s_i^\mathfrak{n}\right)^k.
	\end{equation}
\end{lemma}
In what follows, we assume that interceptors communicate over an undirected graph. Unlike the leader-follower cooperative guidance\cite{Zhao2019NoDy,SUN20141202,9000526,9266104}, residency of the mission over a single interceptor is eliminated since there is no leader in the swarm. We also assume that $\sigma(t)$ is generated exogenously and there is a finite number of switching in a finite time interval, i.e.,  Zeno phenomenon is excluded. Note that at any instant of time during engagement, the time remaining till capture (also known as \emph{time-to-go}) is crucial to the cooperative guidance design. One may, then, speculate that if each interceptor's time-to-go value becomes same and decreases at the same rate, a simultaneous interception shall occur. We now present the main problem addressed in this paper.

\begin{problem}
	Ensure that the swarm of interceptors simultaneously capture a target capable of executing various kinds of motions, i.e., the target may either maneuver, or move with a constant speed, or may not move at all.
\end{problem} 
It is further desirable that the time at which interceptors' agree upon a common time-to-go value can be easily preassigned during the guidance design. Consideration of interceptors' switching topologies, the target's motion and strict control over engagement duration make the problem addressed in this paper challenging and interesting at the same time. Note that we have considered the kinematic relations, \eqref{eq:enggeo}, intact in nonlinear form and we shall design the necessary cooperative guidance commands without any approximations. This shall allow the proposed design to maintain its efficacy over a larger operating region. In other words, errors associated with linear approximations are absent in the proposed design.

\section{Cooperative Salvo when Interceptors Communicate over a Fixed Network}\label{sec:fixed}
In a fixed network, the communication topology of the interceptors remains same throughout the engagement. That is, if interceptors $i$ and $j$ are neighbors at the beginning of the engagement, they will continue to remain so throughout the course of cooperative engagement. We now design cooperative guidance commands for a swarm of interceptors against various target motions.

\subsection{Simultaneous Interception of a Mobile Adversary}
Target's motion adds further challenges to the design process. Here, we use deviated pursuit guidance as our baseline cooperative guidance command, primarily because it was developed to intercept a mobile adversary \cite{Shneydor} and has shown satisfactory performance in a time-constrained scenario \cite{doi:10.2514/1.G005455}. For a mobile target, the time-to-go for $i$\textsuperscript{th} interceptor guided by deviated pursuit, with a fixed look angle, $\delta_i$, is 
\begin{align}
	t_{\mathrm{go}_i} = \dfrac{r_i \sec\delta_i}{v_i^2-v_\mathrm{T}^2}\left[v_i + v_\mathrm{T}\cos\left(\gamma_\mathrm{T}-\theta_i+\delta_i\right)\right].\label{eq:tgodev}
\end{align}
It is worth noting from \eqref{eq:tgodev} that $t_{\mathrm{go}_i} = 0\iff r_i=0$ regardless of the target's maneuver since $v_i>v_\mathrm{T}$ and neither $\sec\delta_i$ nor $v_i + v_\mathrm{T}\cos\left(\gamma_\mathrm{T}-\theta_i+\delta_i\right)$ is zero. This leads to a speculation that an agreement in $t_{\mathrm{go}_i}$ followed by its nullification is sufficient to ensure a simultaneous target capture even if the target maneuvers. For $i$\textsuperscript{th} interceptor, we define the error, $\xi_i$, in the common time of interception, $T_f$, as
\begin{equation}\label{eq:ei}
	\xi_i =  t_{\mathrm{go}_i} - \left(T_f - t\right),
\end{equation}
where $T_f$ is not commanded to any interceptor in advance but the interceptors implicitly agree upon a common value as the engagement proceeds.

Observe that the time differentiation of the expression in \eqref{eq:ei} yields
\begin{equation}\label{eq:eidot}
	\dot{\xi}_i =  \dfrac{r_i^2\dot{\theta}_i^2\sec^2\delta_i}{v_i^2-v_\mathrm{T}^2} - \dfrac{r_i^2\dot{\theta}_i\sec^2\delta_i}{v_i\left(v_i^2-v_\mathrm{T}^2\right)}a_i - \dfrac{r_i\sin\left(\delta_i+\gamma_\mathrm{T}-\theta_i\right)}{\left(v_i^2-v_\mathrm{T}^2\right)\cos\delta_i}a_\mathrm{T}
\end{equation}
evidencing that the relative degree of $\xi_i$ is one with respect to $a_i$. Further, the information of the target's maneuver is also needed to synthesize the cooperative guidance command, making the design a challenging one.

In practice the upper bound on the target's maneuver can be known, and hence it facilitates us to use a nonlinear disturbance observer \cite{ndob} to estimate $a_\mathrm{T}$ in a reasonable manner by augmenting the baseline controller with an active disturbance-attenuation compensation. Consider the error, $\Xi_i$, in estimation of the target's maneuver as,
\begin{equation}
	\Xi_i = a_\mathrm{T} - \hat{a}_{\mathrm{T}_i};~|\Xi_i|\leq\argmin_{i}\Xi_{i_{\max}},~|\dot{\Xi}_i|\leq \argmin_{i}\dot{\Xi}_{i_{\max}},
\end{equation}
with $\hat{a}_{\mathrm{T}_i}$ being the estimate of the target's maneuver as observed by the $i$\textsuperscript{th} interceptor. Using \eqref{eq:vtheta}, we can express the nonlinear disturbance observer dynamics employed by the $i$\textsuperscript{th} interceptor to estimate $a_\mathrm{T}$ as
\begin{align}
	\dot{\mathscr{w}}_{0_i} =& -\dfrac{2r_i\dot{\theta}_i}{r_i} - \dfrac{\cos\delta_i}{r_i}a_i + \dfrac{\cos\left(\gamma_\mathrm{T}-\theta_i\right)}{r_i}\mathscr{u}_0,\nonumber\\
	\mathscr{u}_0=&-\mathscr{G}_{2_i}\digamma_i^{\frac{1}{3}}|\mathscr{w}_{0_i}-\dot{\theta}_i|^\frac{2}{3}\sign(\mathscr{w}_{0_i}-\dot{\theta}_i) - \mathscr{H}_{2_i}\left(\mathscr{w}_{0_i}-\dot{\theta}_i\right)+\mathscr{w}_{1_i},\nonumber\\
	\dot{\mathscr{w}}_{1_i}=&~\mathscr{u}_{1_i} = -\mathscr{G}_{1_i}\digamma_i^{\frac{1}{2}}|\mathscr{w}_{1_i}-\mathscr{u}_{0_i}|^\frac{1}{2}\sign(\mathscr{w}_{1_i}-\mathscr{u}_{0_i}) - \mathscr{H}_{1_i}\left(\mathscr{w}_{1_i}-\mathscr{u}_{0_i}\right)+\mathscr{w}_{2_i},\nonumber\\
	\dot{\mathscr{w}}_{2_i} =& -\mathscr{G}_{0_i}\digamma_i\sign(\mathscr{w}_{2_i}-\mathscr{u}_{1_i}) - \mathscr{H}_{0_i}\left(\mathscr{w}_{2_i}-\mathscr{u}_{1_i}\right),\nonumber
\end{align}
where $|\dot{a}_\mathrm{T}|\leq\argmin_{i}\digamma_i$, $\mathscr{G}_{2_i}>\mathscr{G}_{1_i}>\mathscr{G}_{0_i}>0$ and $\mathscr{H}_{2_i}>\mathscr{H}_{1_i}>\mathscr{H}_{0_i}>0$, while $\mathscr{w}_{1_i}$ yields an estimate of the target's maneuver, $\hat{a}_{\mathrm{T}_i}$, that is used by the $i$\textsuperscript{th} interceptor in its own cooperative guidance command. Notice that as $\Xi_i\to 0$, interceptors agree upon a common value of $a_\mathrm{T}$ in a finite-time. Put differently, the target's maneuver is estimated in a finite-time using the above nonlinear disturbance observer. We are now equipped to present the cooperative guidance command for the $i$\textsuperscript{th} interceptor that ensures simultaneous target capture.
\begin{theorem}\label{thm:mantargetstatic}
	Consider the cooperative target engagement whose kinematics is governed by \eqref{eq:enggeo}, the time-to-go expression, \eqref{eq:tgodev}, and the error in the time of simultaneous interception, as given in \eqref{eq:ei}. The cooperative guidance command for the $i$\textsuperscript{th} interceptor,
\begin{align}
		a_i =& v_i\dot{\theta}_i + \dfrac{n^{2-d}v_i\left(v_i^2-v_\mathrm{T}^2\right)\cos^2\delta_i}{\lambda_2(\mathcal{L}(\mathcal{G}))T_s\mathscr{c}r_i^2\dot{\theta}_i}\left[\left|\sum_{j\in\mathcal{N}_i(\mathcal{G})}\left(\xi_j-\xi_i\right)\right|^{-1} \exp\left\{\left|\sum_{j\in\mathcal{N}_i(\mathcal{G})}\left(\xi_j-\xi_i\right)\right|^{\mathscr{c}}\right\}\left|\sum_{j\in\mathcal{N}_i(\mathcal{G})}\left(\xi_j-\xi_i\right)\right|^{2-\mathscr{c}}+\epsilon\right]\nonumber\\
		&-\dfrac{v_i\sin\left(\delta_i+\gamma_\mathrm{T}-\theta_i\right)}{r_i\dot{\theta}_i\sec\delta_i}\hat{a}_{\mathrm{T}_i};~\mathscr{c}\in(0,1],\label{eq:aidev}
	\end{align}
	with
	\begin{equation}\label{eq:epsilon}
		\epsilon>\argmax_{i}\left(\dfrac{r_i}{v_i^2-v_\mathrm{T}^2}\argmin_{i}\Xi_{i_{\max}}\right),
	\end{equation}
	where $d\geq 1$, $\mathcal{G}$ is a connected (fixed) graph and $\lambda_2(\mathcal{L}(\mathcal{G}))$ is its associated algebraic connectivity, ensures that the $i$\textsuperscript{th} interceptor converges to the deviated pursuit course within a predefined time, $T_s$, specified prior to homing. This essentially means that the interceptors' time-to-go values achieve an agreement within a predefined-time, $T_s$, eventually leading to a simultaneous interception of the maneuvering target.
\end{theorem}
\begin{proof}
	First, observe that
	\begin{equation}
		\sum_{j\in\mathcal{N}_i}\left(\xi_j-\xi_i\right) = \sum_{j\in\mathcal{N}_i}\left(t_{\mathrm{go}_j}-t_{\mathrm{go}_i}\right),
	\end{equation}
	since $T_f$ is a constant, and that the cooperative guidance command, \eqref{eq:aidev}, can be expressed as
	\begin{equation}\label{eq:aicompact}
		a_i = v_i\dot{\theta}_i - \dfrac{v_i\left(v_i^2-v_\mathrm{T}^2\right)\cos^2\delta_i}{r_i^2\dot{\theta}_i}\mathscr{U}_i-\dfrac{v_i\sin\left(\delta_i+\gamma_\mathrm{T}-\theta_i\right)}{r_i\dot{\theta}_i\sec\delta_i}\hat{a}_{\mathrm{T}_i},
	\end{equation}
	where 
	\begin{equation}\label{eq:Uifirst}
		\mathscr{U}_i = -\mathscr{B}_i \zeta_i^{-1}\Omega(|\zeta_i|)-\epsilon
	\end{equation} 
	with $\mathscr{B}_i>0$, $\epsilon$ as given in \eqref{eq:epsilon}, and
	\begin{equation}
		\zeta_i=\sum_{j\in\mathcal{N}_i(\mathcal{G})}(\xi_j-\xi_i).
	\end{equation}
	In \eqref{eq:Uifirst}, the function,
	\begin{equation}
		\Omega(|\zeta_i|)=\dfrac{1}{\mathscr{c}} \exp(|\zeta_i|^{\mathscr{c}})|\zeta_i|^{2-\mathscr{c}}
	\end{equation}
	is a predefined-time consensus function with $\beta(n)=\dfrac{1}{n}$, $\hat{\Omega}(\ell)=\Omega(\ell)$ and $d\geq 1$ (refer \Cref{lem:omega}). In vector form, $\mathbf{\mathscr{U}} = [\mathscr{U}_1,\mathscr{U}_2,\ldots,\mathscr{U}_n]^\top$.
	
	Let us define $\mathbf{\xi}=[\xi_1,\xi_2,\ldots,\xi_n]^\top$ and the agreement subspace as $\mathscr{C}(\mathbf{\xi})=\{\mathbf{\xi}:\xi_1=\xi_2=\cdots=\xi_n\}$. We now consider a radially unbounded Lyapunov function candidate, \begin{equation}\label{eq:Vstatic}
		\mathscr{V}_1(\xi) = \sqrt{\lambda_2(\mathcal{L}(\mathcal{G}))}\beta(n)\sqrt{\mathbf{\xi}^\top \mathcal{L}(\mathcal{G}) \mathbf{\xi}}.
	\end{equation}
	For brevity, we drop the arguments of the variables when it is clear from the context. Note that $\mathscr{V}_1(0) = 0$ if and only if $\mathbf{\xi}\in\mathscr{C}(\mathbf{\xi})$.

	Substituting \eqref{eq:aicompact} in \eqref{eq:eidot}, one obtains
	\begin{equation}\label{eq:integratorsys}
		\dot{\xi}_i = \mathscr{U}_i -\dfrac{r_i\sin\left(\delta_i+\gamma_\mathrm{T}-\theta_i\right)}{\left(v_i^2-v_\mathrm{T}^2\right)\cos\delta_i}\left({a}_{\mathrm{T}_i}-\hat{a}_{\mathrm{T}_i}\right).
	\end{equation}
	
	On differentiating $\mathscr{V}_1$ with respect to time, one may obtain
	\begin{align}
		\dot{\mathscr{V}}_1 =& \dfrac{\sqrt{\lambda_2(\mathcal{L})}\beta(n)}{\sqrt{\mathbf{\xi}^\top \mathcal{L} \mathbf{\xi}}}\mathbf{\xi}^\top\mathcal{L} \dot{\mathbf{\xi}}	=\dfrac{\sqrt{\lambda_2(\mathcal{L})}\beta(n)}{\sqrt{\mathbf{\xi}^\top \mathcal{L} \mathbf{\xi}}}\mathbf{\xi}^\top\mathcal{L} \left[\mathbf{\mathscr{U}}-\dfrac{\mathbf{r}\sin\left(\mathbf{\delta}+\gamma_\mathrm{T}\mathbf{1}_n-\mathbf{\theta}\right)}{\left(\mathbf{v}^2-v_\mathrm{T}^2\mathbf{1}_n\right)\cos\mathbf{\delta}}\mathbf{\Xi}\right]
	\end{align}
	where
	\begin{equation}
		\dfrac{\mathbf{r}\sin\left(\mathbf{\delta}+\gamma_\mathrm{T}\mathbf{1}_n-\mathbf{\theta}\right)}{\left(\mathbf{v}^2-v_\mathrm{T}^2\mathbf{1}_n\right)\cos\mathbf{\delta}}\mathbf{\Xi} = 
		\begin{pmatrix}
			\dfrac{r_1\sin\left(\delta_1+\gamma_\mathrm{T}-\theta_1\right)}{\left(v_1^2-v_\mathrm{T}^2\right)\cos\delta_1}\Xi_1\\
			\dfrac{r_2\sin\left(\delta_2+\gamma_\mathrm{T}-\theta_2\right)}{\left(v_2^2-v_\mathrm{T}^2\right)\cos\delta_2}\Xi_2\\
			\vdots\\
			\dfrac{r_n\sin\left(\delta_n+\gamma_\mathrm{T}-\theta_n\right)}{\left(v_n^2-v_\mathrm{T}^2\right)\cos\delta_n}\Xi_n
		\end{pmatrix}.
	\end{equation}
	Therefore,	
	\begin{align}
		\dot{\mathscr{V}}_1  =&-\lambda_2(\mathcal{L})\beta(n)^{2-d}\mathscr{V}_1^{-1}\sum_{i=1}^{n}\beta(n)^d\dfrac{\mathscr{B}_i}{\mathscr{c}}\exp(|\zeta_i|^{\mathscr{c}})|\zeta_i|^{2-\mathscr{c}} \nonumber\\
		&- \lambda_2(\mathcal{L})\beta(n)^{2}\mathscr{V}_1^{-1} \left[\epsilon \sum_{i=1}^{n}|\zeta_i|+\dfrac{\mathbf{\zeta}^\top\mathbf{r}\sin\left(\mathbf{\delta}+\gamma_\mathrm{T}\mathbf{1}_n-\mathbf{\theta}\right)}{\left(\mathbf{v}^2-v_\mathrm{T}^2\mathbf{1}_n\right)\cos\mathbf{\delta}}\mathbf{\Xi}\right] \nonumber\\
		=& -\lambda_2(\mathcal{L})\beta(n)^{2-d}\mathscr{V}_1^{-1}\sum_{i=1}^{n}\beta(n)^d\mathscr{B}_i\Omega(|\zeta_i|) \nonumber\\
		&- \sqrt{\lambda_2(\mathcal{L})}\beta(n) \left[\dfrac{\epsilon }{\sqrt{\mathbf{\xi}^\top \mathcal{L} \mathbf{\xi}}}\sum_{i=1}^{n}|\zeta_i|+\dfrac{\mathbf{\zeta}^\top}{\sqrt{\mathbf{\xi}^\top \mathcal{L} \mathbf{\xi}}}\left(\dfrac{\mathbf{r}\sin\left(\mathbf{\delta}+\gamma_\mathrm{T}\mathbf{1}_n-\mathbf{\theta}\right)}{\left(\mathbf{v}^2-v_\mathrm{T}^2\mathbf{1}_n\right)\cos\mathbf{\delta}}\mathbf{\Xi}\right)\right].\label{eq:Vdotstep1}
	\end{align}
	It transpires from \Cref{lem:norm} that $\|\mathbf{\zeta}\|_1 = \sum_{i=1}^{n}|\zeta_i|$, $\|\mathbf{\zeta}\|_2 = \sqrt{\mathbf{\zeta}^\top\mathbf{\zeta}} = \sqrt{\mathbf{\xi}^\top \mathcal{L}^2 \mathbf{\xi}}$, and $\|\mathbf{\zeta}\|_1 \geq \|\mathbf{\zeta}\|_2$. Further, it readily follows from the results in \Cref{lem:ac} that $\mathbf{\xi}^\top \mathcal{L}^2 \mathbf{\xi} \geq \lambda_2(\mathcal{L})\mathbf{\xi}^\top \mathcal{L} \mathbf{\xi}$. Knowing that $\Omega(|\zeta_i|)$ is a predefined-time consensus function and letting $\mathscr{B}=\argmin_{i}\mathscr{B}_i$, the results in \Cref{lem:omega} and the above observations lead us to simplify \eqref{eq:Vdotstep1}  as
	\begin{align}
		\dot{\mathscr{V}}_1 \leq& -\lambda_2(\mathcal{L})\beta(n)^{2-d}\mathscr{V}_1^{-1}\mathscr{B}\sum_{i=1}^{n}\beta(n)^d\Omega(|\zeta_i|)  - \beta(n) \lambda_2^{\frac{3}{2}}(\mathcal{L})\left[\dfrac{ \|\mathbf{\zeta}\|_1}{\|\mathbf{\zeta}\|_2}\epsilon-\dfrac{\mathbf{\zeta}^\top }{\|\mathbf{\zeta}\|_2 }\left(\dfrac{\mathbf{r}\sin\left(\mathbf{\delta}+\gamma_\mathrm{T}\mathbf{1}_n-\mathbf{\theta}\right)}{\left(\mathbf{v}^2-v_\mathrm{T}^2\mathbf{1}_n\right)\cos\mathbf{\delta}}\mathbf{\Xi}\right)\right]\nonumber\\
		\leq& -\lambda_2(\mathcal{L})\beta(n)^{2-d}\mathscr{V}_1^{-1}\mathscr{B}\hat{\Omega}\left(\beta(n)\|\mathbf{\zeta}\|_2\right) - \beta(n) \lambda_2^{\frac{3}{2}}(\mathcal{L})\left[\epsilon-\argmax_{i}\left(\dfrac{r_i}{v_i^2-v_\mathrm{T}^2}\argmin_{i}\Xi_{i_{\max}}\right)\right] \nonumber\\
		\leq& -\lambda_2(\mathcal{L})\beta(n)^{2-d}\mathscr{V}_1^{-1}\mathscr{B}\hat{\Omega}\left(\beta(n)\sqrt{\mathbf{\xi}^\top\mathcal{L}^2\mathbf{\xi}}\right) - \beta(n) \lambda_2^{\frac{3}{2}}(\mathcal{L})\left[\epsilon-\argmax_{i}\left(\dfrac{r_i}{v_i^2-v_\mathrm{T}^2}\argmin_{i}\Xi_{i_{\max}}\right)\right]\nonumber\displaybreak\\
		\leq& -\lambda_2(\mathcal{L})\beta(n)^{2-d}\mathscr{V}_1^{-1}\mathscr{B}\hat{\Omega}\left(\beta(n)\sqrt{\lambda_2(\mathcal{L})}\sqrt{\mathbf{\xi}^\top\mathcal{L}\mathbf{\xi}}\right)  \nonumber\\
		&-\beta(n) \lambda_2^{\frac{3}{2}}(\mathcal{L})\left[\epsilon-\argmax_{i}\left(\dfrac{r_i}{v_i^2-v_\mathrm{T}^2}\argmin_{i}\Xi_{i_{\max}}\right)\right].
	\end{align}	
	Using the results in \Cref{lem:omega} in the above expression, we can further simplify $\dot{\mathscr{V}}_1$ as
	\begin{align}
		\dot{\mathscr{V}}_1 \leq& -\lambda_2(\mathcal{L})\beta(n)^{2-d}\mathscr{V}_1^{-1}\mathscr{B}\hat{\Omega}\left(\beta(n)\beta(n)^{-1}\mathscr{V}_1\right)  - \beta(n) \lambda_2^{\frac{3}{2}}(\mathcal{L})\left[\epsilon-\argmax_{i}\left(\dfrac{r_i}{v_i^2-v_\mathrm{T}^2}\argmin_{i}\Xi_{i_{\max}}\right)\right]\nonumber\\
		\leq& -\lambda_2(\mathcal{L})\beta(n)^{2-d}\mathscr{V}_1^{-1}\mathscr{B}\hat{\Omega}\left(\mathscr{V}_1\right)  - \beta(n) \lambda_2^{\frac{3}{2}}(\mathcal{L})\left[\epsilon-\argmax_{i}\left(\dfrac{r_i}{v_i^2-v_\mathrm{T}^2}\argmin_{i}\Xi_{i_{\max}}\right)\right]\nonumber\\
		\leq& -\lambda_2(\mathcal{L})\beta(n)^{2-d}\mathscr{V}_1^{-1}\mathscr{B}\mathscr{V}\psi\left(\mathscr{V}_1\right) - \beta(n) \lambda_2^{\frac{3}{2}}(\mathcal{L})\left[\epsilon-\argmax_{i}\left(\dfrac{r_i}{v_i^2-v_\mathrm{T}^2}\argmin_{i}\Xi_{i_{\max}}\right)\right] \nonumber\\
		\leq& -\mathscr{B}\lambda_2(\mathcal{L})\beta(n)^{2-d}\psi\left(\mathscr{V}_1\right) - \beta(n) \lambda_2^{\frac{3}{2}}(\mathcal{L})\left[\epsilon-\argmax_{i}\left(\dfrac{r_i}{v_i^2-v_\mathrm{T}^2}\argmin_{i}\Xi_{i_{\max}}\right)\right].			\label{eq:Vdotsetp2}
	\end{align}	
	If we let  $\mathscr{B} = \dfrac{1}{\lambda_2(\mathcal{L})\beta(n)^{2-d}T_s}$, $\beta(n)=\dfrac{1}{n}$, and $\epsilon$ satisfies \eqref{eq:epsilon}, then \eqref{eq:Vdotsetp2} can be written as
	\begin{align}
		\dot{\mathscr{V}}_1 \leq& -\dfrac{1}{T_s}\psi\left(\mathscr{V}_1\right) ,
	\end{align}
	and $a_i$ can be expressed in the form given by \eqref{eq:aidev}. Hence, we infer according to \Cref{lem:predeftime} that an agreement in $\xi_i$ and hence in $t_{\mathrm{go}_i}$ is achieved in a predefined-time, $T_s$, specified directly in the cooperative guidance command during design.  This concludes the proof.
\end{proof}
It is worth observing from \Cref{thm:mantargetstatic} that $a_i$ requires the knowledge of $\lambda_2(\mathcal{L})$, whose lower bound can be computed using well-known algorithms (for example, see \cite{ARAGUES20143253}).

If the target moves with a constant speed, that is, $a_\mathrm{T}=0$, then the cooperative guidance command to intercept such target can be deduced from \Cref{thm:mantargetstatic}, as discussed next.
\begin{corollary}\label{corr:mantargetstatic}
	Consider the cooperative target engagement whose kinematics is governed by \eqref{eq:enggeo}, the time-to-go expression, \eqref{eq:tgodev}, and the error in the time of simultaneous interception, as given in \eqref{eq:ei}. The cooperative guidance command for the $i$\textsuperscript{th} interceptor,
	\begin{equation}
		a_i = v_i\dot{\theta}_i + \dfrac{n^{2-d}v_i\left(v_i^2-v_\mathrm{T}^2\right)\cos^2\delta_i}{\lambda_2(\mathcal{L}(\mathcal{G}))T_s\mathscr{c}r_i^2\dot{\theta}_i}\left|\sum_{j\in\mathcal{N}_i(\mathcal{G})}\left(\xi_j-\xi_i\right)\right|^{-1} \exp\left\{\left|\sum_{j\in\mathcal{N}_i(\mathcal{G})}\left(\xi_j-\xi_i\right)\right|^{\mathscr{c}}\right\}\left|\sum_{j\in\mathcal{N}_i(\mathcal{G})}\left(\xi_j-\xi_i\right)\right|^{2-\mathscr{c}},\label{eq:aidevcvt}
	\end{equation}
	where $\mathscr{c}\in(0,1]$, $d\geq 1$, $\mathcal{G}$ is a connected (fixed) graph and $\lambda_2(\mathcal{L}(\mathcal{G}))$ is its associated algebraic connectivity, ensures that the $i$\textsuperscript{th} interceptor converges to the deviated pursuit course within a predefined time, $T_s$, specified prior to homing. This essentially means that the interceptors' time-to-go values achieve an agreement within a predefined-time, $T_s$, eventually leading to a simultaneous interception of the constant speed target.
\end{corollary}
\begin{proof}
	We omit the detailed proof of \Cref{corr:mantargetstatic} since the proof follows exactly same as that in \Cref{thm:mantargetstatic} by letting $a_\mathrm{T}=0$ or $\Xi_i=0$ as the target does not maneuver. 
\end{proof}
In \Cref{thm:mantargetstatic} and \Cref{corr:mantargetstatic}, $a_i$ is composed of a nominal deviated pursuit term, $v_i\dot{\theta}_i$, needed to maintain the interceptors on the deviated pursuit course. The additional terms are responsible for consensus in interceptors' time-to-go, and compensate for the target's maneuver (for a maneuvering target). As the nonlinear disturbance observer yields an estimate of the target's maneuver that resembles the actual value of the same within a finite-time, the term compensating for target's maneuver vanishes. The terms responsible for consensus in interceptors' time-to-go also become zero within a predefined-time, eventually making $a_i\approxeq v_i\dot{\theta}_i$ in the endgame, which is nothing but the deviated pursuit guidance command for the $i$\textsuperscript{th} interceptor.

It is also interesting to note that for a constant speed target, the expression for the dynamics of the look angle for the $i$\textsuperscript{th} interceptor,
\begin{equation}
	\dot{\delta}_i = -\dfrac{v_i\left(v_i^2-v_\mathrm{T}^2\right)\cos^2\delta_i}{r_i^2\dot{\theta}_i}\mathscr{U}_i,
\end{equation}
becomes zero when consensus in interceptors' time-to-go is achieved. That is, $\mathscr{U}_i=0 \implies \dot{\delta}_i =0 \implies \delta_i=$ constant, implying that the swarm of interceptors simultaneously capture the constant speed target with a fixed look angle.

While the commands proposed above are good against a mobile adversary, intercepting a stationary target using the deviated pursuit scheme may lead to increasingly large value of $\theta_i$ near interception \cite{Shneydor}. Consequently, this will place a huge demand on the interceptors' lateral acceleration requirement. To design cooperative guidance commands against a stationary target, we seek a technique capable of ensuring target interception without burdening the lateral acceleration demand.

\subsection{Simultaneous Interception of a Stationary Adversary}
Designing efficient cooperative guidance laws for simultaneous target interception requires accurate control over engagement duration regardless of the target's motion. That said, the time-to-go for the cooperative engagement is crucial to the design process. For a stationary target, $v_\mathrm{T}=0$, reducing the relative velocity components in the cooperative target engagement, \eqref{eq:enggeo}, to $\dot{r}_i =- v_{i}\cos\delta_i$ and $r_i\dot{\theta}_i =- v_{i}\sin\delta_i$. Without assuming interceptors' small heading angle errors, the time-to-go for the $i$\textsuperscript{th} interceptor against a stationary target \cite{9000526} is given by
\begin{equation}\label{eq:tgost}
	t_{\mathrm{go}_i} = \dfrac{r_i}{v_i}\left(1+\dfrac{\sin^2\delta_i}{4N_i -2}\right),~N_i\geq 3, 
\end{equation}
such that $t_{\mathrm{go}_i} =0 \iff r_i=0$. This inference ultimately leads to the same speculation that an agreement in $t_{\mathrm{go}_i}$ along with making $t_{\mathrm{go}_i}\to 0$ eventually will guarantee a simultaneous interception. Therefore, we consider the same error variable, $\xi_i$, as that in \eqref{eq:ei} to proceed in this case also.

Observe that using \eqref{eq:ei}, and differentiating the expression in \eqref{eq:tgost} with respect to time yields
\begin{equation}\label{eq:eidotst}
	\dot{\xi}_i =  1 -\cos\delta_i\left(1-\dfrac{\sin^2\delta_i}{4N_i-2}\right)+\left(\dfrac{r_i\sin2\delta_i}{v_i^2\left(4N_i-2\right)}\right)a_i,
\end{equation}
evidencing that the relative degree of $\xi_i$ is one with respect to $a_i$. These inferences are similar to those in the case of a mobile target interception. Hence, we proceed in a similar manner as discussed previously. We present the cooperative guidance command, $a_i$, for the $i$\textsuperscript{th} interceptor in the following theorem without detailed proof since the essence of \Cref{thm:mantargetstatic} also remains valid in this case.
\begin{theorem}\label{thm:sttargetstatic}
	Consider the cooperative target engagement whose kinematics is governed by \eqref{eq:enggeo}, the time-to-go expression, \eqref{eq:tgost}, and the error in the time of simultaneous interception, as given in \eqref{eq:ei}. The cooperative guidance command for the $i$\textsuperscript{th} interceptor,
	\begin{align}
		a_i = & \dfrac{v_i^2\left(4N_i-2\right)}{r_i\sin2\delta_i}\left[-\dfrac{n^{2-d}}{\lambda_2(\mathcal{L}(\mathcal{G}))T_s\mathscr{c}}\left|\sum_{j\in\mathcal{N}_i(\mathcal{G})}\left(\xi_j-\xi_i\right)\right|^{-1}\right.\exp\left\{\left|\sum_{j\in\mathcal{N}_i(\mathcal{G})}\left(\xi_j-\xi_i\right)\right|^{\mathscr{c}}\right\}\left|\sum_{j\in\mathcal{N}_i(\mathcal{G})}\left(\xi_j-\xi_i\right)\right|^{2-\mathscr{c}}\nonumber\\
		&\left.-1 + \cos\delta_i \left(1-\dfrac{\sin^2\delta_i}{4N_i-2}\right)\right]	;~\mathscr{c}\in(0,1],\label{eq:aist}		
	\end{align}
	where $d\geq 1$, $\mathcal{G}$ is a connected fixed undirected graph and $\lambda_2(\mathcal{L}(\mathcal{G}))$ is its associated algebraic connectivity, ensures that the time-to-go of the $i$\textsuperscript{th} interceptor achieves consensus with that of its neighbors within a predefined time, $T_s$, specified prior to homing. This eventually leads to a simultaneous interception of the stationary target.
\end{theorem} 
\begin{proof}
	Expressing \eqref{eq:aist} as
	\begin{equation}\label{eq:aicompactst}
		a_i = \dfrac{v_i^2\left(4N_i-2\right)}{r_i\sin2\delta_i}\left[\mathscr{U}_i-1 + \cos\delta_i \left(1-\dfrac{\sin^2\delta_i}{4N_i-2}\right)\right],
	\end{equation} 
	where $\mathscr{U}_i$ is defined in \Cref{thm:mantargetstatic},	and substituting \eqref{eq:aicompactst} in \eqref{eq:eidotst}, one obtains $\dot{\xi}_i = \mathscr{U}_i$. Then, we can proceed in the same manner as done in the proof of \Cref{thm:mantargetstatic} by considering the same Lyapunov function, \eqref{eq:Vstatic}, and thereafter ignoring the term containing the target's maneuver. 
\end{proof}
It is interesting to note that on $\mathscr{C}(\mathbf{\xi})$, the look angle for the $i$\textsuperscript{th} interceptor decreases monotonically to zero after consensus in interceptors' time-to-go, and
\begin{align}
	r_i|_{\mathscr{C}(\mathbf{\xi})} =&  \dfrac{r_i(0)\left(\left\vert \cos\delta_i(0)-1\right\vert\right)^{-1/\left(2N_i-2\right)}}{\left(\left\vert \cos\delta_i^2(0)+\cos\delta_i(0)-4N_i+2\right\vert\right)^{(2N_i-1)/\left(2-2N_i\right)}}\left(\frac{\left\vert 2\cos\delta_i(0)+1-\sqrt{16N_i-7}\right\vert}{\left\vert 2\cos\delta_i(0)+1+\sqrt{16N_i-7}\right\vert}\right)^{\frac{2N_i-1}{4\left(2N_i-2\right)\left(\sqrt{16N_i-7}\right)}}\nonumber\\
	&\times \frac{\left(\left\vert \cos\delta_i-1\right\vert\right)^{1/\left(2N_i-2\right)}}{\left(\left\vert \cos\delta_i^2+\cos\delta_i-K\right\vert\right)^{(2N_i-1)/\left(2N_i-2\right)}}\left(\frac{\left\vert 2\cos\delta_i+1+\sqrt{16N_i-7}\right\vert}{\left\vert 2\cos\delta_i+1-\sqrt{16N_i-7}\right\vert}\right)^{\frac{2N_i-1}{4\left(2N_i-2\right)\left(\sqrt{16N_i-7}\right)}}.\label{eq:rsigma}
\end{align}
This can be readily verified from \eqref{eq:aist} as $\delta_i\dot{\delta}_i<0$ for all $\delta_i\in(-\pi/2,\pi/2)$ on $\mathscr{C}(\mathbf{\xi})$ and $\delta_i=0$ only at $r_i=0$, as evident from \eqref{eq:rsigma}.

\section{Cooperative Salvo when Interceptors Communicate over Switching Networks}\label{sec:switching}
Having designed the cooperative guidance commands when interceptors communicate over a fixed network, we now turn our attention to extend the proposed scheme when the interceptors' communication topology changes. Usually, sensing and communication limitations may lead to changes in the interceptors' communication topology. Guaranteeing a simultaneous target interception even when the interaction topology changes is much more challenging. 

To this aim, we now extend the proposed scheme to the case when interceptors communicate over switching networks. We begin by showing that the commands proposed in the previous section can guarantee only fixed-time consensus in time-to-go in case of switching topologies. Since the Lyapunov function candidate used in \Cref{thm:mantargetstatic,thm:sttargetstatic} is composed of the Laplacian of a fixed network, additional (non-smooth Lyapunov) analysis is needed to show that the same inferences extend to the case when interceptors communicate over switching networks. This is discussed next.
\begin{theorem}\label{thm:mantargetdynamic}
	Consider the cooperative target engagement whose kinematics is governed by \eqref{eq:enggeo}, the time-to-go expression, \eqref{eq:tgodev}, and the error in the time of simultaneous interception, as given in \eqref{eq:ei}. If the interceptors communicate over a switched dynamic network, $\mathcal{G}_{\sigma(t)}$, the cooperative guidance command for the $i$\textsuperscript{th} interceptor, 
	\begin{align}
		a_i =& v_i\dot{\theta}_i + \dfrac{\mathscr{B}_iv_i\left(v_i^2-v_\mathrm{T}^2\right)\cos^2\delta_i}{\mathscr{c}r_i^2\dot{\theta}_i}\left[\left|\sum_{j\in\mathcal{N}_i(\mathcal{G}_{\sigma(t)})}\left(\xi_j-\xi_i\right)\right|^{-1}\exp\left\{\left|\sum_{j\in\mathcal{N}_i(\mathcal{G}_{\sigma(t)})}\left(\xi_j-\xi_i\right)\right|^{\mathscr{c}}\right\}\right.\nonumber\\
		&\left.\times \left|\sum_{j\in\mathcal{N}_i(\mathcal{G}_{\sigma(t)})}\left(\xi_j-\xi_i\right)\right|^{2-\mathscr{c}}+\epsilon\right]-\dfrac{v_i\sin\left(\delta_i+\gamma_\mathrm{T}-\theta_i\right)}{r_i\dot{\theta}_i\sec\delta_i}\hat{a}_{\mathrm{T}_i};~\mathscr{B}_i>0,\label{eq:aidev2}
	\end{align}
	where $\epsilon$ satisfies \eqref{eq:epsilon}, ensures that the $i$\textsuperscript{th} interceptor converges to the deviated pursuit course within a fixed-time,
	\begin{equation}\label{eq:ftdynamic}
		T_c \leq \dfrac{1}{\mathscr{B}\lambda_2 \beta(n)^{2-d}}.
	\end{equation}
	This essentially means that the interceptors' time-to-go values achieve an agreement within a fixed-time, \eqref{eq:ftdynamic}, eventually leading to a simultaneous interception of the maneuvering target.
\end{theorem}
\begin{proof}
	Similar to \eqref{eq:aicompact}, we can write the command, \eqref{eq:aidev2}, as
	\begin{align}
		a_i =& v_i\dot{\theta}_i - \dfrac{v_i\left(v_i^2-v_\mathrm{T}^2\right)\cos^2\delta_i}{r_i^2\dot{\theta}_i}\mathscr{U}_i-\dfrac{v_i\sin\left(\delta_i+\gamma_\mathrm{T}-\theta_i\right)}{r_i\dot{\theta}_i\sec\delta_i}\hat{a}_{\mathrm{T}_i},\label{eq:aidev2compact}
	\end{align}
	where 
	\begin{align}
		\mathscr{U}_i =& -\mathscr{B}_i \zeta_i^{-1}\Omega(|\zeta_i|)-\epsilon \nonumber\\
		=& -\mathscr{B}_i \zeta_i^{-1} \left|\sum_{j\in\mathcal{N}_i(\mathcal{G}_{\sigma(t)})}\left(\xi_j-\xi_i\right)\right|^{-1} \exp\left\{\left|\sum_{j\in\mathcal{N}_i(\mathcal{G}_{\sigma(t)})}\left(\xi_j-\xi_i\right)\right|^{\mathscr{c}}\right\}\left|\sum_{j\in\mathcal{N}_i(\mathcal{G}_{\sigma(t)})}\left(\xi_j-\xi_i\right)\right|^{2-\mathscr{c}} -\epsilon.
	\end{align} 
	We recall \eqref{eq:integratorsys} from the proof of \Cref{thm:mantargetstatic} to express
	\begin{equation}\label{eq:integratorsys2}
		\dot{\xi}_i = \mathscr{U}_i -\dfrac{v_i\sin\left(\delta_i+\gamma_\mathrm{T}-\theta_i\right)}{r_i\dot{\theta}_i\sec\delta_i}\left({a}_{\mathrm{T}_i}-\hat{a}_{\mathrm{T}_i}\right).
	\end{equation}		
	Consider a Lipschitz continuous Lyapunov function candidate,
	\begin{equation}\label{eq:Vdynp1}
		\mathscr{V}_2(\mathbf{\xi})  = \argmax_{i}\xi_i - \argmin_{i} \xi_i,
	\end{equation} 
	such that $\mathscr{V}_2$ is positive definite and $\mathscr{V}_2=0 \iff \mathbf{\xi}\in\mathscr{C}(\mathbf{\xi})$.
	
	Time differentiation of \eqref{eq:Vdynp1} allows us to write
	\begin{equation}\label{eq:Vdynp1step2}
		\dot{\mathscr{V}}_2 \leq \argmax_{i}\dot{\xi}_i - \argmin_{i} \dot{\xi}_i
	\end{equation}
	since $\epsilon$ satisfies \eqref{eq:epsilon}. Now suppose $\mathcal{G}_k$ is the active topology over which interceptors are communicating. Without loss of generality, we assume that $\xi_i = \argmax_{i}\xi_i$ while $\xi_j = \argmin_{i}\xi_i$. Hence, one may express \eqref{eq:Vdynp1step2} as
	\begin{equation}\label{eq:Vdynp1step3}
		\dot{\mathscr{V}}_2 \leq -\mathscr{B}_i \zeta_i^{-1}\Omega(|\zeta_i|) + \mathscr{B}_j \zeta_j^{-1}\Omega(|\zeta_j|).
	\end{equation}
	It is worth noting that $\sign(\zeta_j)=-1$ since for some $k\in\mathcal{V}(\mathcal{G}_k)$, $\zeta_j\leq\zeta_k$ whenever $\zeta_j\neq 0$. Thus, $\mathscr{B}_j \zeta_j^{-1}\Omega(|\zeta_j|)\leq 0$. Along similar lines, it is immediate that $\mathscr{B}_i \zeta_i^{-1}\Omega(|\zeta_i|)\geq 0$. 
	
	Consequently, $\dot{\mathscr{V}}_2 \leq 0$ in \eqref{eq:Vdynp1step3} such that $\dot{\mathscr{V}}_2 = 0$ if and only if $\mathbf{\xi}\in\mathscr{C}(\mathbf{\xi})$. Thus, one may readily observe that consensus in time-to-go is achieved at least asymptotically, and this property holds for all $\mathcal{G}_k\,(k=1,2,\ldots,p)$, under arbitrary switching due to $\sigma(t)$, since graphs are connected and $\mathscr{V}_2$ is a common Lyapunov function corresponding to each topology in $\mathcal{G}_{\sigma(t)}$.
	
	However, we have shown in \eqref{eq:Vdotsetp2} in the proof of \Cref{thm:mantargetstatic} that 
	\begin{equation}\label{eq:Vdotstep2other}
		\dot{\mathscr{V}}_1\leq -\mathscr{B}\lambda_2(\mathcal{L})\beta(n)^{2-d}\psi\left(\mathscr{V}_1\right)
	\end{equation} 
	if the network is fixed and the graph is connected. It, thus, follows from \eqref{eq:Vdotstep2other} that $\mathscr{V}_1 \to 0$ within a time, $T_c$, given by \eqref{eq:ftdynamic} in the current topology, $\mathcal{G}_k$.
	
	Notice that $\mathscr{V}_1$ and $\mathscr{V}_2$ are zero only on $\mathscr{C}(\mathbf{\xi})$. Therefore, $\mathscr{V}_2$ must also become zero within the same time, $T_c$, given by \eqref{eq:ftdynamic} since in any given topology, $\mathscr{V}_2$ is always continuous and decreasing. This essentially implies that if the sum of the time intervals in which $\mathcal{G}_k$ remains active is greater than $T_c$, then $\mathscr{V}_2$ is bound to go to zero within the lowest time, $T_c$, under arbitrary switching.
	
	Further, since $T_c$ is independent of the initial engagement parameters, it is immediate that $T_c$ is a fixed-time within which interceptors agree upon their time-to-go values in the case of switching topologies. This concludes the proof.
\end{proof}
In the case of stationary target interception, we can modify \Cref{thm:sttargetstatic} for interceptors' switching topologies using the same arguments as used in \Cref{thm:mantargetdynamic}. Thus, in the next theorem, we present the essence of \Cref{thm:sttargetstatic} without proof when interceptors communicate over switching networks.
\begin{theorem}\label{thm:sttargetdynamic}
	Consider the cooperative target engagement whose kinematics is governed by \eqref{eq:enggeo}, the time-to-go expression, \eqref{eq:tgost}, and the error in the time of simultaneous interception, as given in \eqref{eq:ei}. If the interceptors communicate over a switched dynamic network, $\mathcal{G}_{\sigma(t)}$, the cooperative guidance command for the $i$\textsuperscript{th} interceptor, 
	\begin{align}
		a_{i} =& \dfrac{v_i^2\left(4N_i-2\right)}{r_i\sin2\delta_i}\left[-\dfrac{\mathscr{B}_i}{\mathscr{c}}\left|\sum_{j\in\mathcal{N}_i(\mathcal{G}_{\sigma(t)})}\left(\xi_j-\xi_i\right)\right|^{-1}\right.\exp\left\{\left|\sum_{j\in\mathcal{N}_i(\mathcal{G})}\left(\xi_j-\xi_i\right)\right|^{\mathscr{c}}\right\}\left|\sum_{j\in\mathcal{N}_i(\mathcal{G})}\left(\xi_j-\xi_i\right)\right|^{2-\mathscr{c}}\nonumber\\
		&\left.-1 + \cos\delta_i \left(1-\dfrac{\sin^2\delta_i}{4N_i-2}\right)\right]	;~\mathscr{B}_i>0,\;\mathscr{c}\in(0,1],\label{eq:aistdyn}		
	\end{align}
	ensures that the time-to-go of the $i$\textsuperscript{th} interceptor achieves consensus with that of its neighbors within a fixed-time, $T_c$, given in \eqref{eq:ftdynamic}. This eventually leads to a simultaneous interception of the stationary target.
\end{theorem}
To ensure a predefined-time consensus in time-to-go for a simultaneous target interception irrespective of the interceptors' interaction topology, we shall propose a different strategy in the coming subsections. Our motivation to do so is to prevent overestimation of the controller gains due to a conservative bound on $T_c$ \cite{https://doi.org/10.1002/rnc.4715}, make the design independent of the initial engagement parameters, and to allow a systematic, yet simple way to exercise control over the time of consensus in $t_{\mathrm{go}_i}$ directly via cooperative guidance commands.

\subsection{Simultaneous Interception of a Mobile Adversary}
Similar to the previous design, we shall use deviated pursuit guidance when the target is mobile. The main results for this part are now summarized in the following theorem and corollary.
\begin{theorem}\label{thm:mantargetdynamic2}
	Consider the cooperative target engagement whose kinematics is governed by \eqref{eq:enggeo}, the time-to-go expression, \eqref{eq:tgodev}, and the error in the time of simultaneous interception, as given in \eqref{eq:ei}. If the interceptors communicate over a switched dynamic network, $\mathcal{G}_{\sigma(t)}$, the cooperative guidance command for the $i$\textsuperscript{th} interceptor, 
	\begin{align}
		a_i =& v_i\dot{\theta}_i + \dfrac{\mathscr{P}_iv_i\left(v_i^2-v_\mathrm{T}^2\right)\cos^2\delta_i}{r_i^2\dot{\theta}_i}\sum_{j\in\mathcal{N}_i(\mathcal{G}_{\sigma(t)})}\left[\left(\mathscr{M}\left|\left(\xi_j-\xi_i\right)\right|^\mathfrak{m}+\mathscr{N}\left|\left(\xi_j-\xi_i\right)\right|^\mathfrak{n}\right)^k + \mu\right]\sign\left(\xi_j-\xi_i\right)\nonumber\\
		&-\dfrac{v_i\sin\left(\delta_i+\gamma_\mathrm{T}-\theta_i\right)}{r_i\dot{\theta}_i\sec\delta_i}\hat{a}_{\mathrm{T}_i},\label{eq:aidevdyn}
	\end{align}
	such that
	\begin{equation}\label{eq:PiconditionTs}
		\mathscr{P}_i\geq \dfrac{\Gamma\left(\dfrac{1-k\mathfrak{m}}{\mathfrak{n}-\mathfrak{m}}\right)\Gamma\left(\dfrac{k\mathfrak{n}-1}{\mathfrak{n}-\mathfrak{m}}\right)}{T_s\mathscr{M}^k \Gamma(k)\left(\mathfrak{n}-\mathfrak{m}\right)}\left(\dfrac{\mathscr{M}}{\mathscr{N}}\right)^{\frac{1-k\mathfrak{m}}{\mathfrak{n}-\mathfrak{m}}}\left(\dfrac{\munderbar{E}}{\munderbar{\lambda}_2}\right);~\munderbar{E}=\argmin_{k}\left|\mathcal{E}(\mathcal{G}_k)\right|,\,\munderbar{\lambda}_2 = \argmin_{k}\lambda_2(\mathcal{L}(\mathcal{G}_k)),
	\end{equation}
	and 
	\begin{equation}\label{eq:mu}
		\mu > \dfrac{1}{\argmin_{i}\mathscr{P}_i\sqrt{\munderbar{\lambda}}_2}\argmax_{i}\left[\left(\dfrac{r_i}{v_i^2-v_\mathrm{T}^2}\right)\argmin_{i}\Xi_{i_{\max}}\right]
	\end{equation}
	where $\mathscr{M},\mathscr{N},\mathfrak{m},\mathfrak{n},k,\Gamma(\cdot)$ are defined in \Cref{lem:pdt2}, ensures that the $i$\textsuperscript{th} interceptor converges to the deviated pursuit course within a predefined time, $T_s$, specified prior to homing. This essentially means that the interceptors' time-to-go values achieve an agreement within a predefined-time, $T_s$, eventually leading to a simultaneous interception of the maneuvering target.
\end{theorem}
\begin{proof}
	If we express the cooperative guidance command, \eqref{eq:aidevdyn}, as
	\begin{equation}\label{eq:aicompactdyn}
		a_i = v_i\dot{\theta}_i - \dfrac{v_i\left(v_i^2-v_\mathrm{T}^2\right)\cos^2\delta_i}{r_i^2\dot{\theta}_i}\mathscr{U}_i-\dfrac{v_i\sin\left(\delta_i+\gamma_\mathrm{T}-\theta_i\right)}{r_i\dot{\theta}_i\sec\delta_i}\hat{a}_{\mathrm{T}_i},
	\end{equation}
	where 
	\begin{equation}
		\mathscr{U}_i = -\mathscr{P}_i\sum_{j\in\mathcal{N}_i(\mathcal{G}_{\sigma(t)})}\left[\left(\mathscr{M}\left|\left(\xi_j-\xi_i\right)\right|^\mathfrak{m}+\mathscr{N}\left|\left(\xi_j-\xi_i\right)\right|^\mathfrak{n}\right)^k + \mu\right]\sign\left(\xi_j-\xi_i\right),
	\end{equation} 
	then one can write in vector form,
	\begin{equation}\label{eq:aicompactdevdyn}
		\dot{\mathbf{\xi}} = \mathscr{U} -\dfrac{\mathbf{r}\sin\left(\mathbf{\delta}+\gamma_\mathrm{T}\mathbf{1}_n-\mathbf{\theta}\right)}{\left(\mathbf{v}^2-v_\mathrm{T}^2\mathbf{1}_n\right)\cos\mathbf{\delta}}\mathbf{\Xi}.
	\end{equation}
	Letting $\sigma(t)=k$ for some $k\in[0,T_s]$, \eqref{eq:aicompactdevdyn} can also be expressed in terms of incidence matrix as
	\begin{equation}
		\dot{\mathbf{\xi}} = \mathcal{F}(\mathcal{G}_k)\varrho\left(\mathcal{F}^\top(\mathcal{G}_k)\mathbf{\xi}\right) -\dfrac{\mathbf{r}\sin\left(\mathbf{\delta}+\gamma_\mathrm{T}\mathbf{1}_n-\mathbf{\theta}\right)}{\left(\mathbf{v}^2-v_\mathrm{T}^2\mathbf{1}_n\right)\cos\mathbf{\delta}}\mathbf{\Xi},
	\end{equation}
	where, for convenience, we let $\mathfrak{f}=\mathcal{F}(\mathcal{G}_k)^\top\mathbf{\xi} = [\mathfrak{f}_1,\mathfrak{f}_2,\ldots,\mathfrak{f}_E]^\top$ such that
	\begin{equation}
		\varrho(\mathfrak{f}) = \begin{pmatrix}
			-\mathscr{P}_1 \left[\left(\mathscr{M}\left|\mathfrak{f}_1\right|^\mathfrak{m}+\mathscr{N}\left|\mathfrak{f}_1\right|^\mathfrak{n}\right)^k + \mu\right]\sign\left(\mathfrak{f}_1\right) \\
			-\mathscr{P}_2 \left[\left(\mathscr{M}\left|\mathfrak{f}_2\right|^\mathfrak{m}+\mathscr{N}\left|\mathfrak{f}_2\right|^\mathfrak{n}\right)^k + \mu\right]\sign\left(\mathfrak{f}_2\right) \\
			\vdots \\
			-\mathscr{P}_E \left[\left(\mathscr{M}\left|\mathfrak{f}_E\right|^\mathfrak{m}+\mathscr{N}\left|\mathfrak{f}_E\right|^\mathfrak{n}\right)^k + \mu\right]\sign\left(\mathfrak{f}_E\right)
		\end{pmatrix}.
	\end{equation}
	Consider a Lyapunov function candidate,
	\begin{equation}\label{eq:V3}
		\mathscr{V}_3 (\mathbf{\xi})= \dfrac{\sqrt{\munderbar{\lambda}_2}}{\munderbar{E}}\sqrt{\mathbf{\xi}^\top\mathbf{\xi}}
	\end{equation}
	such that $\mathscr{V}_3(0)\iff \mathbf{\xi}\in\mathscr{C}(\mathbf{\xi})$. To prove that the command, \eqref{eq:aidevdyn}, leads to simultaneous target interception by means of predefined-time consensus in interceptors' time-to-go when their interaction topology switches, we need to show that \eqref{eq:V3} is a common Lyapunov function corresponding to each topology in $\mathcal{G}_{\sigma(t)}$.
	
	One may observe that time differentiation of \eqref{eq:V3} yields
	\begin{align}
		\dot{\mathscr{V}}_3 =& \dfrac{\sqrt{\munderbar{\lambda}_2}}{\munderbar{E}\sqrt{\mathbf{\xi}^\top\mathbf{\xi}}}\mathbf{\xi}^\top\dot{\mathbf{\xi}} = \dfrac{\sqrt{\munderbar{\lambda}_2}}{\munderbar{E}\sqrt{\mathbf{\xi}^\top\mathbf{\xi}}}\mathbf{\xi}^\top\left[\mathcal{F}(\mathcal{G}_k)\varrho\left(\mathcal{F}^\top(\mathcal{G}_k)\mathbf{\xi}\right) -\dfrac{\mathbf{r}\sin\left(\mathbf{\delta}+\gamma_\mathrm{T}\mathbf{1}_n-\mathbf{\theta}\right)}{\left(\mathbf{v}^2-v_\mathrm{T}^2\mathbf{1}_n\right)\cos\mathbf{\delta}}\mathbf{\Xi}\right]\nonumber\\
		=& \dfrac{\sqrt{\munderbar{\lambda}_2}}{\munderbar{E}\|\mathbf{\xi}\|_2}\left[\mathbf{\xi}^\top\mathcal{F}(\mathcal{G}_k)\varrho\left(\mathcal{F}^\top(\mathcal{G}_k)\mathbf{\xi}\right) -\mathbf{\xi}^\top\left(\dfrac{\mathbf{r}\sin\left(\mathbf{\delta}+\gamma_\mathrm{T}\mathbf{1}_n-\mathbf{\theta}\right)}{\left(\mathbf{v}^2-v_\mathrm{T}^2\mathbf{1}_n\right)\cos\mathbf{\delta}}\mathbf{\Xi}\right)\right]\nonumber\\
		=&\dfrac{\sqrt{\munderbar{\lambda}_2}}{\munderbar{E}\|\mathbf{\xi}\|_2}\left[\mathfrak{f}^\top\varrho\left(\mathfrak{f}\right) -\mathbf{\xi}^\top\left(\dfrac{\mathbf{r}\sin\left(\mathbf{\delta}+\gamma_\mathrm{T}\mathbf{1}_n-\mathbf{\theta}\right)}{\left(\mathbf{v}^2-v_\mathrm{T}^2\mathbf{1}_n\right)\cos\mathbf{\delta}}\mathbf{\Xi}\right)\right].
	\end{align}
	On further simplification of $\dot{\mathscr{V}}_3$, one obtains 
	\begin{align}
		\dot{\mathscr{V}}_3 =&\dfrac{\sqrt{\munderbar{\lambda}_2}}{\munderbar{E}\|\mathbf{\xi}\|_2}\left[-\sum_{i=1}^{E}\mathscr{P}_i|\mathfrak{f}_i|\left(\mathscr{M}\left|\mathfrak{f}_i\right|^\mathfrak{m}+\mathscr{N}\left|\mathfrak{f}_i\right|^\mathfrak{n}\right)^k - \mu\sum_{i=1}^{E}\mathscr{P}_i|\mathfrak{f}_i| -\mathbf{\xi}^\top\left(\dfrac{\mathbf{r}\sin\left(\mathbf{\delta}+\gamma_\mathrm{T}\mathbf{1}_n-\mathbf{\theta}\right)}{\left(\mathbf{v}^2-v_\mathrm{T}^2\mathbf{1}_n\right)\cos\mathbf{\delta}}\mathbf{\Xi}\right)\right]\nonumber\\
		\leq&\dfrac{\sqrt{\munderbar{\lambda}_2}}{\munderbar{E}\|\mathbf{\xi}\|_2}\left[-\sum_{i=1}^{E}\mathscr{P}_i|\mathfrak{f}_i|\left(\mathscr{M}\left|\mathfrak{f}_i\right|^\mathfrak{m}+\mathscr{N}\left|\mathfrak{f}_i\right|^\mathfrak{n}\right)^k - \mu\sum_{i=1}^{E}\mathscr{P}_i|\mathfrak{f}_i| +\mathbf{\xi}^\top\argmax_{i}\left(\left(\dfrac{r_i}{v_i^2-v_\mathrm{T}^2}\right)\argmin_{i}\Xi_{i_{\max}}\right)\right].\label{eq:V3dotstep1}
	\end{align}
	By letting $\mathscr{P}=\argmin_{i}\mathscr{P}_i$, and using the results in \Cref{lem:jenson2}, we can further simplify \eqref{eq:V3dotstep1} as
	\begin{align}
		\dot{\mathscr{V}}_3 \leq& \dfrac{\sqrt{\munderbar{\lambda}_2}}{\munderbar{E}\|\mathbf{\xi}\|_2}\left[-\mathscr{P}\sum_{i=1}^{E}|\mathfrak{f}_i|\left(\mathscr{M}\left|\mathfrak{f}_i\right|^\mathfrak{m}+\mathscr{N}\left|\mathfrak{f}_i\right|^\mathfrak{n}\right)^k - \mu\mathscr{P}\sum_{i=1}^{E}|\mathfrak{f}_i| +\mathbf{\xi}^\top\argmax_{i}\left(\left(\dfrac{r_i}{v_i^2-v_\mathrm{T}^2}\right)\argmin_{i}\Xi_{i_{\max}}\right)\right] \nonumber\\
		\leq& \dfrac{\sqrt{\munderbar{\lambda}_2}}{\munderbar{E}\|\mathbf{\xi}\|_2}\left\{-E\mathscr{P}\left(\dfrac{1}{E}	\sum_{i=1}^{E}|\mathfrak{f}_i|\right)\left[\mathscr{M}\left(\dfrac{1}{E}	\sum_{i=1}^{E}|\mathfrak{f}_i|\right)^\mathfrak{m}+\mathscr{N}\left(\dfrac{1}{E}	\sum_{i=1}^{E}|\mathfrak{f}_i|\right)^\mathfrak{n}\right]^k\right.\nonumber\\ &\left.- \mu\mathscr{P}\sum_{i=1}^{E}|\mathfrak{f}_i| +\mathbf{\xi}^\top\argmax_{i}\left(\left(\dfrac{r_i}{v_i^2-v_\mathrm{T}^2}\right)\argmin_{i}\Xi_{i_{\max}}\right)\right\}
	\end{align}
	which, according to \Cref{lem:ac,lem:norm}, can be written as
	\begin{align}
		\dot{\mathscr{V}}_3 \leq& \dfrac{\sqrt{\munderbar{\lambda}_2}}{\munderbar{E}}\left\{-\dfrac{E\mathscr{P}}{\|\mathbf{\xi}\|_2}\left(\dfrac{1}{E}	\|\mathbf{\xi}\|_1\right)\left[\mathscr{M}\left(\dfrac{1}{E}	\|\mathbf{\xi}\|_1\right)^\mathfrak{m}+\mathscr{N}\left(\dfrac{1}{E}	\|\mathbf{\xi}\|_1\right)^\mathfrak{n}\right]^k \right.\nonumber\\ &\left.- \dfrac{\mu\mathscr{P}}{\|\mathbf{\xi}\|_2}\|\mathbf{\xi}\|_1+\argmax_{i}\left(\left(\dfrac{r_i}{v_i^2-v_\mathrm{T}^2}\right)\argmin_{i}\Xi_{i_{\max}}\right)\right\}\nonumber \\
		\leq& \dfrac{\sqrt{\munderbar{\lambda}_2}}{\munderbar{E}}\left\{-\dfrac{E\mathscr{P}}{\|\mathbf{\xi}\|_2}\left(\dfrac{1}{E}	\|\mathbf{\xi}\|_1\right)\left[\mathscr{M}\left(\dfrac{1}{E}	\|\mathbf{\xi}\|_1\right)^\mathfrak{m}+\mathscr{N}\left(\dfrac{1}{E}	\|\mathbf{\xi}\|_1\right)^\mathfrak{n}\right]^k\right.\nonumber\\
		&\left.- \dfrac{\mu\mathscr{P}}{\|\mathbf{\xi}\|_2}\|\mathbf{\xi}\|_1+\argmax_{i}\left(\left(\dfrac{r_i}{v_i^2-v_\mathrm{T}^2}\right)\argmin_{i}\Xi_{i_{\max}}\right)\right\}\nonumber\displaybreak\\
		\leq& \dfrac{\sqrt{\munderbar{\lambda}_2}}{\munderbar{E}}\left\{-\dfrac{E\mathscr{P}}{\|\mathbf{\xi}\|_2}\left(\dfrac{\sqrt{\munderbar{\lambda}}_2}{E}	\|\mathbf{\xi}\|_2\right)\left[\mathscr{M}\left(\dfrac{\sqrt{\munderbar{\lambda}}_2}{E}	\|\mathbf{\xi}\|_2\right)^\mathfrak{m}+\mathscr{N}\left(\dfrac{\sqrt{\munderbar{\lambda}}_2}{E}	\|\mathbf{\xi}\|_2\right)^\mathfrak{n}\right]^k \right.\nonumber\\
		&\left.- \left[\mu\mathscr{P}\sqrt{\munderbar{\lambda}}_2-\argmax_{i}\left(\left(\dfrac{r_i}{v_i^2-v_\mathrm{T}^2}\right)\argmin_{i}\Xi_{i_{\max}}\right)\right] \right\}\nonumber\\
		\leq& \dfrac{\sqrt{\munderbar{\lambda}_2}}{\munderbar{E}}\left\{-\mathscr{P}\sqrt{\munderbar{\lambda}}_2\left[\mathscr{M}\left(\dfrac{\munderbar{\lambda}_2}{\munderbar{E}}	\|\mathbf{\xi}\|_2\right)^\mathfrak{m}+\mathscr{N}\left(\dfrac{\munderbar{\lambda}_2}{\munderbar{E}}	\|\mathbf{\xi}\|_2\right)^\mathfrak{n}\right]^k \right.\nonumber\\
		&\left.- \left[\mu\mathscr{P}\sqrt{\munderbar{\lambda}}_2-\argmax_{i}\left(\left(\dfrac{r_i}{v_i^2-v_\mathrm{T}^2}\right)\argmin_{i}\Xi_{i_{\max}}\right)\right] \right\}.\label{eq:V3dotstep2}
	\end{align}
	If $\mu$ satisfies \eqref{eq:mu}, then \eqref{eq:V3dotstep2} reduces to
	\begin{align}
		\dot{\mathscr{V}}_3 \leq& -\dfrac{\mathscr{P}{\munderbar{\lambda}_2}}{\munderbar{E}}\left[\mathscr{M}\left(\dfrac{\munderbar{\lambda}_2}{\munderbar{E}}	\|\mathbf{\xi}\|_2\right)^\mathfrak{m}+\mathscr{N}\left(\dfrac{\munderbar{\lambda}_2}{\munderbar{E}}	\|\mathbf{\xi}\|_2\right)^\mathfrak{n}\right]^k \leq  -\dfrac{\mathscr{P}{\munderbar{\lambda}_2}}{\munderbar{E}}\left[\mathscr{M}\mathscr{V}_3^\mathfrak{m}+\mathscr{N}\mathscr{V}_3^\mathfrak{n}\right]^k\nonumber\\
		\leq&- \dfrac{\Gamma\left(\dfrac{1-k\mathfrak{m}}{\mathfrak{n}-\mathfrak{m}}\right)\Gamma\left(\dfrac{k\mathfrak{n}-1}{\mathfrak{n}-\mathfrak{m}}\right)}{T_s\mathscr{M}^k \Gamma(k)\left(\mathfrak{n}-\mathfrak{m}\right)}\left(\dfrac{\mathscr{M}}{\mathscr{N}}\right)^{\frac{1-k\mathfrak{m}}{\mathfrak{n}-\mathfrak{m}}}\left[\mathscr{M}\mathscr{V}_3^\mathfrak{m}+\mathscr{N}\mathscr{V}_3^\mathfrak{n}\right]^k
	\end{align}
	whenever \eqref{eq:PiconditionTs} holds. 
	
	The above result leads us to infer that  interceptors achieve a consensus in $\xi_i$, and hence, in their time-to-go values within a predefined-time, $T_s$, irrespective of their interaction topology since the above result holds for any $\mathcal{G}_k\in\mathcal{G}_{\sigma(t)}$. This completes the proof.
\end{proof}
If the target moves with a constant speed, then similar to \Cref{corr:mantargetstatic}, we can also deduce the cooperative guidance command for interceptors communicating over switching topologies by letting $a_\mathrm{T}=0$, as presented in the following result.

\begin{corollary}\label{corr:mantargetdynamic2}
	Consider the cooperative target engagement whose kinematics is governed by \eqref{eq:enggeo}, the time-to-go expression, \eqref{eq:tgodev}, and the error in the time of simultaneous interception, as given in \eqref{eq:ei}. If the interceptors communicate over a switched dynamic network, $\mathcal{G}_{\sigma(t)}$, the cooperative guidance command for the $i$\textsuperscript{th} interceptor, 
	\begin{align}
		a_i =& v_i\dot{\theta}_i + \dfrac{\mathscr{P}_iv_i\left(v_i^2-v_\mathrm{T}^2\right)\cos^2\delta_i}{r_i^2\dot{\theta}_i}\sum_{j\in\mathcal{N}_i(\mathcal{G}_{\sigma(t)})}\left[\left(\mathscr{M}\left|\left(\xi_j-\xi_i\right)\right|^\mathfrak{m}+\mathscr{N}\left|\left(\xi_j-\xi_i\right)\right|^\mathfrak{n}\right)^k \right]\sign\left(\xi_j-\xi_i\right),\label{eq:aidevcvtdyn}
	\end{align}
	where $\mathscr{P}_i$ is defined in \eqref{eq:PiconditionTs}, ensures that the $i$\textsuperscript{th} interceptor converges to the deviated pursuit course within a predefined time, $T_s$, specified prior to homing. This essentially means that the interceptors' time-to-go values achieve an agreement within a predefined-time, $T_s$, eventually leading to a simultaneous interception of the constant speed target.
\end{corollary}
We omit the proof of \Cref{corr:mantargetdynamic2} since it readily follows from the proof of \Cref{thm:mantargetdynamic2} by letting $a_\mathrm{T}$ and correspondingly $\mathbf{\Xi}$ to zero.

Notice that the structure of the cooperative guidance commands are similar in the case of fixed and switching topologies after consensus in time-to-go values is established. This essentially means that the inferences concerning the behavior of $\delta_i$ presented in the previous section are also valid when the interceptors communicate over switching topologies.

\subsection{Simultaneous Interception of a Stationary Adversary}
We finally turn our attention to the case of simultaneous capture of a stationary target when the interceptors exchange the information of their time-to-go over switching topologies.
\begin{theorem}\label{thm:sttargetdynamic2}
	Consider the cooperative target engagement whose kinematics is governed by \eqref{eq:enggeo}, the time-to-go expression, \eqref{eq:tgost}, and the error in the time of simultaneous interception, as given in \eqref{eq:ei}. If the interceptors communicate over a dynamic network, $\mathcal{G}_{\sigma(t)}$, the cooperative guidance command for the $i$\textsuperscript{th} interceptor, 
	\begin{align}
		a_i = & \dfrac{v_i^2\left(4N_i-2\right)}{r_i\sin2\delta_i}\left\{-\dfrac{\mathscr{P}_iv_i\left(v_i^2-v_\mathrm{T}^2\right)\cos^2\delta_i}{r_i^2\dot{\theta}_i}\sum_{j\in\mathcal{N}_i(\mathcal{G}_{\sigma(t)})}\left[\left(\mathscr{M}\left|\left(\xi_j-\xi_i\right)\right|^\mathfrak{m}+\mathscr{N}\left|\left(\xi_j-\xi_i\right)\right|^\mathfrak{n}\right)^k \right]\sign\left(\xi_j-\xi_i\right)\right.\nonumber\\
		&\left.-1 + \cos\delta_i \left(1-\dfrac{\sin^2\delta_i}{4N_i-2}\right)\right\},\label{eq:aistdyn2}		
	\end{align}
	where $\mathscr{P}_i$ is defined in \eqref{eq:PiconditionTs}, ensures that the time-to-go of the $i$\textsuperscript{th} interceptor achieves consensus with that of its neighbors within a predefined time, $T_s$, specified prior to homing. This eventually leads to a simultaneous interception of the stationary target.
\end{theorem}
\begin{proof}
	Notice that \eqref{eq:aistdyn2} can be expressed in the compact form, \eqref{eq:aicompactst}, for
	\begin{equation}\label{eq:Uistdyn}
		\mathscr{U}_i = -\mathscr{P}_i\sum_{j\in\mathcal{N}_i(\mathcal{G}_{\sigma(t)})}\left[\left(\mathscr{M}\left|\left(\xi_j-\xi_i\right)\right|^\mathfrak{m}+\mathscr{N}\left|\left(\xi_j-\xi_i\right)\right|^\mathfrak{n}\right)^k \right]\sign\left(\xi_j-\xi_i\right).
	\end{equation} 
	Thereafter, similar to the proof of \Cref{thm:sttargetstatic}, it follows that $\dot{\xi}_i = \mathscr{U}_i$ where $ \mathscr{U}_i$ is given in \eqref{eq:Uistdyn}. 
	
	On considering the Lyapunov function, \eqref{eq:V3}, and proceeding in the same manner as that in the proof of \Cref{thm:mantargetdynamic2} while ignoring the term containing the target's maneuver, it is immediate that consensus in ${\xi}_i$, and therefore in $t_{\mathrm{go}_i}$ is achieved within a predefined-time, $T_s$, facilitating a simultaneous target capture. This completes the proof.
\end{proof}

\section{Simulations and Validations}\label{sec:simulations}
In this section, we show the performance of the proposed cooperative guidance scheme subject to various engagement scenarios. In each trajectory plot that follows, square and circle markers denote the initial position of the vehicles while a `$\times$' represents interception. We have used a nonlinear finite-time disturbance observer to estimate the target's maneuver, and the estimated value of the target's maneuver is used in the cooperative guidance command. The observer gains are chosen as $\mathscr{G}_{0_i} = 0.01$, $\mathscr{G}_{1_i} = 0.05$, $\mathscr{G}_{2_i}=1.30$, $\mathscr{H}_{0_i} = 0.005$, $\mathscr{H}_{1_i}=3.25$, $\mathscr{H}_{2_i}=3.3$, and $\argmin_{i}\digamma_i=0.1$.

\subsection{Target Capture when Interceptors Communicate over a Fixed Topology}
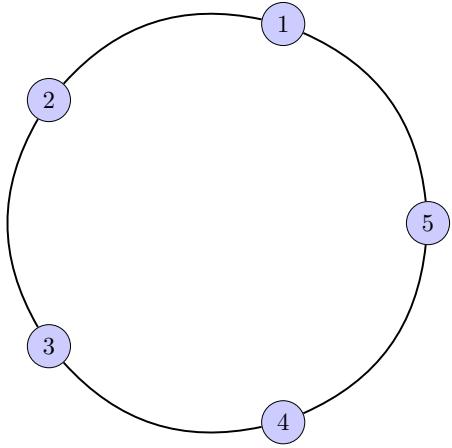
\begin{figure}[h!]
	\centering
	\resizebox{.35\columnwidth}{!}{
		\begin{tikzpicture}
			\tikzstyle{vertex} = [circle,draw=black,fill=blue!20]
			\tikzstyle{edge} = [-,>=stealth',shorten >=0pt, bend left, thick, auto]
			\tikzstyle{dashedEdge} = [->,>=stealth',shorten >=0pt, bend left, thick, dotted, auto]
			\tikzstyle{writeBelow} = [sloped, anchor=center, below]
			\tikzstyle{writeAbove} = [sloped, anchor=center, above]
			\foreach \phi in {1,...,5}{
				\node[vertex] (v_\phi) at (360/5 * \phi:3cm) {$\phi$};
				
			}
			
			\draw[edge]
			(v_2) edge node[writeAbove] {} (v_1) 
			(v_3) edge node[writeAbove] {} (v_2)
			(v_4) edge node[writeBelow] {} (v_3)
			(v_5) edge node[writeBelow] {} (v_4)
			(v_1) edge node[writeAbove] {} (v_5)
			;		
		\end{tikzpicture}%
	}
	\caption{Interceptors connected over an undirected cycle.}
	\label{fig:cycle}
\end{figure}

We now demonstrate the merits of the proposed cooperative guidance commands derived in \Cref{thm:mantargetstatic,thm:sttargetstatic} and in \Cref{corr:mantargetstatic}. We present four cases of simultaneous interception--one against a maneuvering target, two against a constant speed target, and one against a stationary target. In each case, we assume that there are five interceptors ($n=5$) and they communicate over an undirected cycle (shown in \Cref{fig:cycle}). The parameters used in simulation along with the  information of the time-to-go values are provided in \Cref{tb:param}. We assume the target to be initially located at the origin in each case.

\begin{table*}[h!]
	\centering
	\caption{Simulation parameters for the case of fixed topology (for commands \eqref{eq:aidev}, \eqref{eq:aidevcvt} and \eqref{eq:aist}).}
	\label{tb:param}
	\resizebox{\textwidth}{!}{
		\begin{tabular}{cccc}
			\toprule
			Parameter & Maneuvering Target& Constant Speed Target  & Stationary Target \\ 
			\midrule
			$v_i$ & 400 m/s & 400 m/s  & 200 m/s \\
			$v_\mathrm{T}$ & 200 m/s & 300 m/s  & 0 \\
			$a_\mathrm{T}$ & $10 \left[1 + \sin\dfrac{\pi}{10}t\right]$ m/s$^2$ & 0 &  0 \\
			$r_i(0)$ & 10 km & 10 km  & 10 km \\
			$\theta_i(0)$ & $[35^\circ,25^\circ,20^\circ,30^\circ,10^\circ]$ &  $[0^\circ,-10^\circ,-20^\circ,-165^\circ,200^\circ]$ &  $[60^\circ,150^\circ,30^\circ,-60^\circ,45^\circ]$ \\
			$\gamma_i(0)$ & $[0^\circ,10^\circ,30^\circ,10^\circ,15^\circ]$ & $[0^\circ,0^\circ,0^\circ,180^\circ,190^\circ]$  &  $[30^\circ,70^\circ,90^\circ,-30^\circ,45^\circ]$ \\
			$\gamma_\mathrm{T}$ & $120^\circ$ & $100^\circ$  & --  \\
			$N_i$ & --& -- &  3\\
			$d$ & 4 & 4 &  4 \\
			$T_s$ & 7 s& 5 s &  1 s \\
			$\mathscr{c}$ & 0.0125 &  0.0125 & 0.0125 \\
			$\lambda_2(\mathcal{L})$ & 1.3820  & 1.3820 & 1.3820 \\
			$t_{\mathrm{go}_i}(0)$ & [53.77 s, 37.50 s, 28.05 s, 41.53 s, 26.39 s]  & [49.70 s, 36.26 s, 25.87 s, 43.98 s, 43.14 s]  & [51.25 s, 54.84 s, 53.75 s, 51.25 s, 50.00 s]  \\
			$a_i^{\max}$ & 20 g & 20 g & 20 g \\
			$T_f$ & $\approxeq$ 38 s & $\approxeq$ 39 s &  $\approxeq$ 52 s \\
			\bottomrule
		\end{tabular}%
	}
\end{table*}
\begin{figure}[h!]
	\begin{subfigure}[t]{0.32\linewidth}
		\centering
		\includegraphics[width=1.1\linewidth]{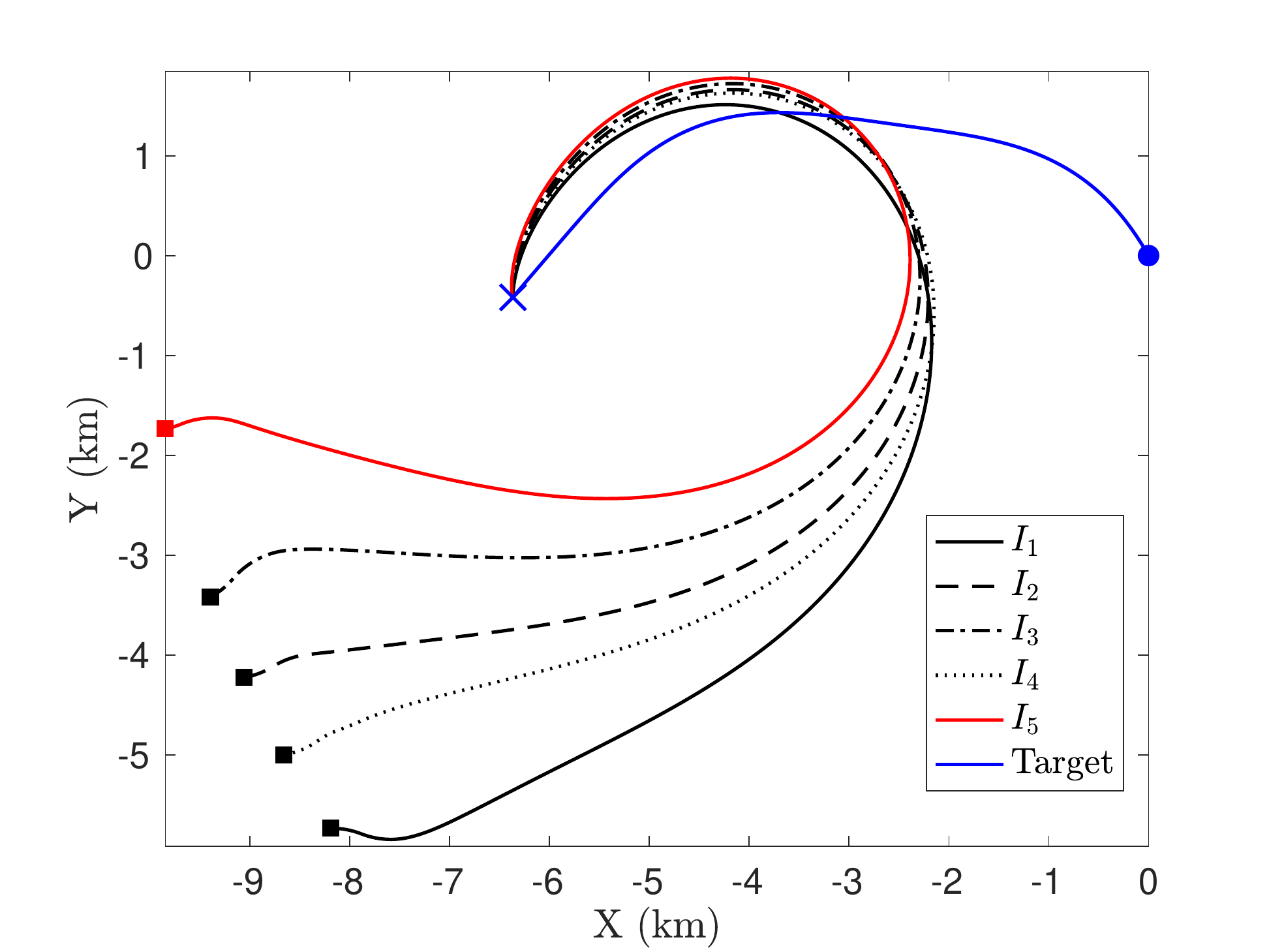}
		\caption{Trajectories.}
		\label{fig:manstrajectory}
	\end{subfigure}
	\begin{subfigure}[t]{0.32\linewidth}
		\centering
		\includegraphics[width=1.1\linewidth]{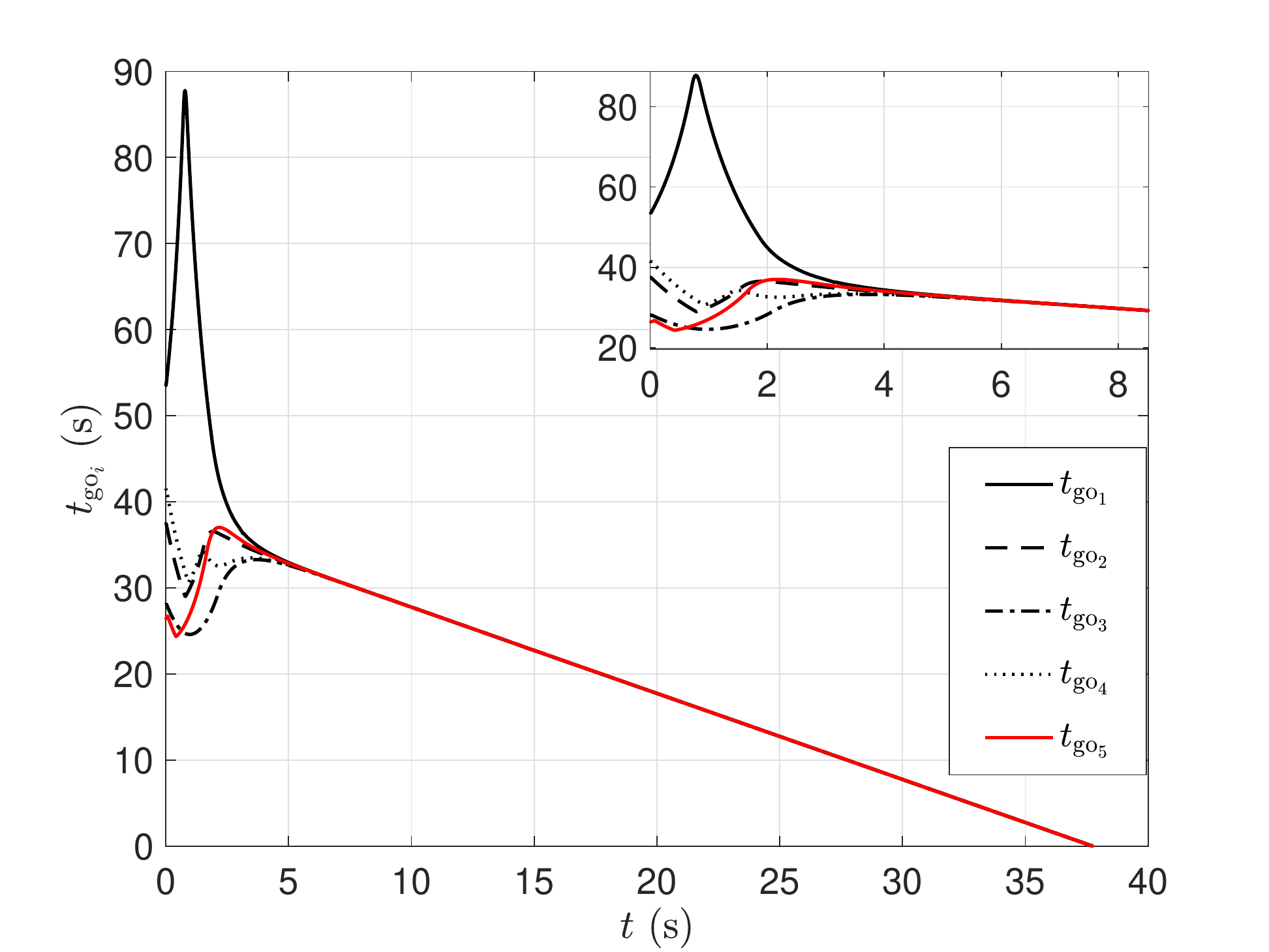}
		\caption{Time-to-go.}
		\label{fig:manstgo}
	\end{subfigure}
	\begin{subfigure}[t]{0.32\linewidth}
		\centering
		\includegraphics[width=1.1\linewidth]{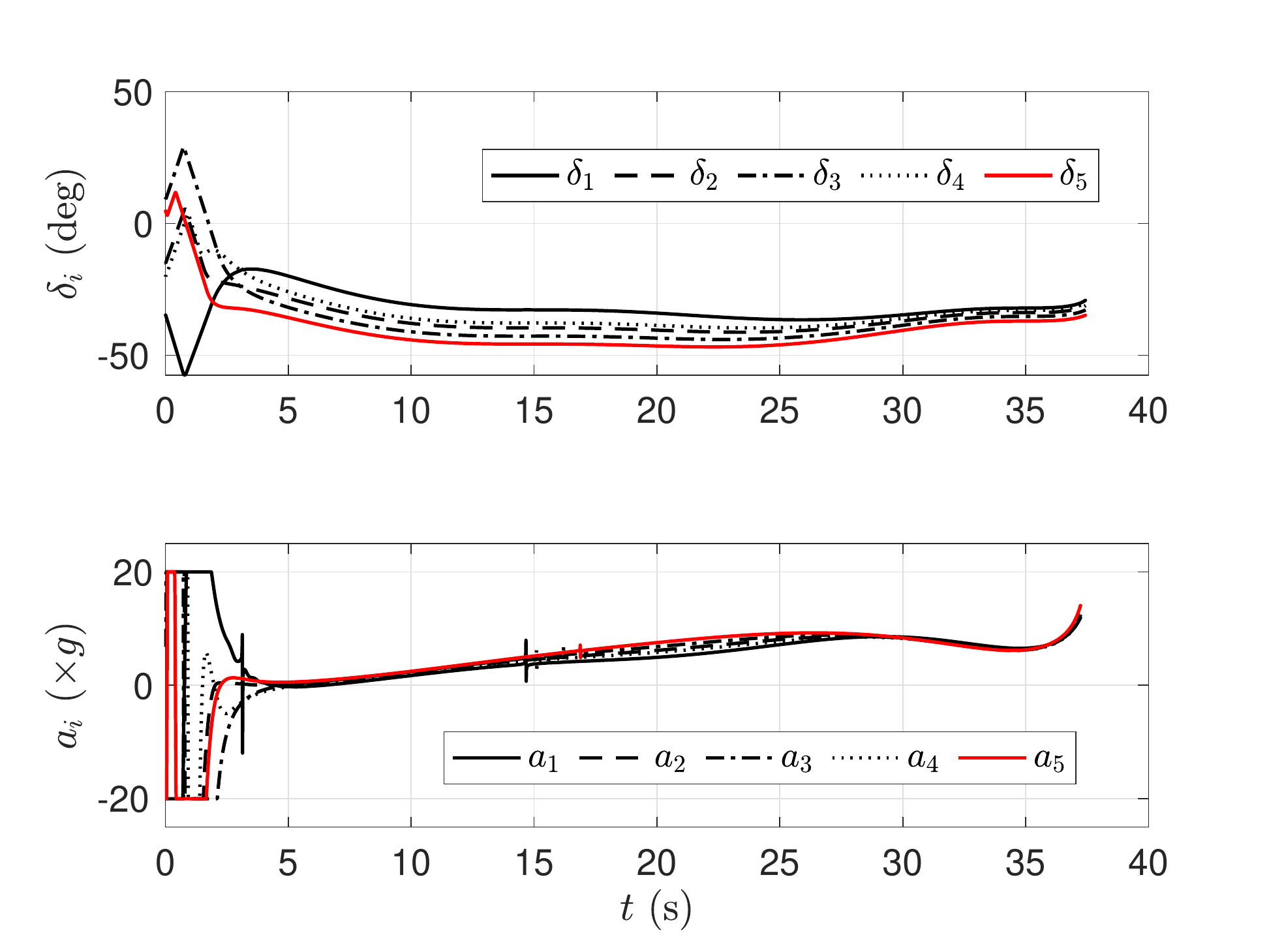}
		\caption{Look angles and lateral accelerations.}
		\label{fig:manssigma}
	\end{subfigure}
	\caption{Simultaneous interception of a maneuvering target using command \eqref{eq:aidev}.}
	\label{fig:mans}
\end{figure}	
\begin{figure}[h!]
	\begin{subfigure}[t]{0.32\linewidth}
		\centering
		\includegraphics[width=1.1\linewidth]{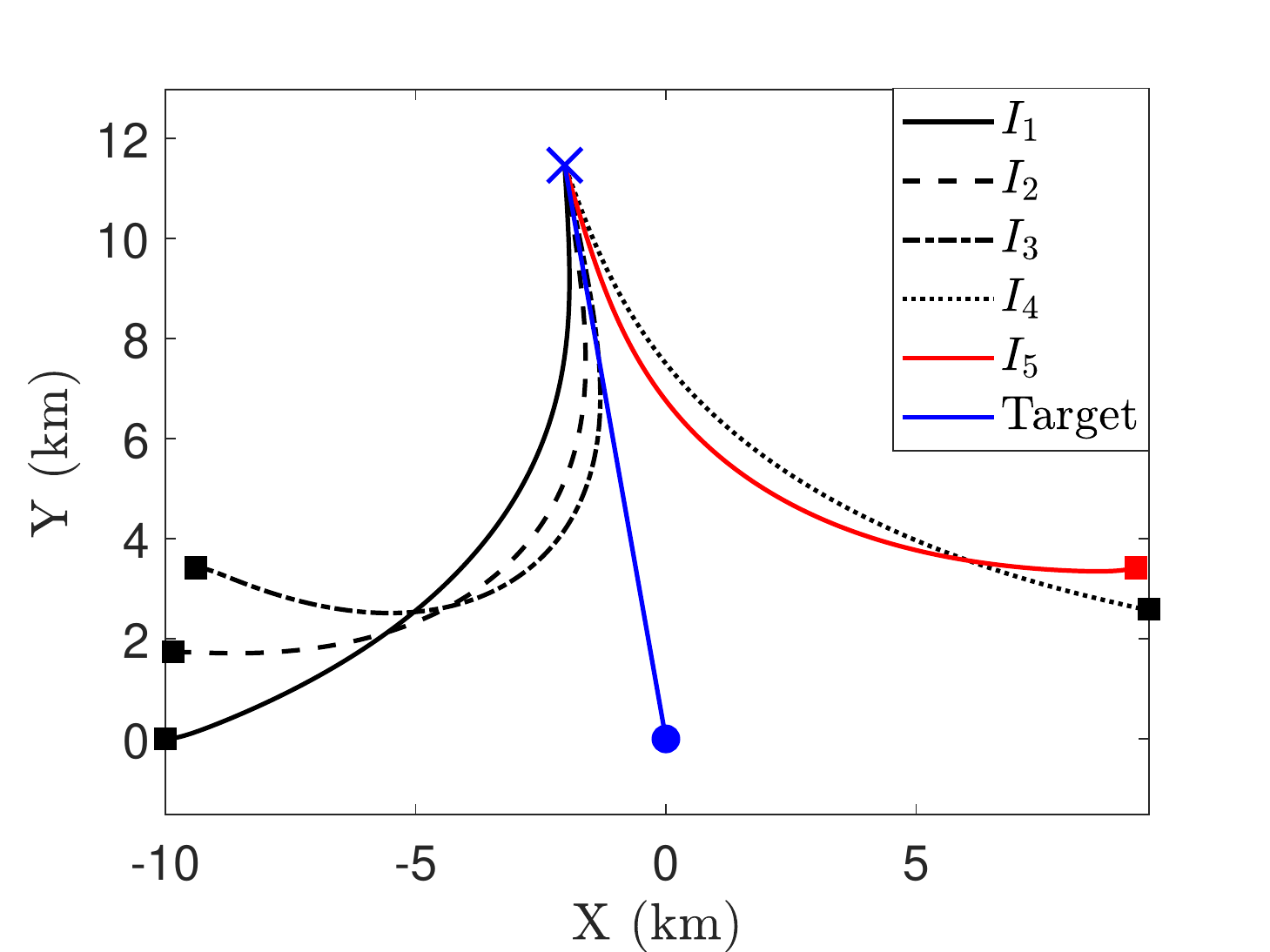}
		\caption{Trajectories.}
		\label{fig:cvtstrajectory}
	\end{subfigure}
	\begin{subfigure}[t]{0.32\linewidth}
		\centering
		\includegraphics[width=1.1\linewidth]{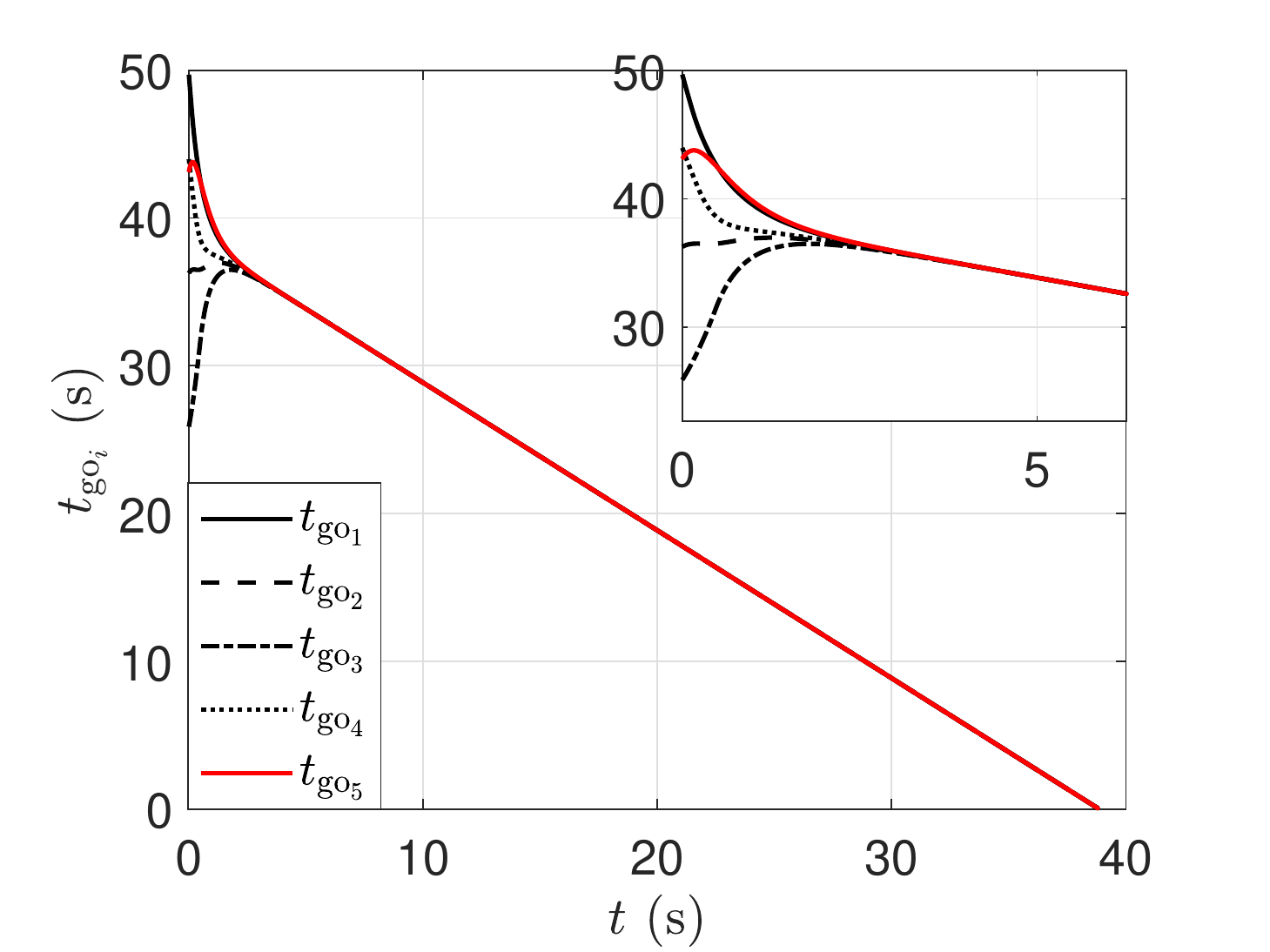}
		\caption{Time-to-go.}
		\label{fig:cvtstgo}
	\end{subfigure}
	\begin{subfigure}[t]{0.32\linewidth}
		\centering
		\includegraphics[width=1.1\linewidth]{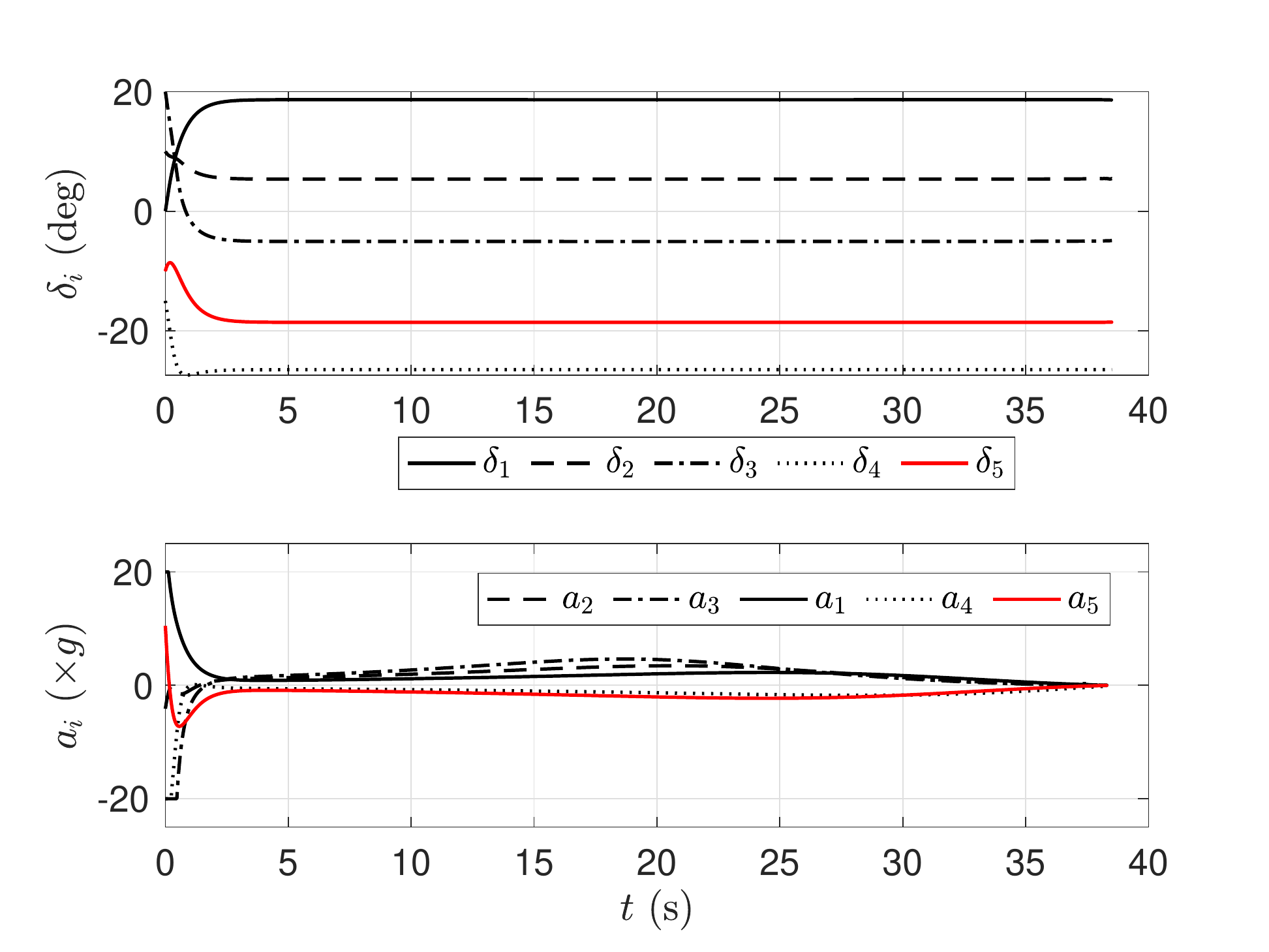}
		\caption{Look angles and lateral accelerations.}
		\label{fig:cvtssigma}
	\end{subfigure}
	\caption{Simultaneous interception of a constant speed target using command \eqref{eq:aidevcvt}.}
	\label{fig:cvts}
\end{figure}
\begin{figure}[h!]
	\begin{subfigure}[t]{0.32\linewidth}
		\centering
		\includegraphics[width=1.1\linewidth]{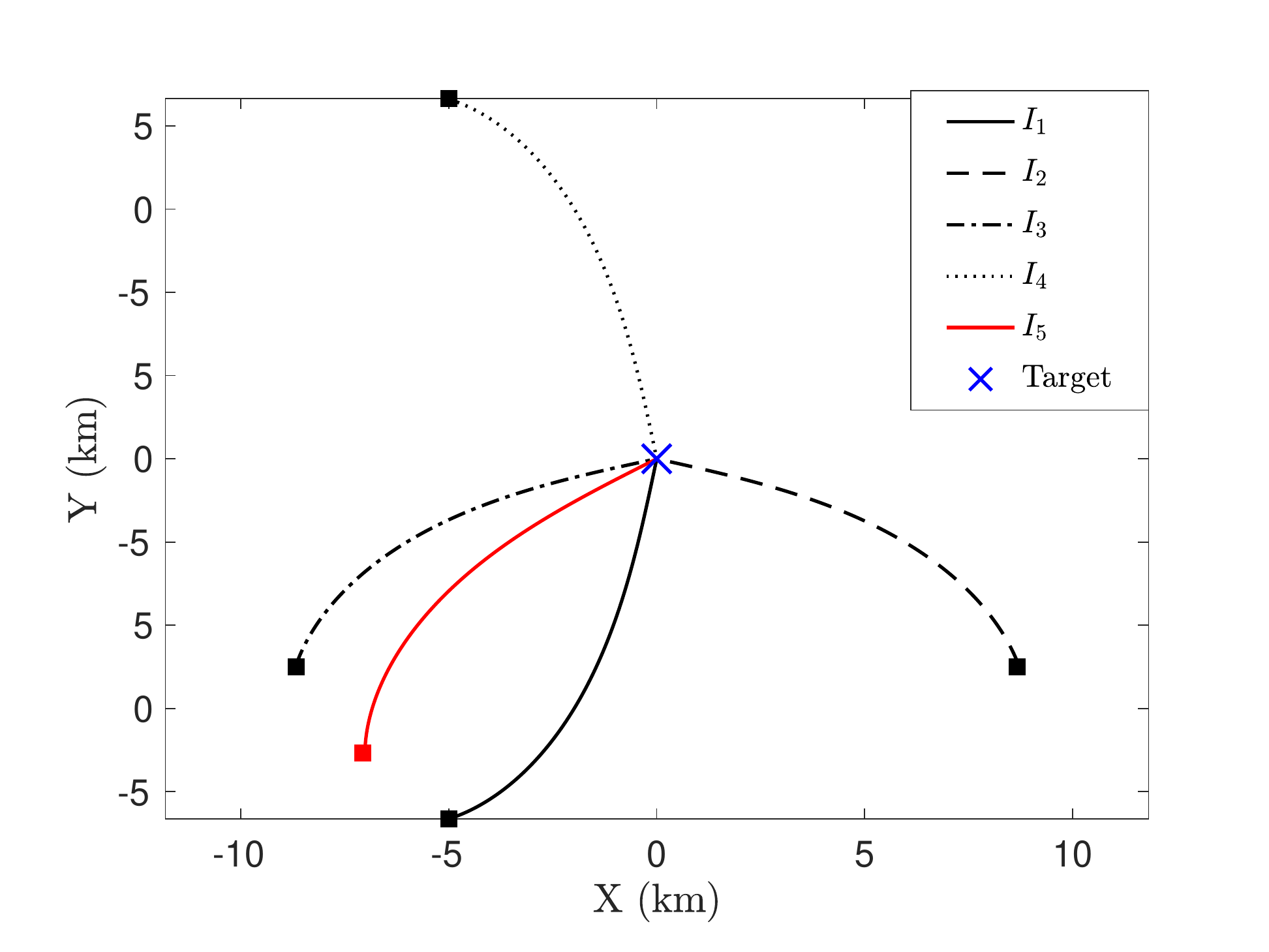}
		\caption{Trajectories.}
		\label{fig:ststrajectory}
	\end{subfigure}
	\begin{subfigure}[t]{0.32\linewidth}
		\centering
		\includegraphics[width=1.1\linewidth]{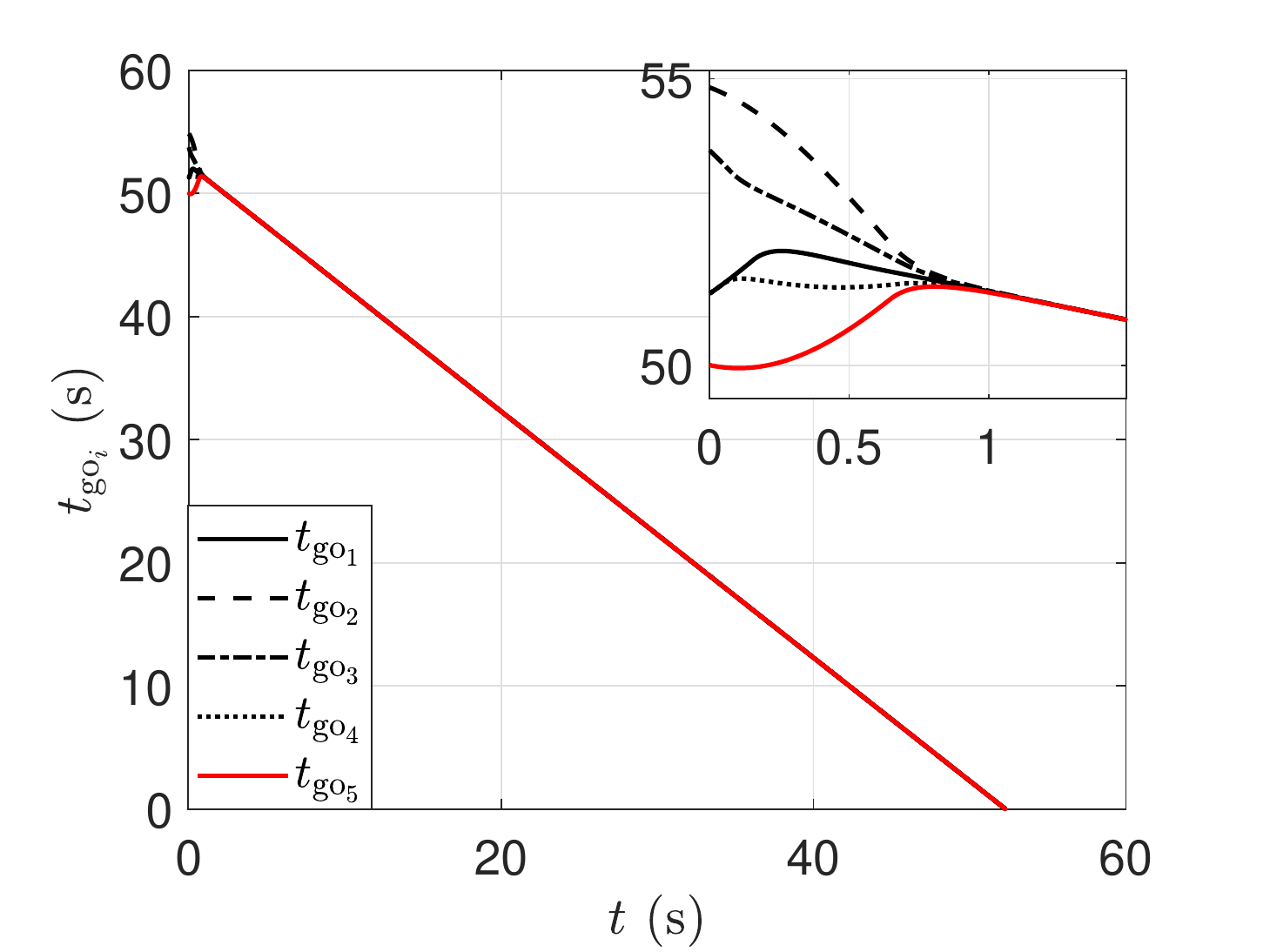}
		\caption{Time-to-go.}
		\label{fig:ststgo}
	\end{subfigure}
	\begin{subfigure}[t]{0.32\linewidth}
		\centering
		\includegraphics[width=1.1\linewidth]{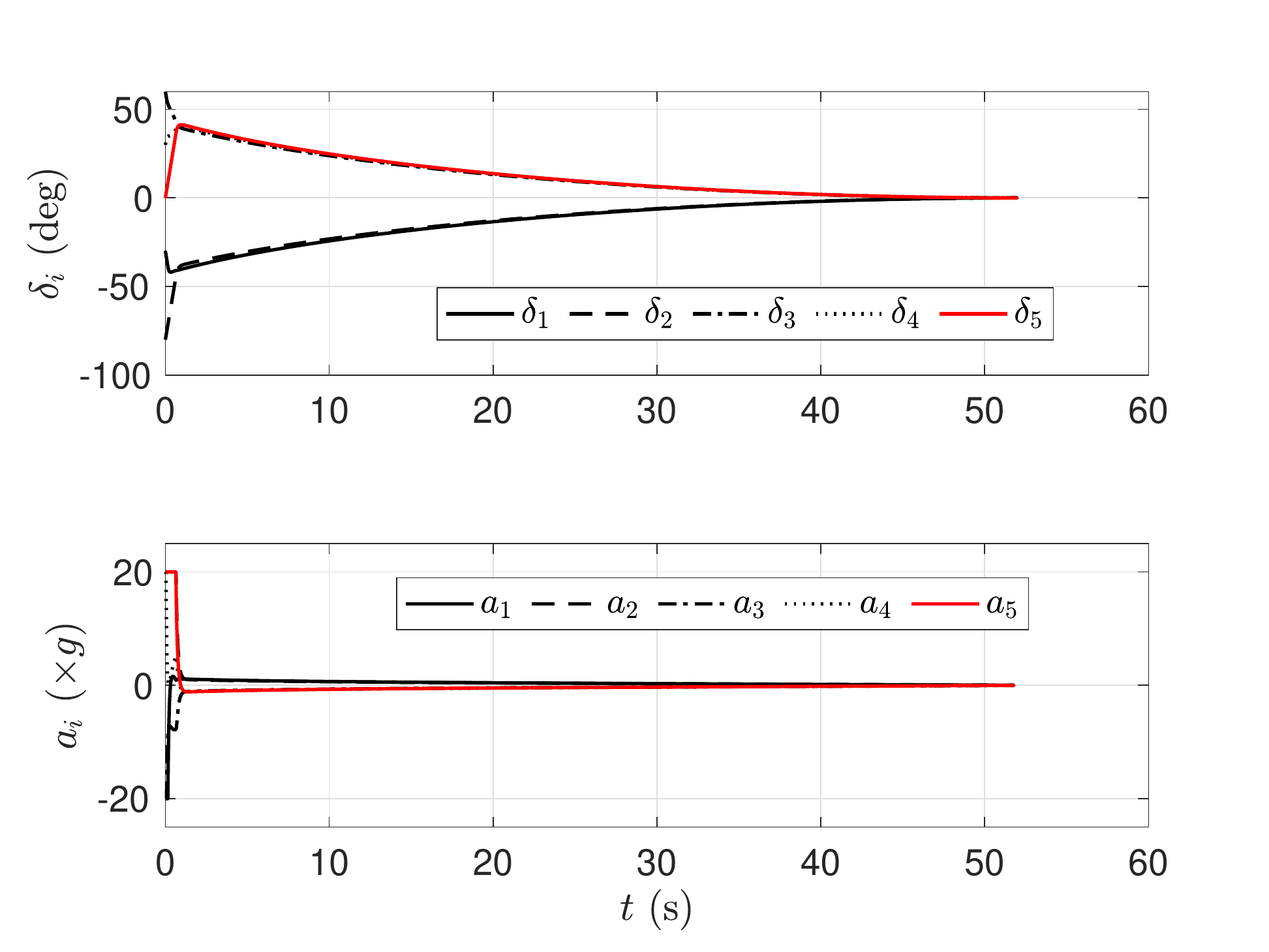}
		\caption{Look angles and lateral accelerations.}
		\label{fig:stssigma}
	\end{subfigure}
	\caption{Simultaneous interception of a stationary target using command \eqref{eq:aist}.}
	\label{fig:sts}
\end{figure}

\Cref{fig:mans} depicts simultaneous interception of a maneuvering target where interceptors start from different locations and arrive at the target in same time. It can be observed from \Cref{fig:manstrajectory} that interceptors tail-chase the maneuvering target in the endgame. Consensus in their time-to-go values, as shown in \Cref{fig:manstgo}, is established in nearly 6 s, which is within the chosen predefined-time, $T_s=7$ s. During the transient phase, the target's maneuver is estimated and interceptors adjust their trajectories accordingly. This causes the  lateral acceleration demand to become high initially (shown in \Cref{fig:manssigma}), but then the demand reduces after time-to-go values come in agreement. It is also observed from \Cref{fig:manssigma} that initially, the look angles vary to place the interceptors on the suitable collision courses, and then the variation in the magnitudes of the look angles becomes less as the engagement proceeds.

Simultaneous interception of a constant speed target using deviated pursuit, under the action of the cooperative guidance command, \eqref{eq:aidevcvt}, is shown in \Cref{fig:cvts}. Similar to the case of maneuvering target interception, the interceptors {tail-chase} the constant speed target to simultaneously capture it (see \Cref{fig:cvtstrajectory}). Although the predefined upper bound on the consensus time is chosen as 5 s, interceptors agree upon a common time-to-go value slightly before that time, as evident from \Cref{fig:cvtstgo}. We also observe from \Cref{fig:cvtssigma} that compared to \Cref{fig:manssigma}, some interceptors in the swarm require lesser control effort to achieve the objective of simultaneous interception. This may be attributed to the exact nature of the expression, \eqref{eq:tgodev}, when the target does not maneuver. We also observe that the look angles become fixed once consensus in time-to-go is achieved (see \Cref{fig:cvtssigma}), and the interceptors capture the constant speed target simultaneously with constant look angles. This is different from \Cref{fig:mans} where the magnitudes of the look angles exhibit a tendency of slow variation in the endgame but do not become constant.

\Cref{fig:sts} depicts simultaneous interception of a stationary target where interceptors start from different locations and arrive at the target in same time. It can be observed from \Cref{fig:ststrajectory} that after a short while, interceptors exhibit similar trajectories. This is expected since after consensus in their time-to-go values within a predefined-time (chosen as 1 s, shown in \Cref{fig:ststgo}), each interceptor needs same time to reach the target's location. Although $T_s$ is chosen as 1 s, the interceptors' time-to-go values achieve consensus around 0.7 s, which is quite early in the engagement. This allows the interceptors sufficient time to alter their trajectories for a simultaneous target interception. To place the interceptors on the requisite collision paths, the lateral acceleration demand is high in the transient phase (shown in \Cref{fig:stssigma}), which subsides to zero slowly once the time-to-go values achieve consensus. As mentioned earlier, the look angles first increase to a certain value to allow for trajectory correction, and then they decrease monotonically to zero after consensus in time-to-go, as shown in \Cref{fig:stssigma}. It is also worth noting that the look angle for the $i$\textsuperscript{th} interceptor becomes zero only at the time of interception, as discussed in \eqref{eq:rsigma}.  One may also observe that all the relevant variables agree within 1 s, which is the predefined-time, and exhibit similar behavior thereafter.

\subsection{Target Capture Independent of Interceptors Communication Topology}
The results in the previous part were obtained assuming that the interceptors' interaction topology remains fixed. We now concentrate our focus on scenarios where the interceptors' interaction may change. We assume that at any given point of time during engagement, interceptors may communicate over one of the three possible topologies. The active topology is decided by an arbitrary switching signal, $\sigma(t)$, that can take discrete values from the set $\{1,2,3\}$. Corresponding to the state of the switching signal, a particular topology is active, as illustrated in \Cref{fig:switchingtopologies}.
\begin{figure}[h!]
	\begin{subfigure}[t]{0.32\linewidth}
		\centering
		\resizebox{\linewidth}{!}{
			\begin{tikzpicture}
				\tikzstyle{vertex} = [circle,draw=black,fill=blue!20]
				\tikzstyle{edge} = [-,>=stealth',shorten >=0pt,  thick, auto]
				\tikzstyle{dashedEdge} = [->,>=stealth',shorten >=0pt, bend left, thick, dotted, auto]
				\tikzstyle{writeBelow} = [sloped, anchor=center, below]
				\tikzstyle{writeAbove} = [sloped, anchor=center, above]
				\foreach \phi in {1,...,5}{
					\node[vertex] (v_\phi) at (360/5 * \phi:3cm) {$\phi$};
					
				}
				
				\draw[edge]
				(v_1) edge node[writeAbove] {} (v_3) 
				(v_1) edge node[writeAbove] {} (v_4)
				(v_2) edge node[writeBelow] {} (v_5)
				(v_3) edge node[writeBelow] {} (v_4)
				(v_3) edge node[writeAbove] {} (v_5)
				;		
			\end{tikzpicture}%
		}
		\caption{Topology 1 is active for $\sigma(t)=1$.}
		\label{fig:graph1}
	\end{subfigure}
	\begin{subfigure}[t]{0.32\linewidth}
		\centering
		\resizebox{\linewidth}{!}{
			\begin{tikzpicture}
				\tikzstyle{vertex} = [circle,draw=black,fill=blue!20]
				\tikzstyle{edge} = [-,>=stealth',shorten >=0pt,  thick, auto]
				\tikzstyle{dashedEdge} = [->,>=stealth',shorten >=0pt, bend left, thick, dotted, auto]
				\tikzstyle{writeBelow} = [sloped, anchor=center, below]
				\tikzstyle{writeAbove} = [sloped, anchor=center, above]
				\foreach \phi in {1,...,5}{
					\node[vertex] (v_\phi) at (360/5 * \phi:3cm) {$\phi$};
					
				}
				
				\draw[edge]
				(v_1) edge node[writeAbove] {} (v_2) 
				(v_1) edge node[writeAbove] {} (v_3)
				(v_1) edge node[writeAbove] {} (v_5)
				(v_2) edge node[writeBelow] {} (v_4)
				(v_2) edge node[writeBelow] {} (v_5)
				(v_3) edge node[writeAbove] {} (v_5)
				;		
			\end{tikzpicture}%
		}
		\caption{Topology 2 is active for $\sigma(t)=2$.}
		\label{fig:graph2}
	\end{subfigure}
	\begin{subfigure}[t]{0.32\linewidth}
		\centering
		\resizebox{\linewidth}{!}{
			\begin{tikzpicture}
				\tikzstyle{vertex} = [circle,draw=black,fill=blue!20]
				\tikzstyle{edge} = [-,>=stealth',shorten >=0pt, bend left, thick, auto]
				\tikzstyle{dashedEdge} = [->,>=stealth',shorten >=0pt, bend left, thick, dotted, auto]
				\tikzstyle{writeBelow} = [sloped, anchor=center, below]
				\tikzstyle{writeAbove} = [sloped, anchor=center, above]
				\foreach \phi in {1,...,5}{
					\node[vertex] (v_\phi) at (360/5 * \phi:3cm) {$\phi$};
					
				}
				
				\draw[edge]
				(v_2) edge node[writeAbove] {} (v_1) 
				(v_3) edge node[writeAbove] {} (v_2)
				(v_4) edge node[writeBelow] {} (v_3)
				(v_5) edge node[writeBelow] {} (v_4)
				(v_1) edge node[writeAbove] {} (v_5)
				;		
			\end{tikzpicture}%
		}
		\caption{Topology 3 is active for $\sigma(t)=3$.}
		\label{fig:cycle3}
	\end{subfigure}
	\caption{Interceptors' various interaction topologies.}
	\label{fig:switchingtopologies}
\end{figure}
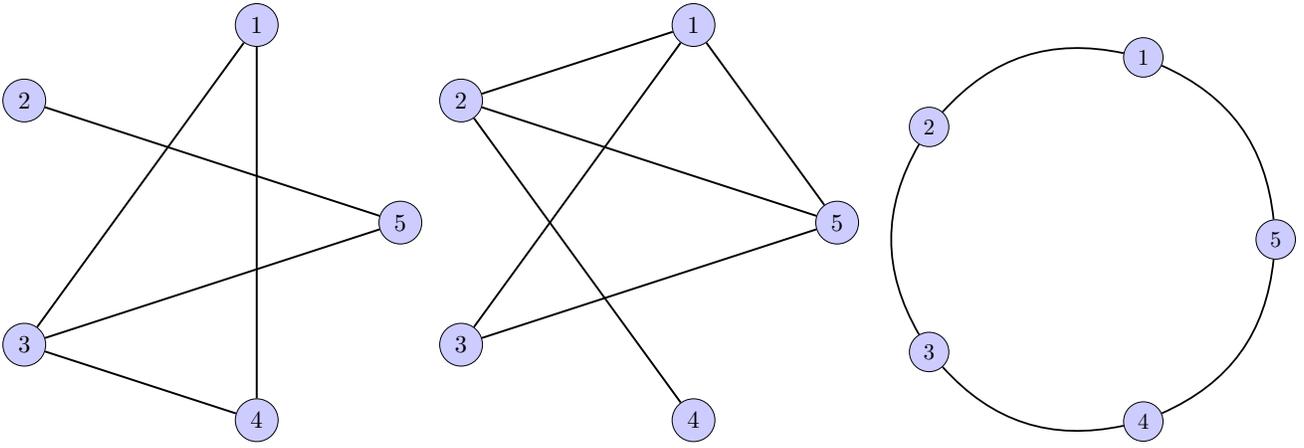
\begin{table*}[h!]
	\centering
	\caption{Simulation parameters for commands \eqref{eq:aidevdyn}, \eqref{eq:aidevcvtdyn} and \eqref{eq:aistdyn2}.}
	\label{tb:param2}
	\resizebox{\textwidth}{!}{
		\begin{tabular}{cccc}
			\toprule
			Parameter & Maneuvering Target& Constant Speed Target   & Stationary Target \\ 
			\midrule
			$v_i$ & 400 m/s & 400 m/s & 400 m/s \\
			$v_\mathrm{T}$ & 200 m/s & 300 m/s  & 0 \\
			$a_\mathrm{T}$ & $10 \left[1 + \sin\dfrac{\pi}{10}t\right]$ m/s$^2$ & 0  & 0 \\
			$r_i(0)$ & 10 km & 10 km  & 10 km \\
			$\theta_i(0)$ & $[35^\circ,25^\circ,20^\circ,30^\circ,10^\circ]$ &  $[-40^\circ,-45^\circ,-35^\circ,-30^\circ,-25^\circ]$ &  $[0^\circ,180^\circ,45^\circ,135^\circ,270^\circ]$ \\
			$\gamma_i(0)$ & $[0^\circ,10^\circ,30^\circ,10^\circ,15^\circ]$ & $[-20^\circ,-40^\circ,-30^\circ,-20^\circ,-30^\circ]$ &   $[30^\circ,150^\circ,60^\circ,120^\circ,-30^\circ]$ \\
			$\gamma_\mathrm{T}$ & $120^\circ$ & $30^\circ$  & --  \\
			$N_i$ & --& --  & 3\\
			$\mathscr{M}$ & $10^{-3}$ & 1  & 1 \\
			$\mathscr{N}$ & 0.1 & 0.5  & 5 \\
			$\mathfrak{m}$ & 0.0566 & 0.0282  & 0.1 \\	
			$\mathfrak{n}$ & 1.1 & 2  & 2 \\
			$k$ & 1.1 & 2  & 2 \\
			$T_s$ & 2 s& 5 s  & 3 s \\
			$t_{\mathrm{go}_i}(0)$ & [53.77 s, 37.50 s, 28.05 s, 41.53 s, 26.39 s]  & [35.47 s, 36.36 s, 39.18 s, 39.63 s, 44.21 s]  &  [25.62 s, 25.62 s, 25.16 s, 25.16 s, 25.87 s]  \\
			$a_i^{\max}$ & 20 g & 20 g & 20 g \\
			$T_f$ & $\approxeq$ 37 s & $\approxeq$ 41 s & $\approxeq$ 26 s \\
			\bottomrule
		\end{tabular}%
	}
\end{table*}

As rigorously proved earlier, commands presented in \Crefrange{thm:mantargetstatic}{thm:sttargetdynamic} can only guarantee fixed-time consensus in interceptors' time-to-go values under arbitrary switching topologies. However, we are interested in specifying the time of consensus as a design parameter in the cooperative guidance command itself. To this aim, we first note that for the graphs shown in \Cref{fig:switchingtopologies}, $\munderbar{E}=5$ and $\munderbar{\lambda}_2 = 0.5188$.	Rest of the parameters are presented in \Cref{tb:param2}.

\begin{figure}[h!]
	\begin{subfigure}[t]{0.32\linewidth}
		\centering
		\includegraphics[width=1.1\linewidth]{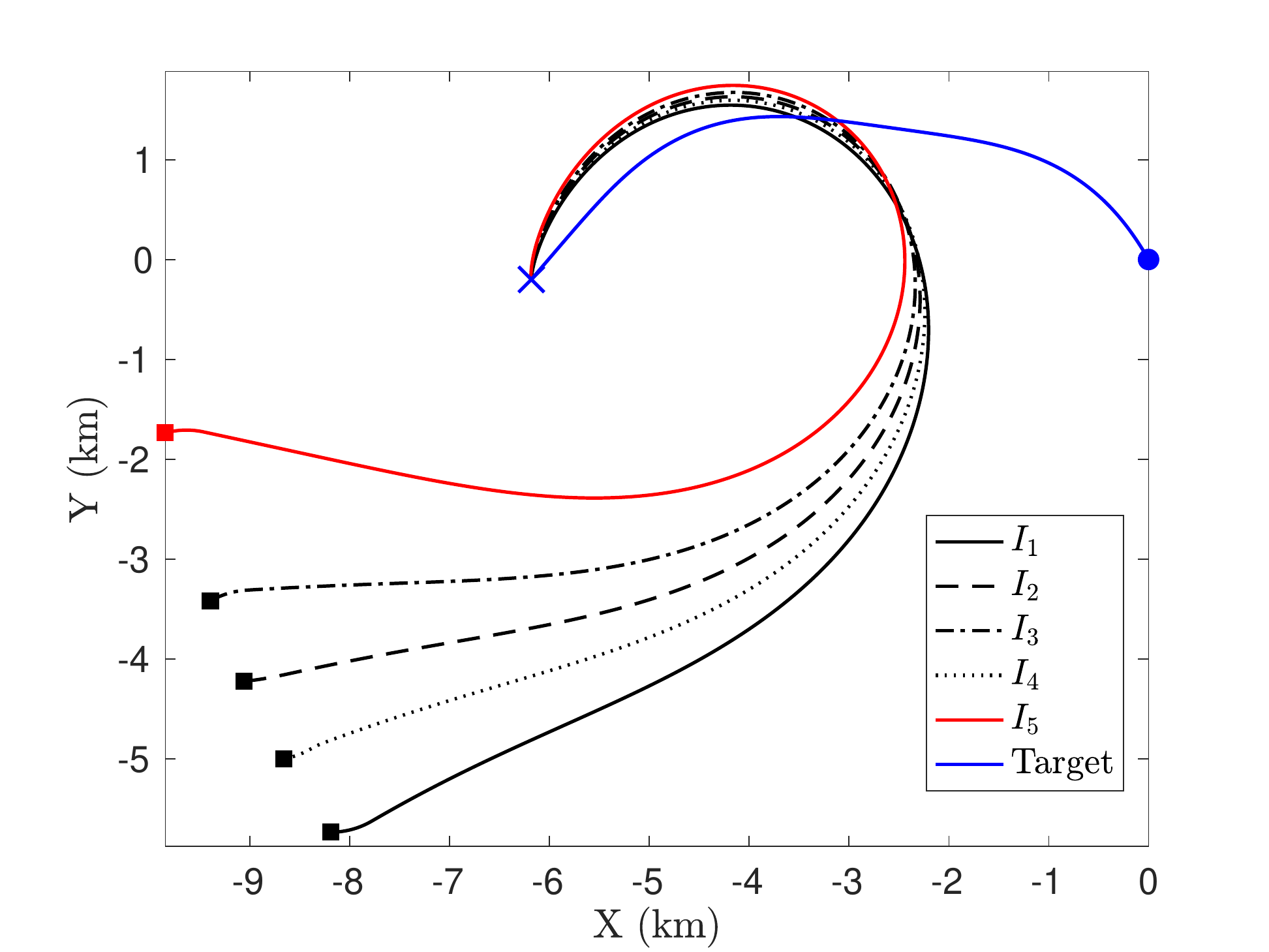}
		\caption{Trajectories.}
		\label{fig:mans2trajectory}
	\end{subfigure}
	\begin{subfigure}[t]{0.32\linewidth}
		\centering
		\includegraphics[width=1.1\linewidth]{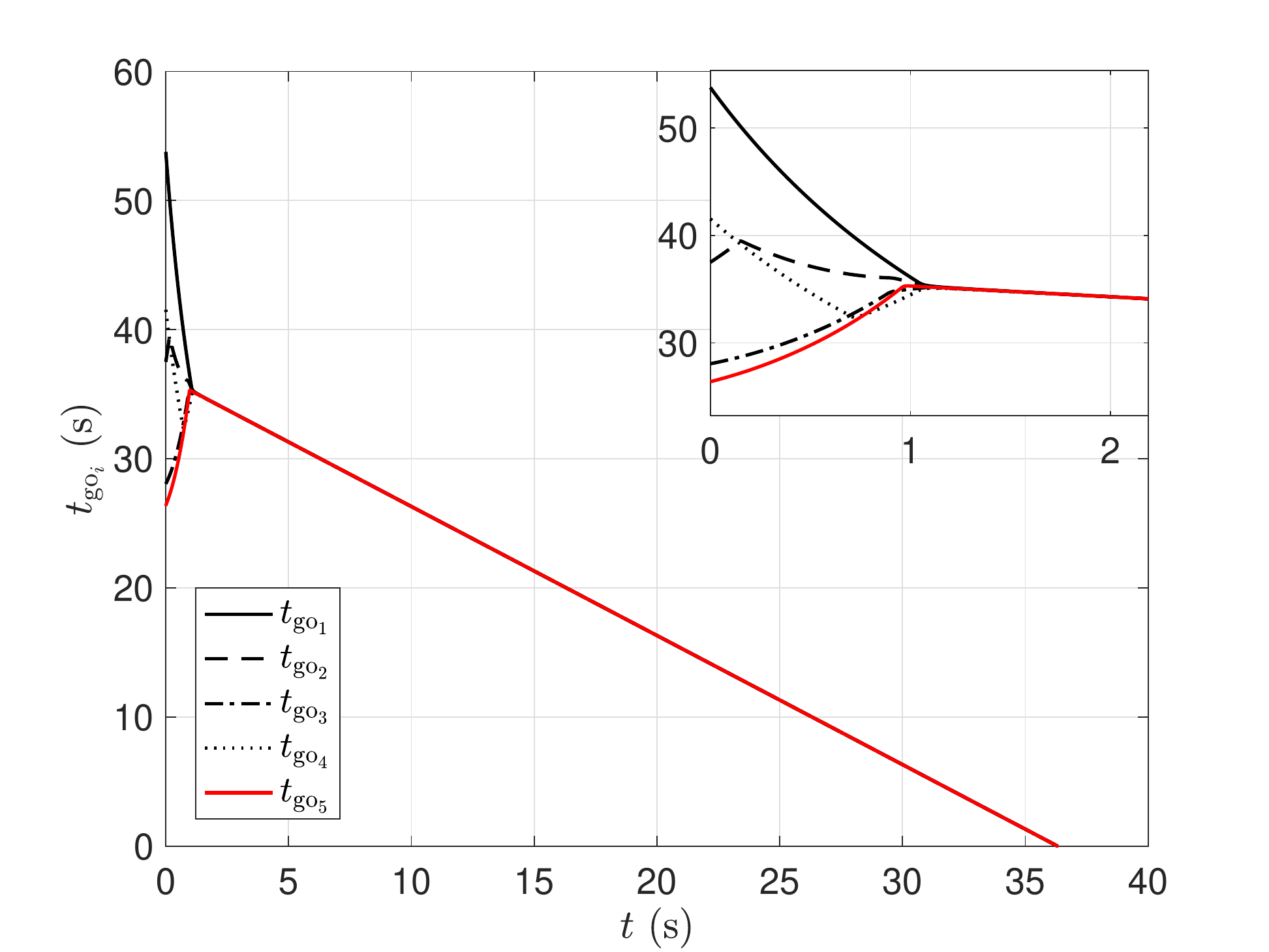}
		\caption{Time-to-go.}
		\label{fig:mans2tgo}
	\end{subfigure}
	\begin{subfigure}[t]{0.32\linewidth}
		\centering
		\includegraphics[width=1.1\linewidth]{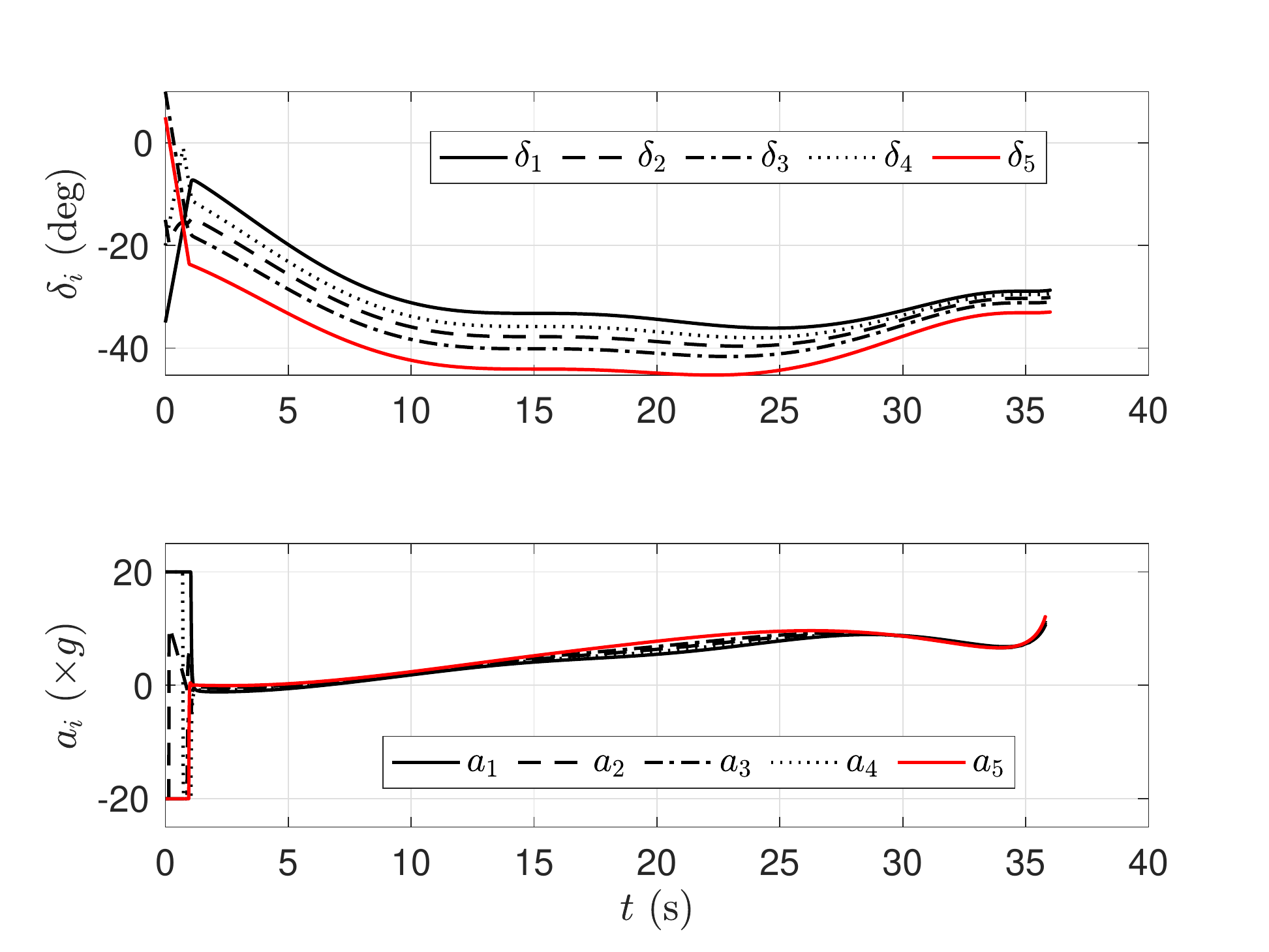}
		\caption{Look angles and lateral accelerations.}
		\label{fig:mans2sigma}
	\end{subfigure}
	\caption{Simultaneous interception of a maneuvering target using command \eqref{eq:aidevdyn} over fixed topology.}
	\label{fig:mans2}
\end{figure}
\begin{figure}[h!]
	\begin{subfigure}[t]{0.32\linewidth}
		\centering
		\includegraphics[width=1.1\linewidth]{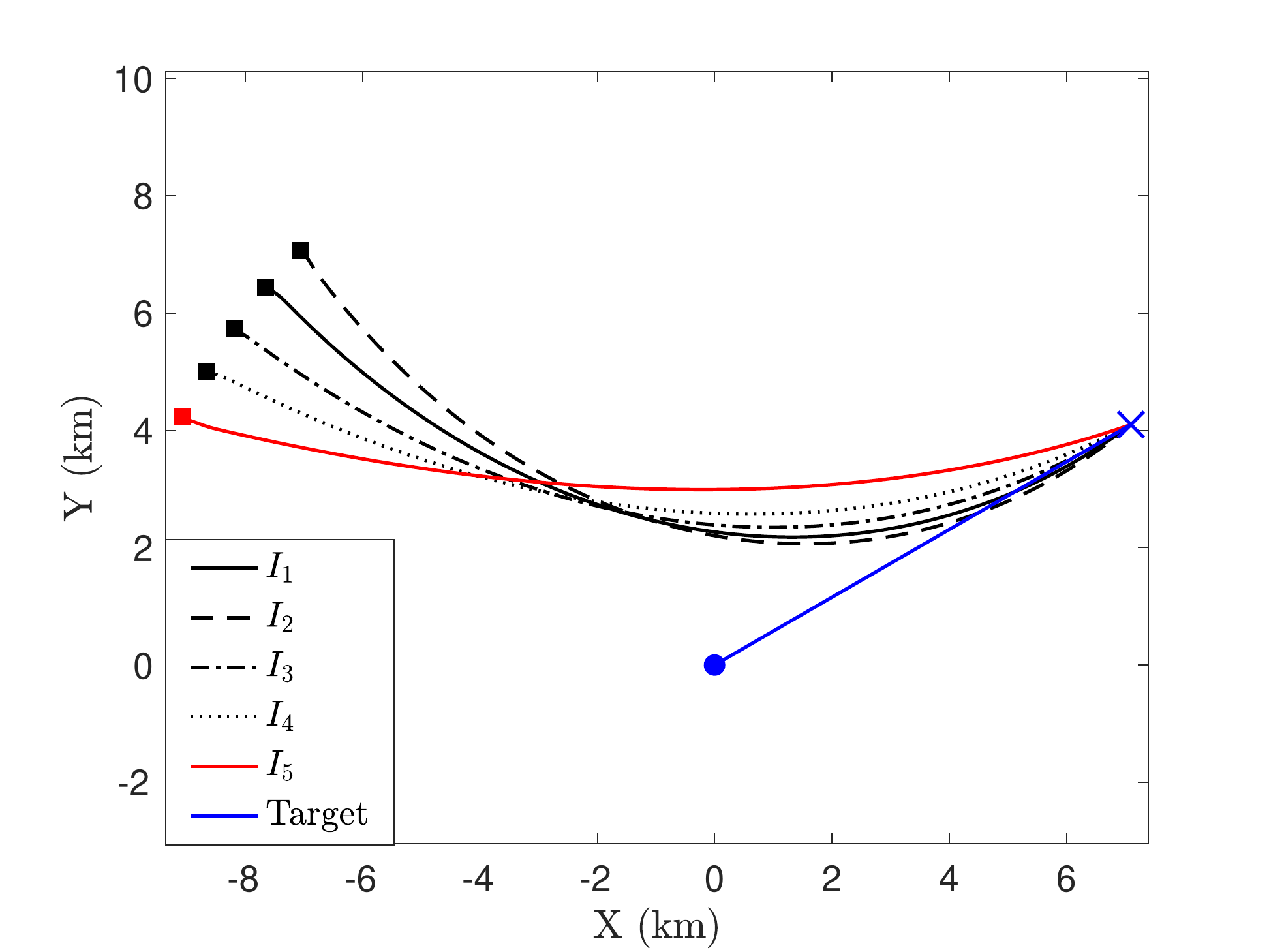}
		\caption{Trajectories.}
		\label{fig:cvts2trajectory}
	\end{subfigure}
	\begin{subfigure}[t]{0.32\linewidth}
		\centering
		\includegraphics[width=1.1\linewidth]{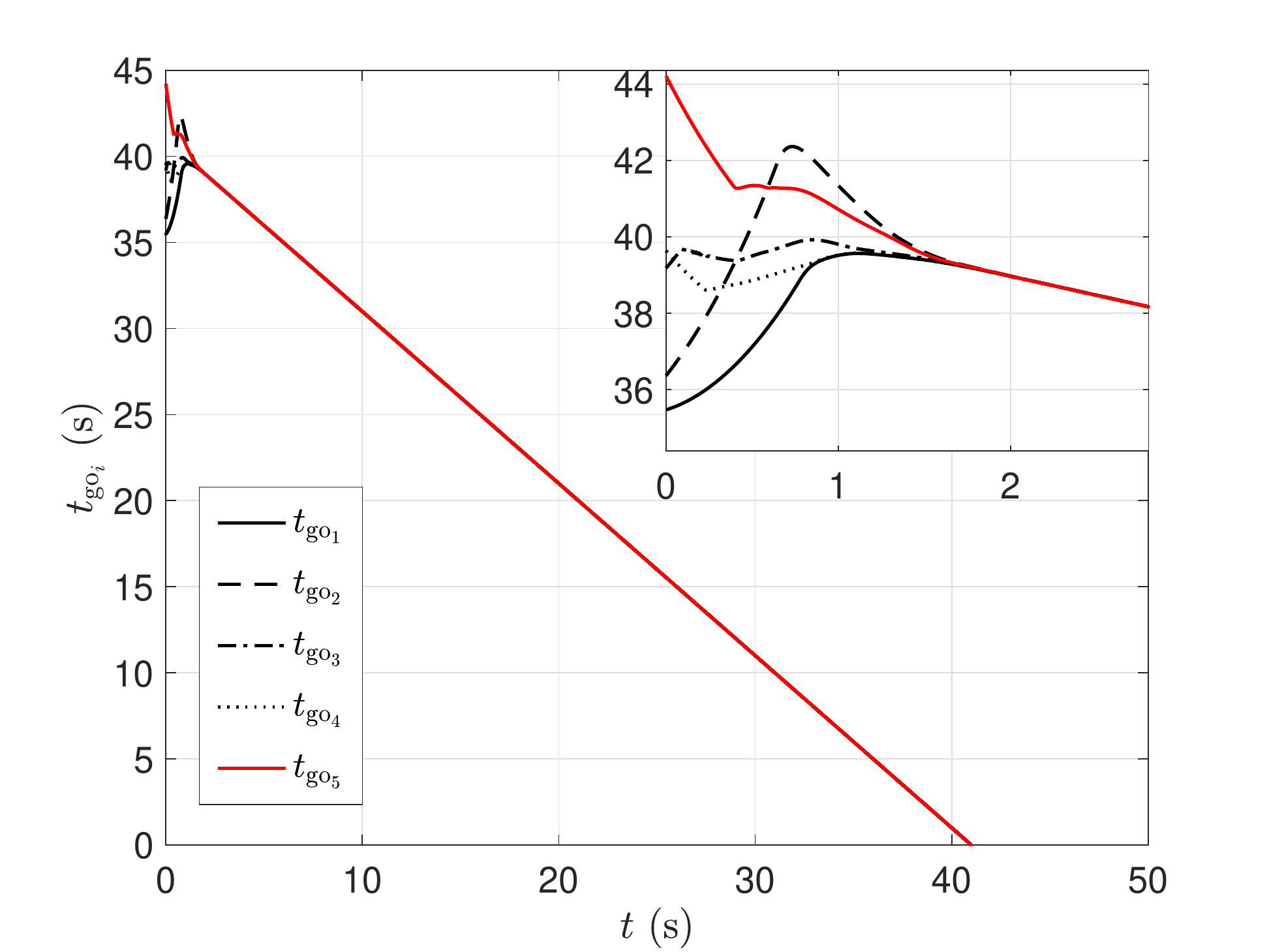}
		\caption{Time-to-go.}
		\label{fig:cvts2tgo}
	\end{subfigure}
	\begin{subfigure}[t]{0.32\linewidth}
		\centering
		\includegraphics[width=1.1\linewidth]{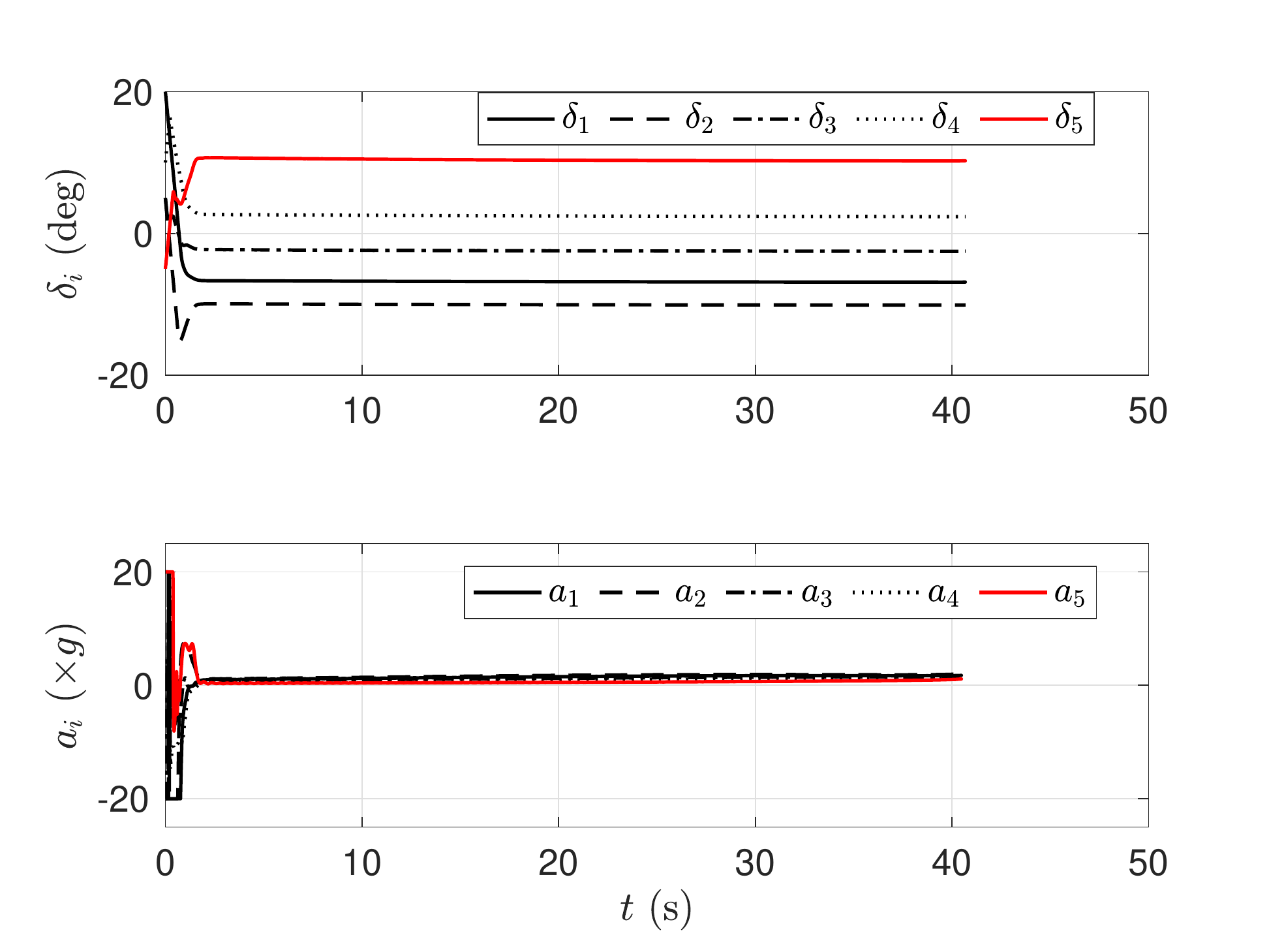}
		\caption{Look angles and lateral accelerations.}
		\label{fig:cvts2sigma}
	\end{subfigure}
	\caption{Simultaneous interception of a constant speed target using command \eqref{eq:aidevcvtdyn} over fixed topology.}
	\label{fig:cvts2}
\end{figure}

\begin{figure}[h!]
	\begin{subfigure}[t]{0.32\linewidth}
		\centering
		\includegraphics[width=1.1\linewidth]{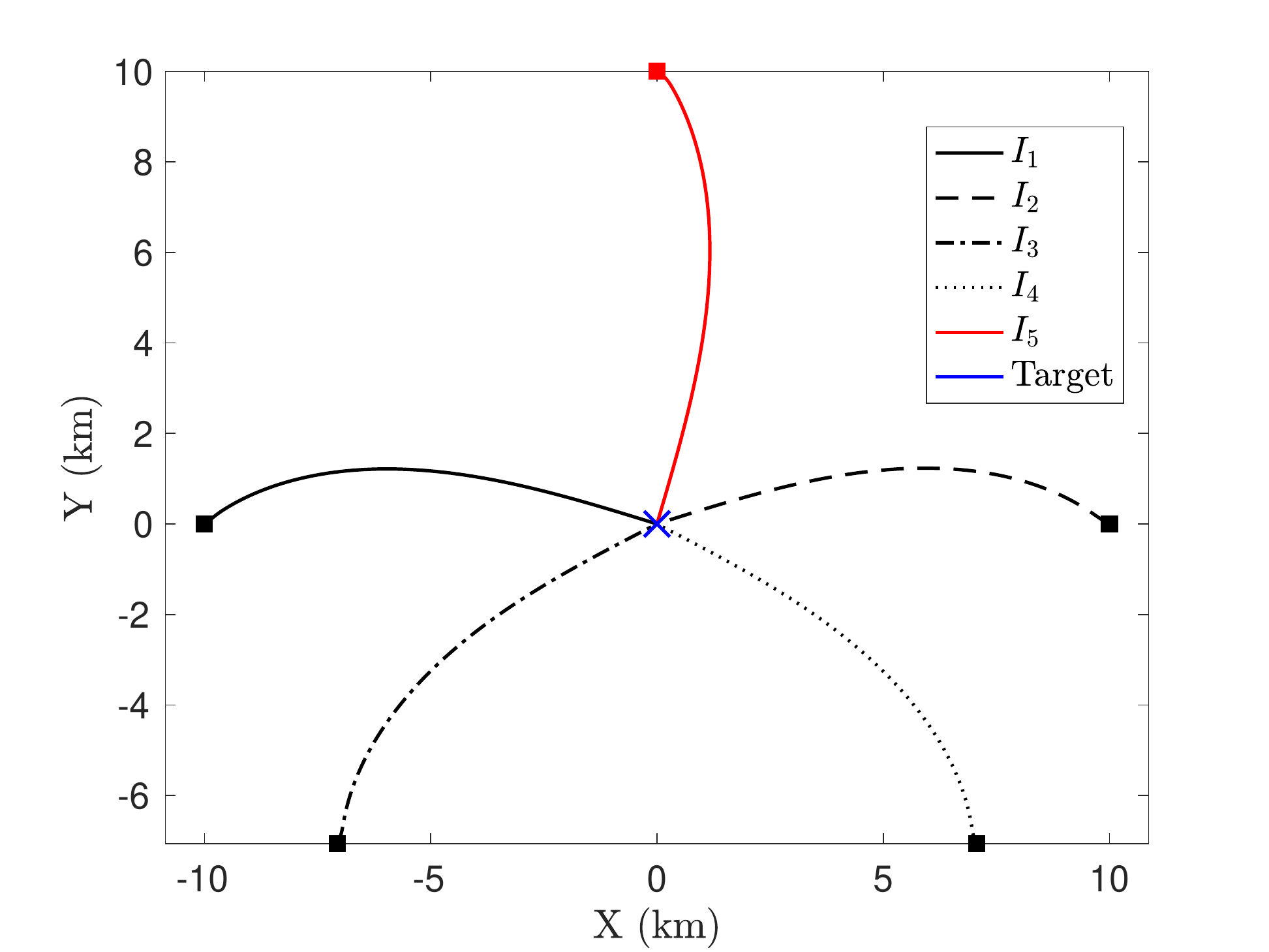}
		\caption{Trajectories.}
		\label{fig:s2trajectory}
	\end{subfigure}
	\begin{subfigure}[t]{0.32\linewidth}
		\centering
		\includegraphics[width=1.1\linewidth]{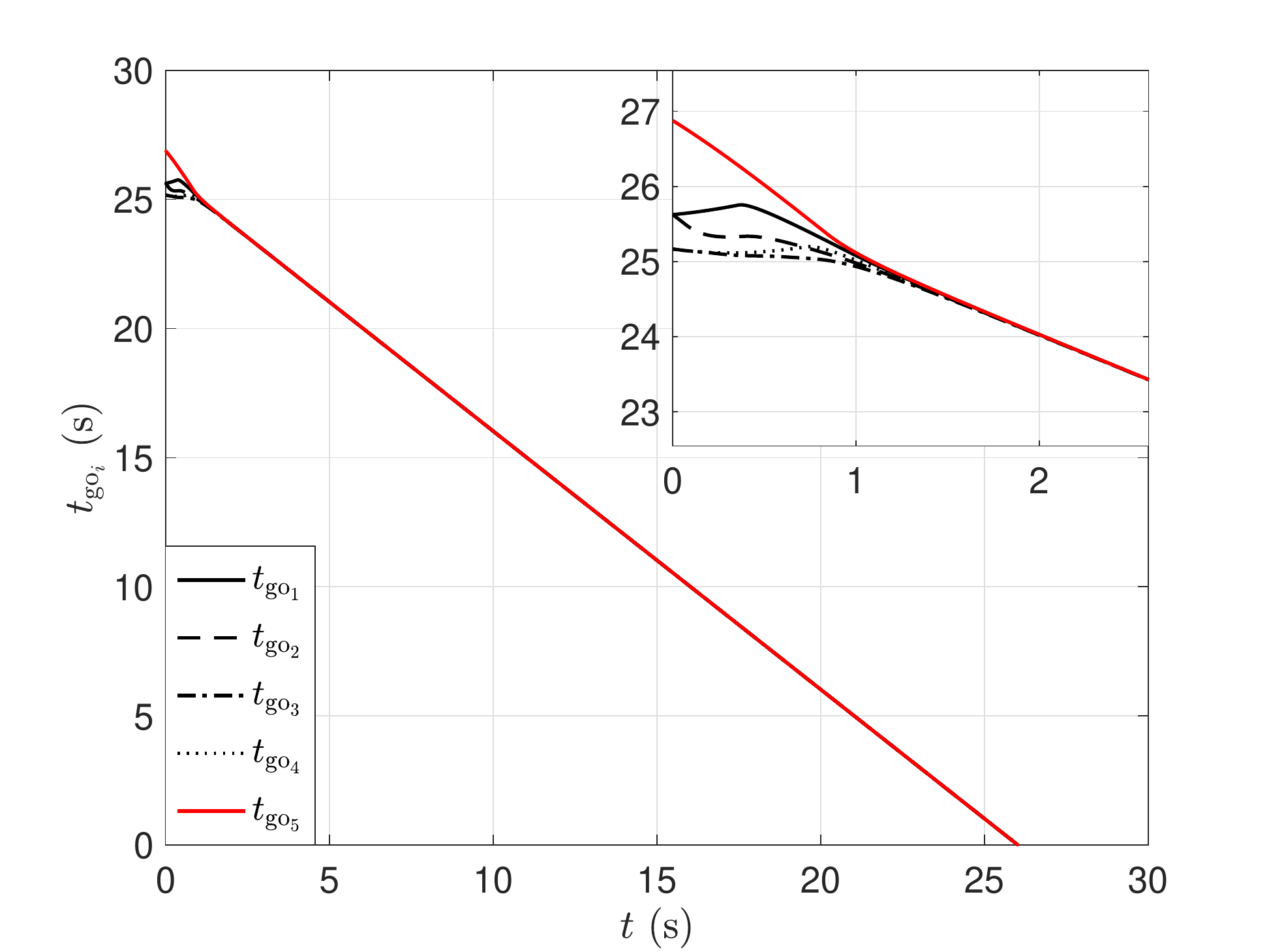}
		\caption{Time-to-go.}
		\label{fig:s2tgo}
	\end{subfigure}
	\begin{subfigure}[t]{0.32\linewidth}
		\centering
		\includegraphics[width=1.1\linewidth]{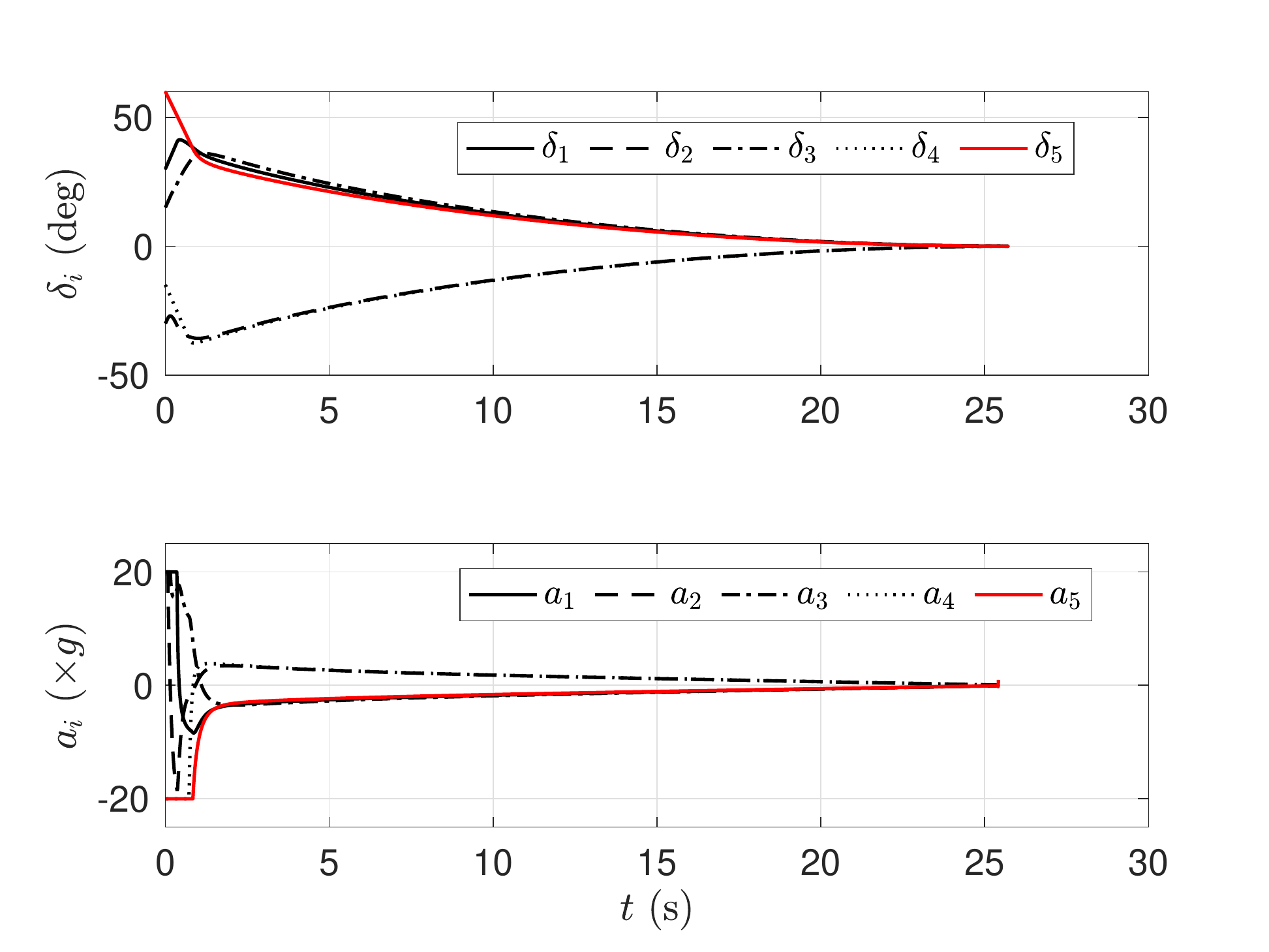}
		\caption{Look angles and lateral accelerations.}
		\label{fig:s2sigma}
	\end{subfigure}
	\caption{Simultaneous interception of a stationary target using command \eqref{eq:aistdyn2} over fixed topology.}
	\label{fig:s2}
\end{figure}

Through \Crefrange{fig:mans2}{fig:s2}, we discuss the case when commands \eqref{eq:aidevdyn}, \eqref{eq:aidevcvtdyn} and \eqref{eq:aistdyn2} are used by the cooperating interceptors to simultaneously capture the target when the interceptors are connected over a fixed topology. We assume a cycle graph, as shown in \Cref{fig:cycle}, when the topology is fixed. Through these results, we demonstrate that irrespective of the interaction topology, a predefined-time consensus in time-to-go can be achieved for simultaneous target interception.  Just as in the previous part, it is observed that the profiles of the engagement variables are similar while respecting the predefined-time, $T_s$.

\begin{figure}[h!]
	\begin{subfigure}[t]{0.475\linewidth}
		\centering
		\includegraphics[width=1.1\linewidth]{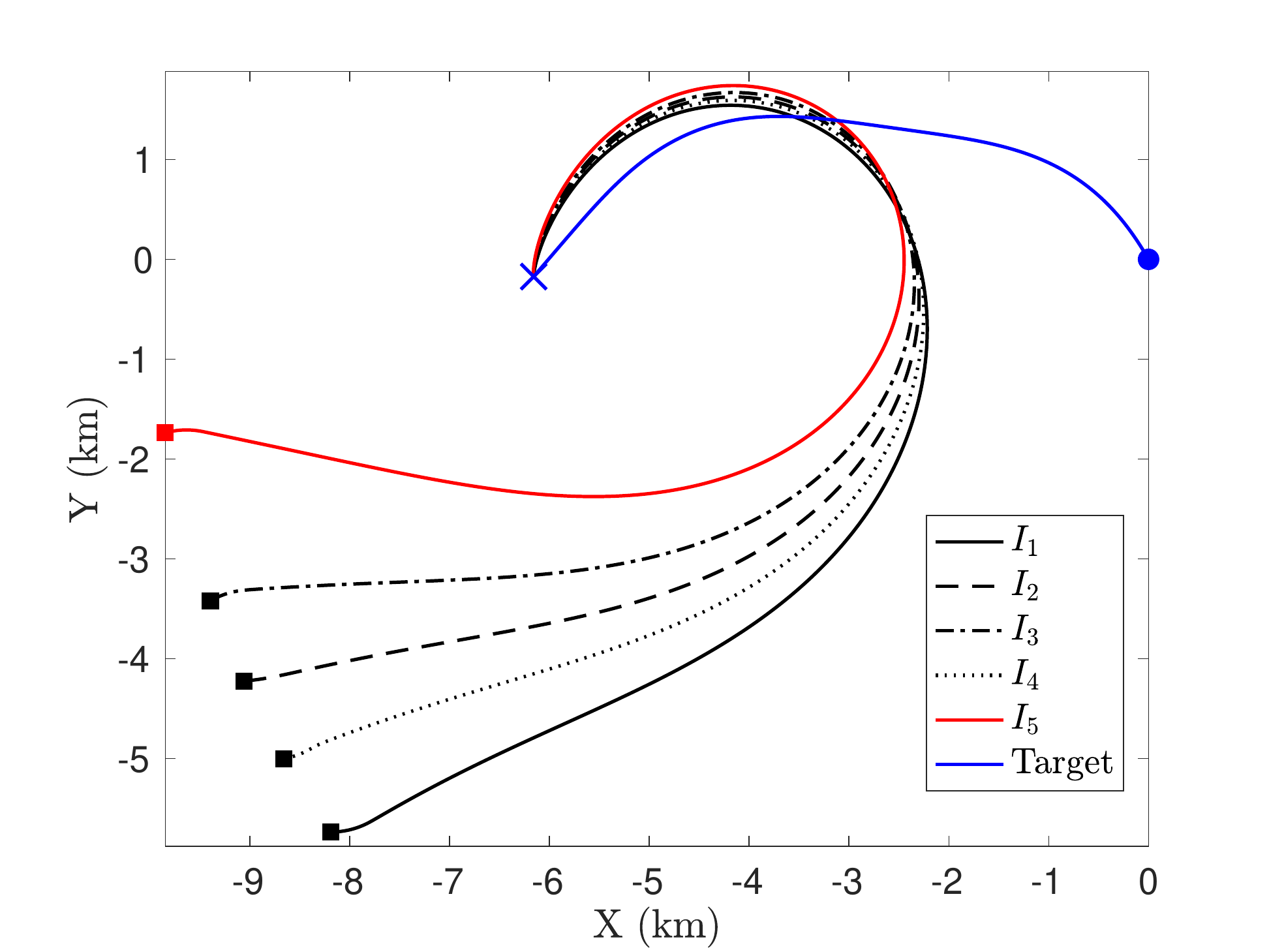}
		\caption{Trajectories.}
		\label{fig:mand2trajectory}
	\end{subfigure}
	\hfill
	\begin{subfigure}[t]{0.475\linewidth}
		\centering
		\includegraphics[width=1.1\linewidth]{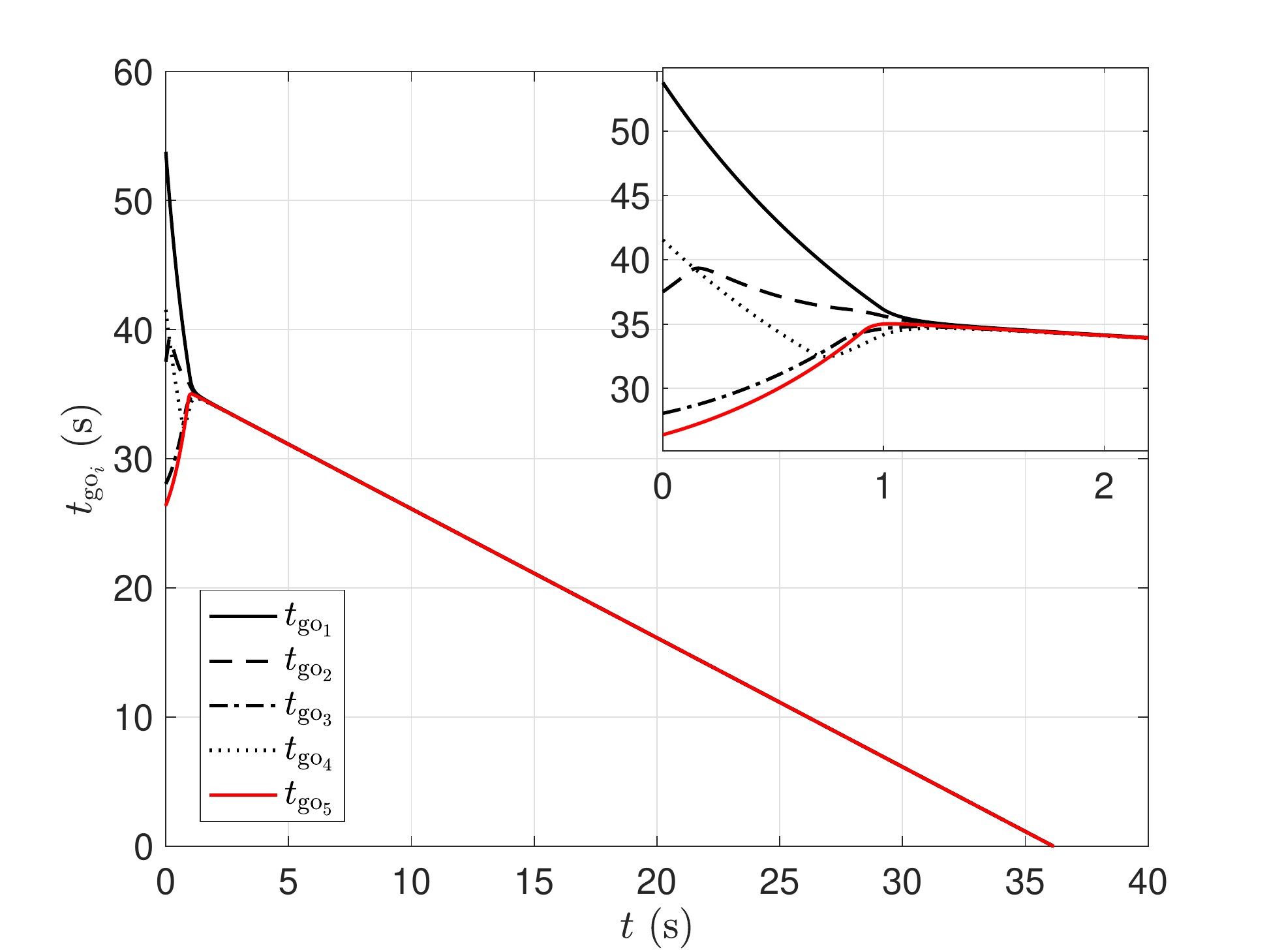}
		\caption{Time-to-go.}
		\label{fig:mand2tgo}
	\end{subfigure}
	\begin{subfigure}[t]{0.475\linewidth}
		\centering
		\includegraphics[width=1.1\linewidth]{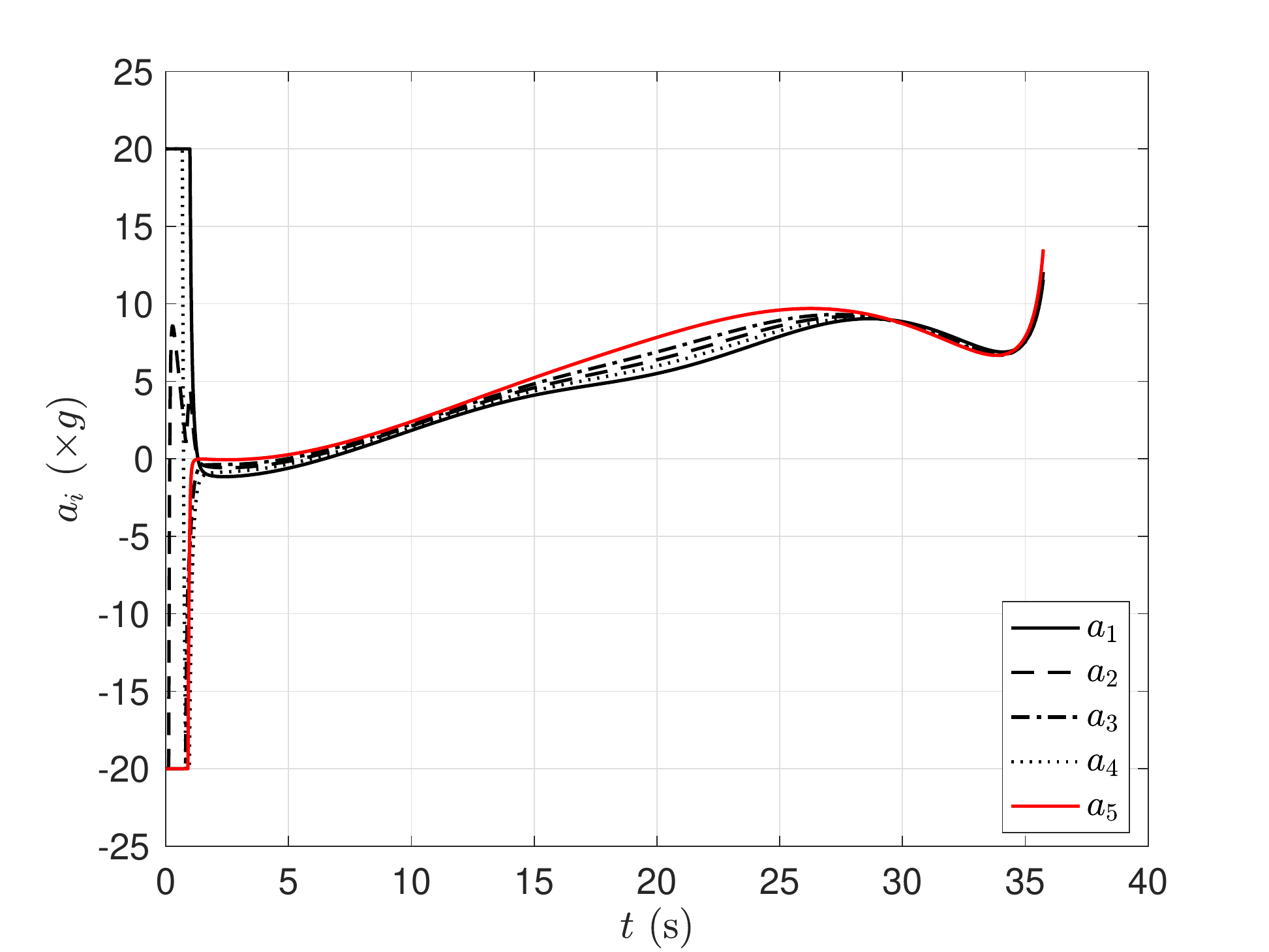}
		\caption{Lateral accelerations.}
		\label{fig:mand2am}
	\end{subfigure}
	\hfill
	\begin{subfigure}[t]{0.475\linewidth}
		\centering
		\includegraphics[width=1.1\linewidth]{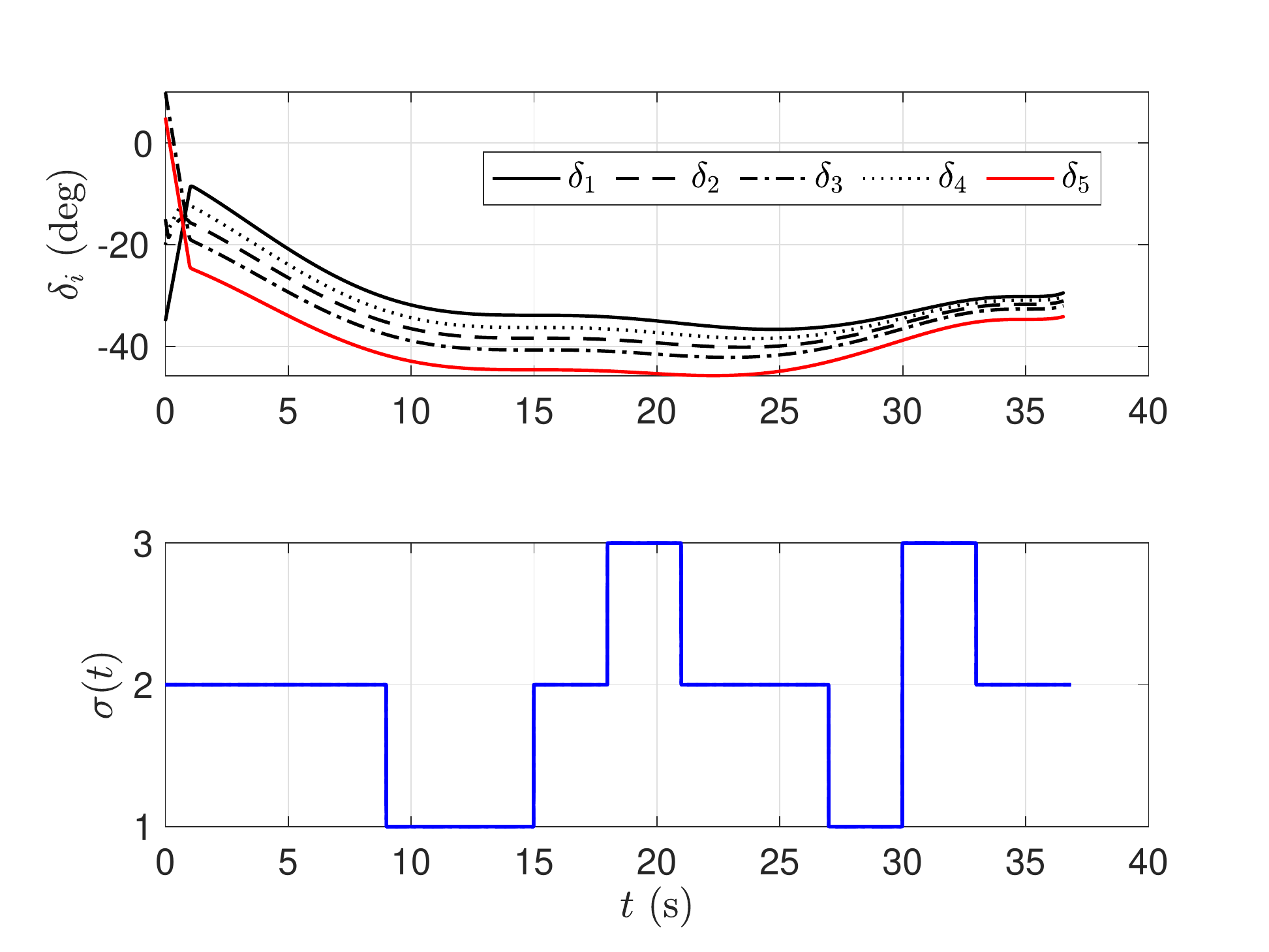}
		\caption{Look angles and the switching signal.}
		\label{fig:mand2sigma}
	\end{subfigure}
	\caption{Simultaneous interception of a maneuvering target (switching topologies).}
	\label{fig:mand2}
\end{figure}
\begin{figure}[h!]
	\begin{subfigure}[t]{0.475\linewidth}
		\centering
		\includegraphics[width=1.1\linewidth]{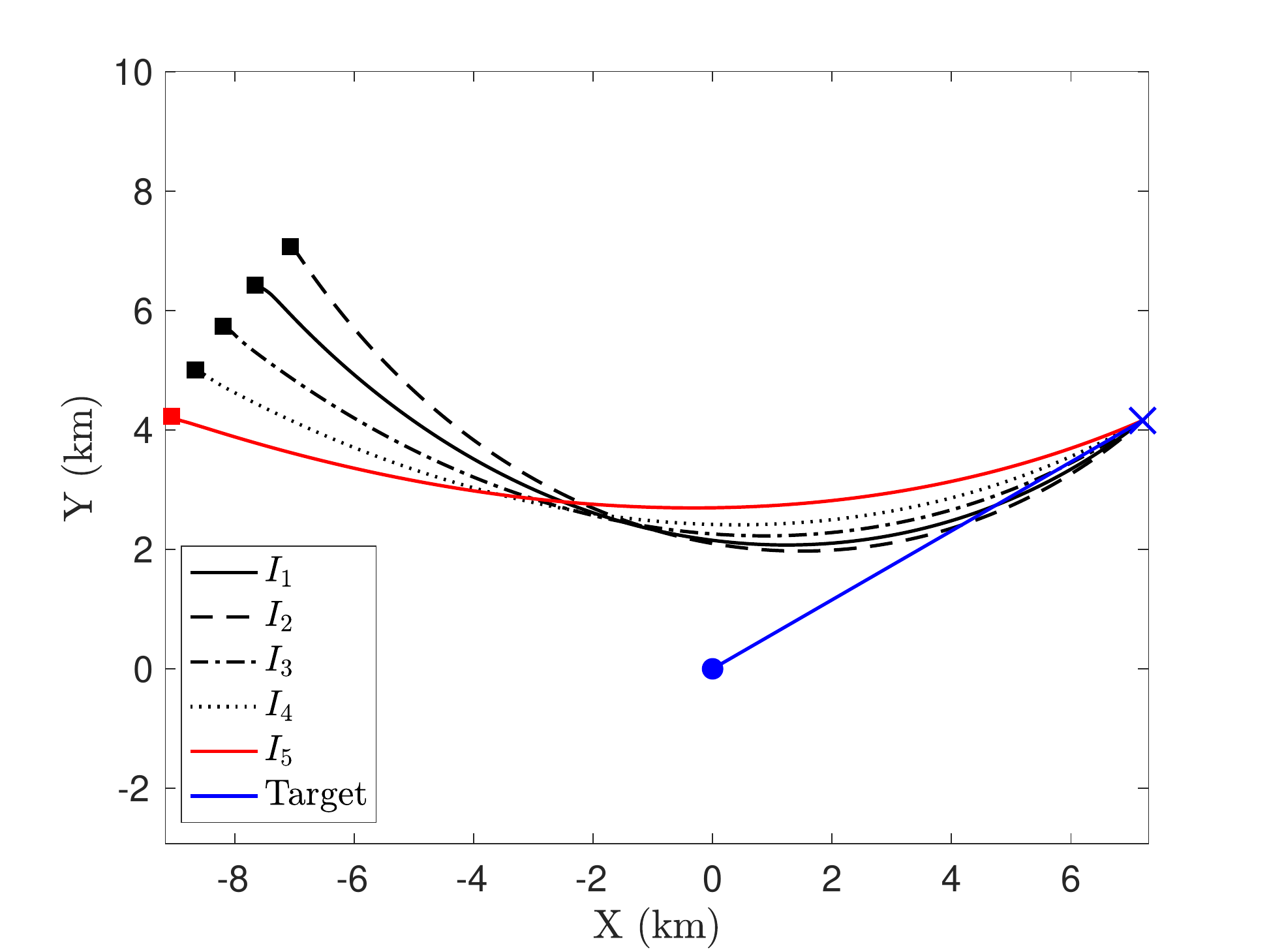}
		\caption{Trajectories.}
		\label{fig:cvtd2trajectory}
	\end{subfigure}
	\hfill
	\begin{subfigure}[t]{0.475\linewidth}
		\centering
		\includegraphics[width=1.1\linewidth]{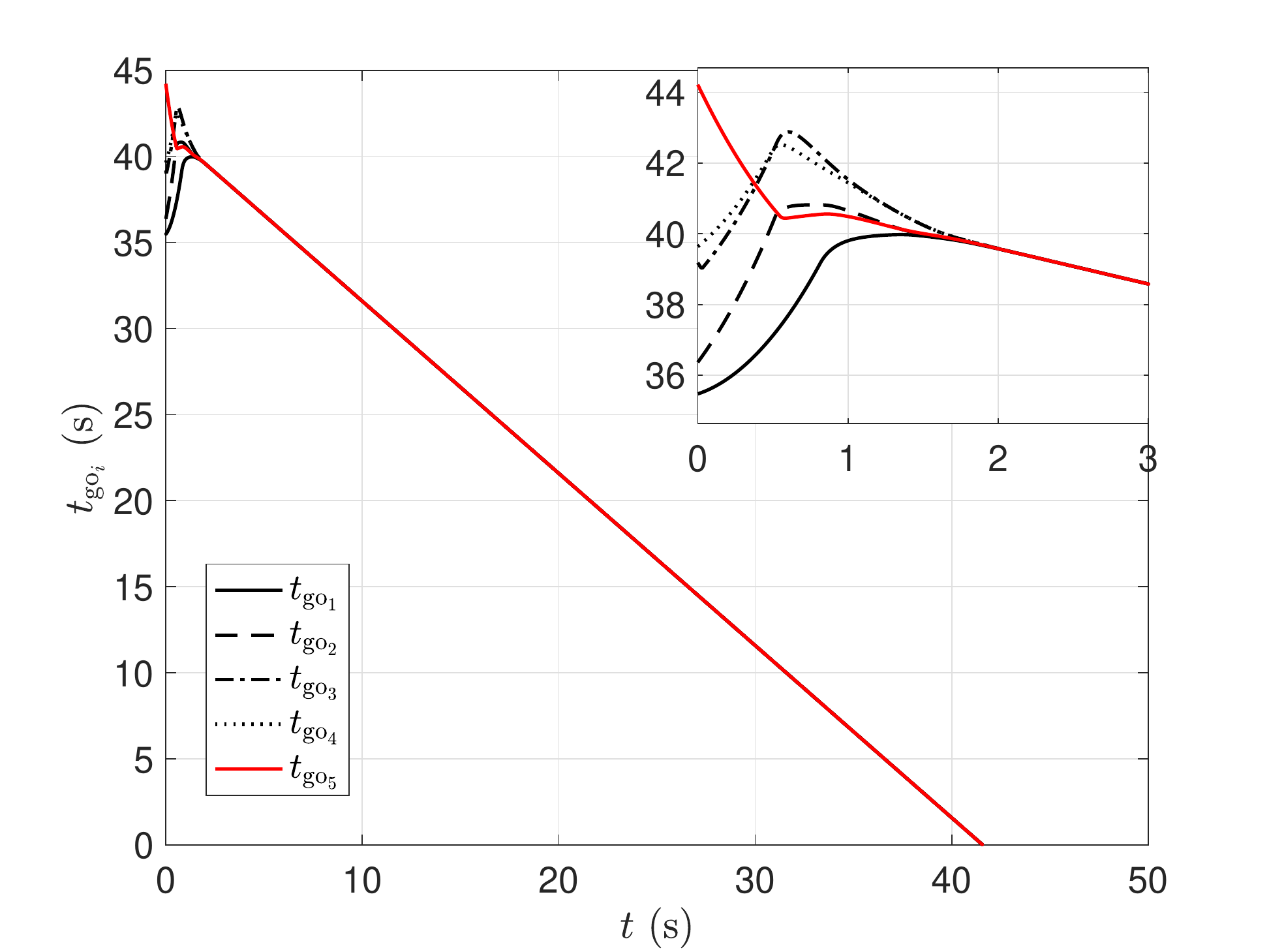}
		\caption{Time-to-go.}
		\label{fig:cvtd2tgo}
	\end{subfigure}
	\begin{subfigure}[t]{0.475\linewidth}
		\centering
		\includegraphics[width=1.1\linewidth]{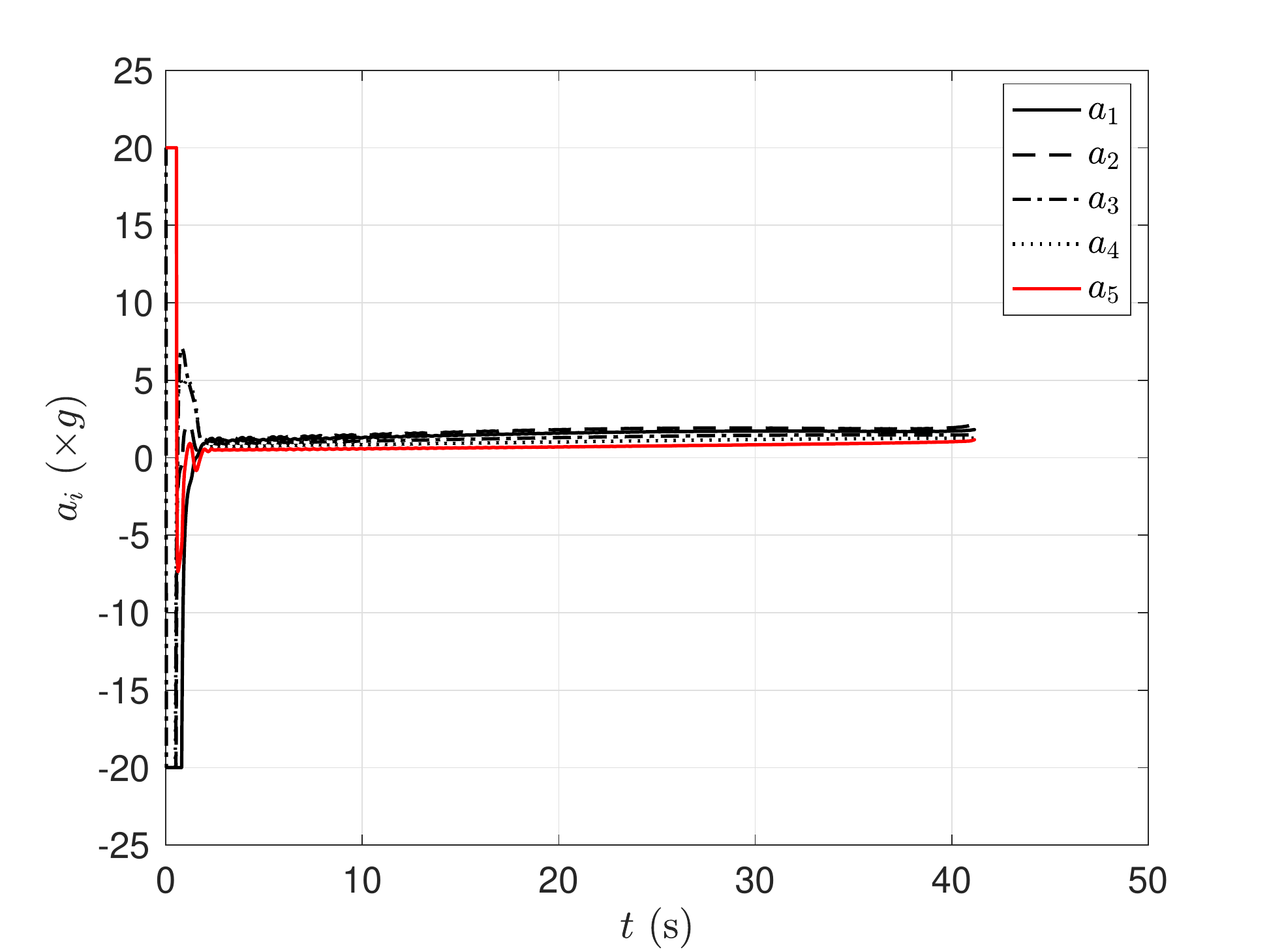}
		\caption{Lateral accelerations.}
		\label{fig:cvtd2am}
	\end{subfigure}
	\hfill
	\begin{subfigure}[t]{0.475\linewidth}
		\centering
		\includegraphics[width=1.1\linewidth]{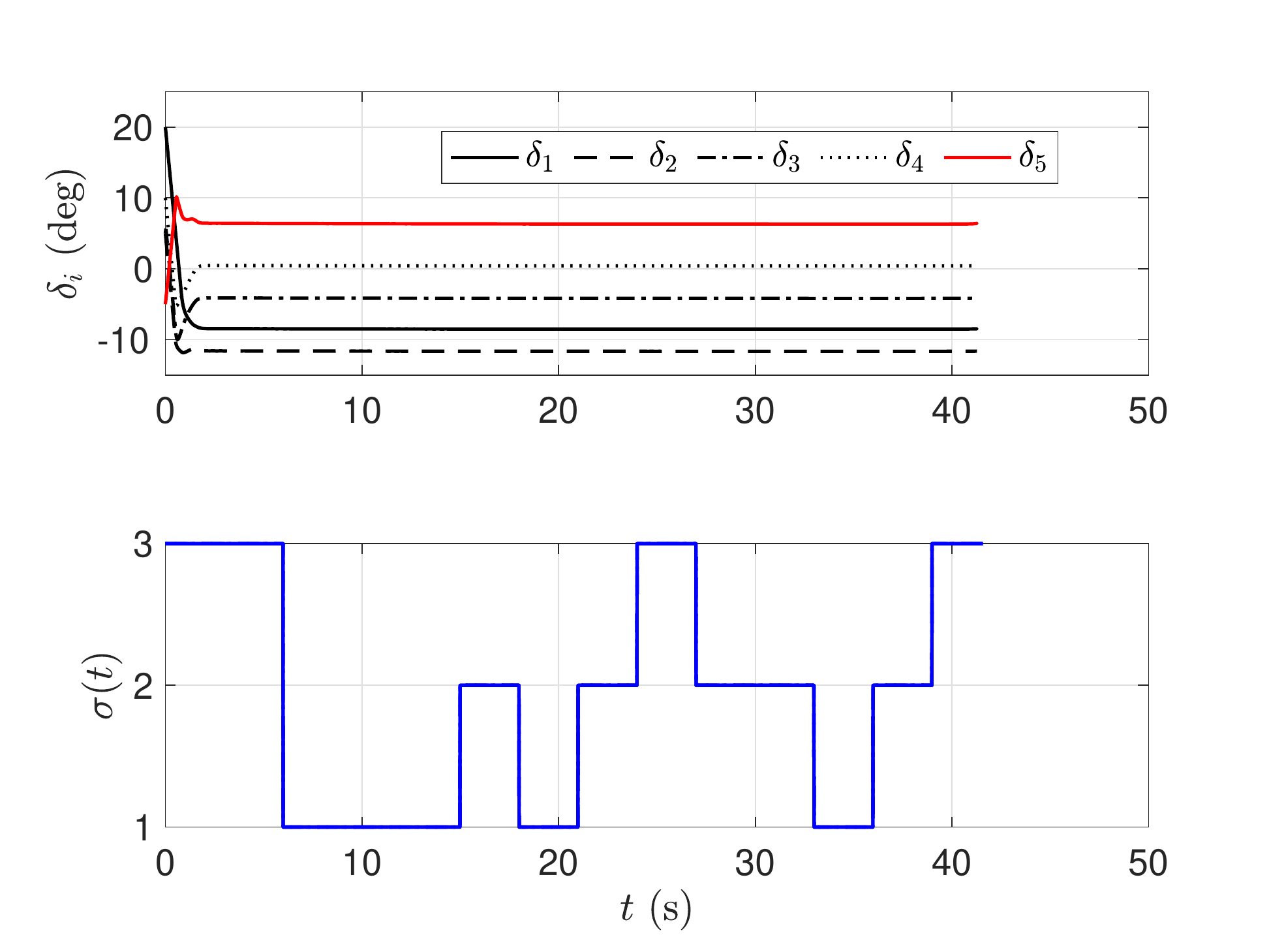}
		\caption{Look angles and the switching signal.}
		\label{fig:cvtd2sigma}
	\end{subfigure}
	\caption{Simultaneous interception of a constant speed target (switching topologies).}
	\label{fig:cvtd2}
\end{figure}
\begin{figure}[h!]
	\begin{subfigure}[t]{0.475\linewidth}
		\centering
		\includegraphics[width=1.1\linewidth]{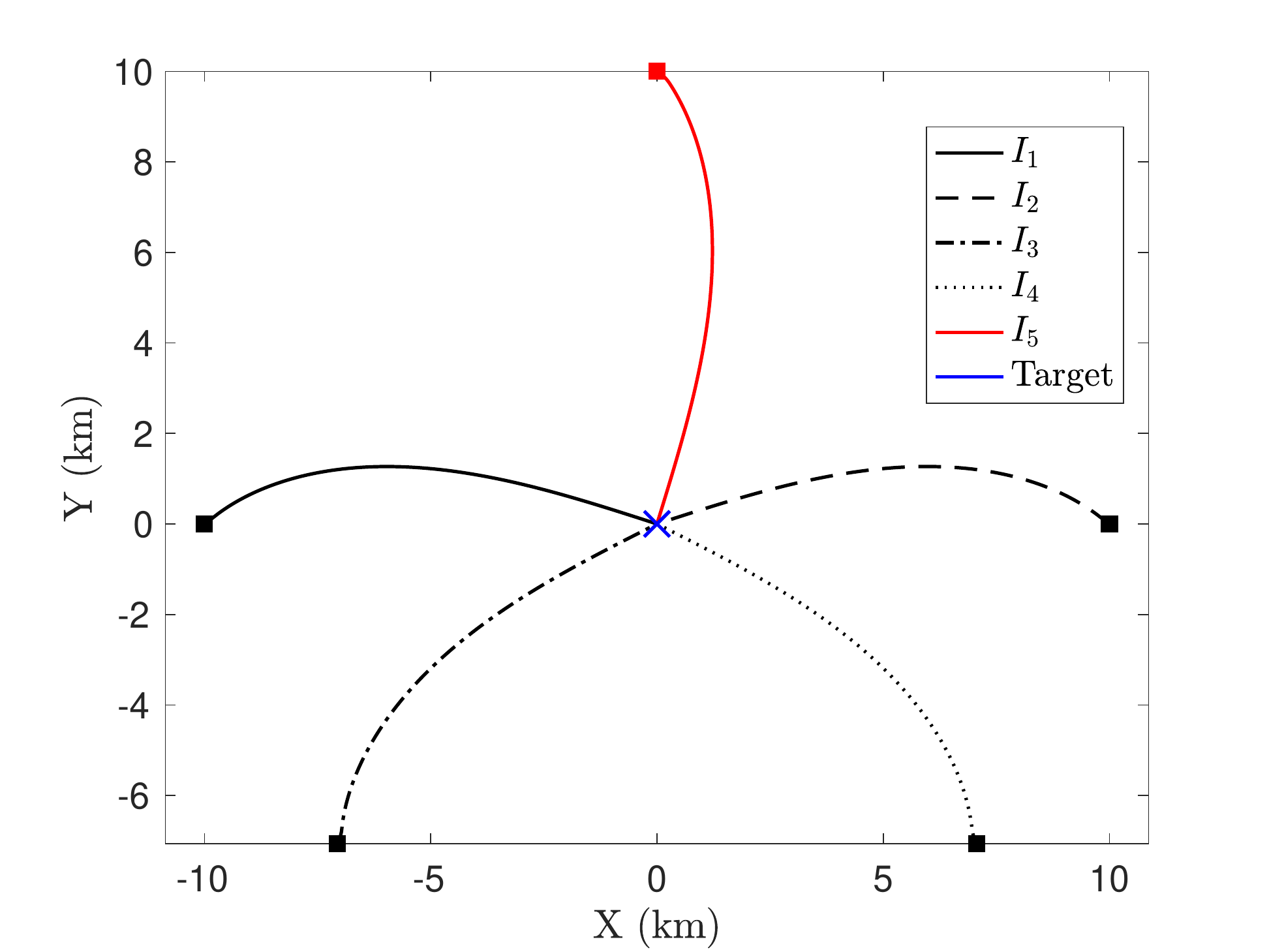}
		\caption{Trajectories.}
		\label{fig:std2trajectory}
	\end{subfigure}
	\hfill
	\begin{subfigure}[t]{0.475\linewidth}
		\centering
		\includegraphics[width=1.1\linewidth]{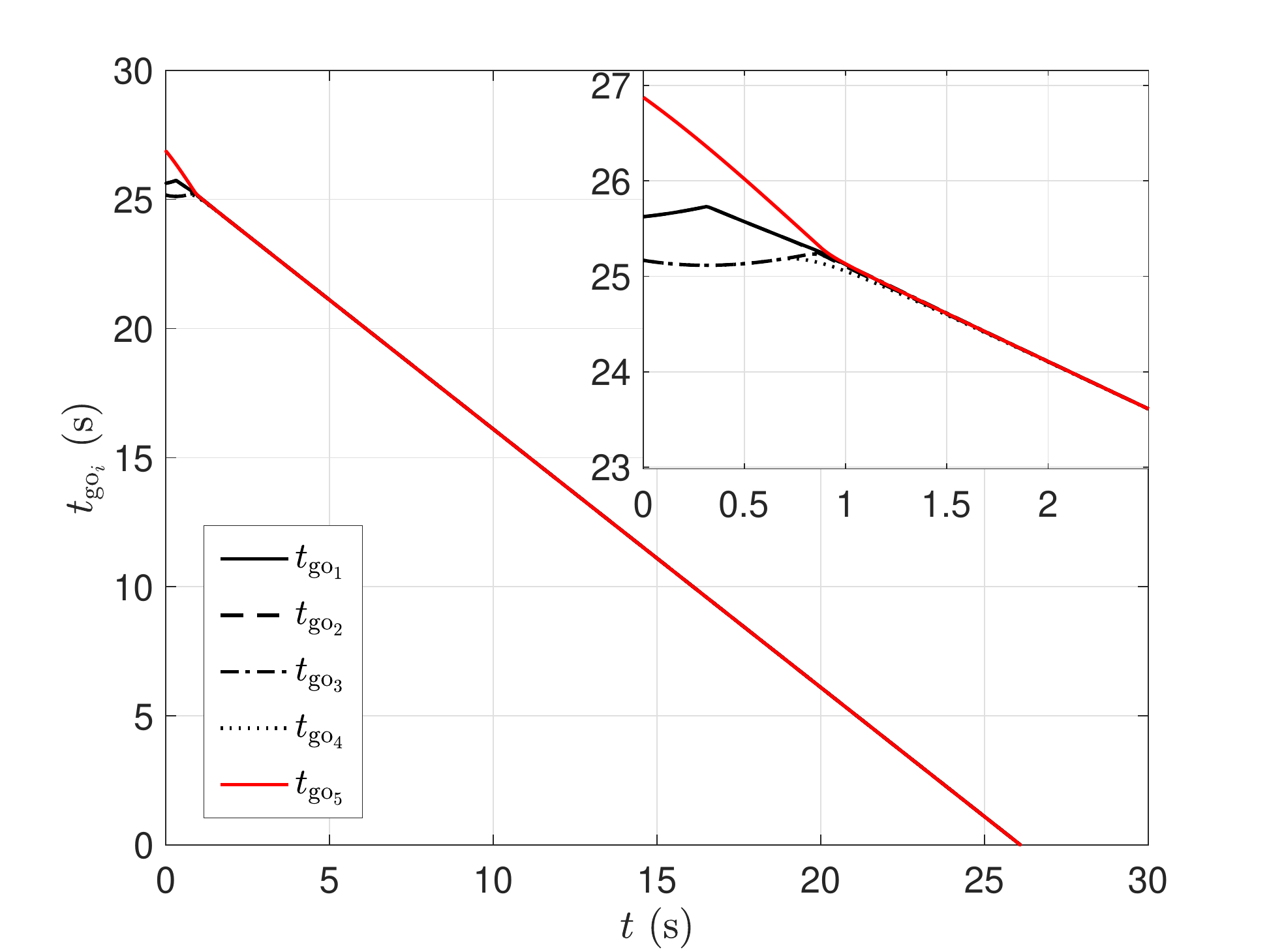}
		\caption{Time-to-go.}
		\label{fig:std2tgo}
	\end{subfigure}
	\begin{subfigure}[t]{0.475\linewidth}
		\centering
		\includegraphics[width=1.1\linewidth]{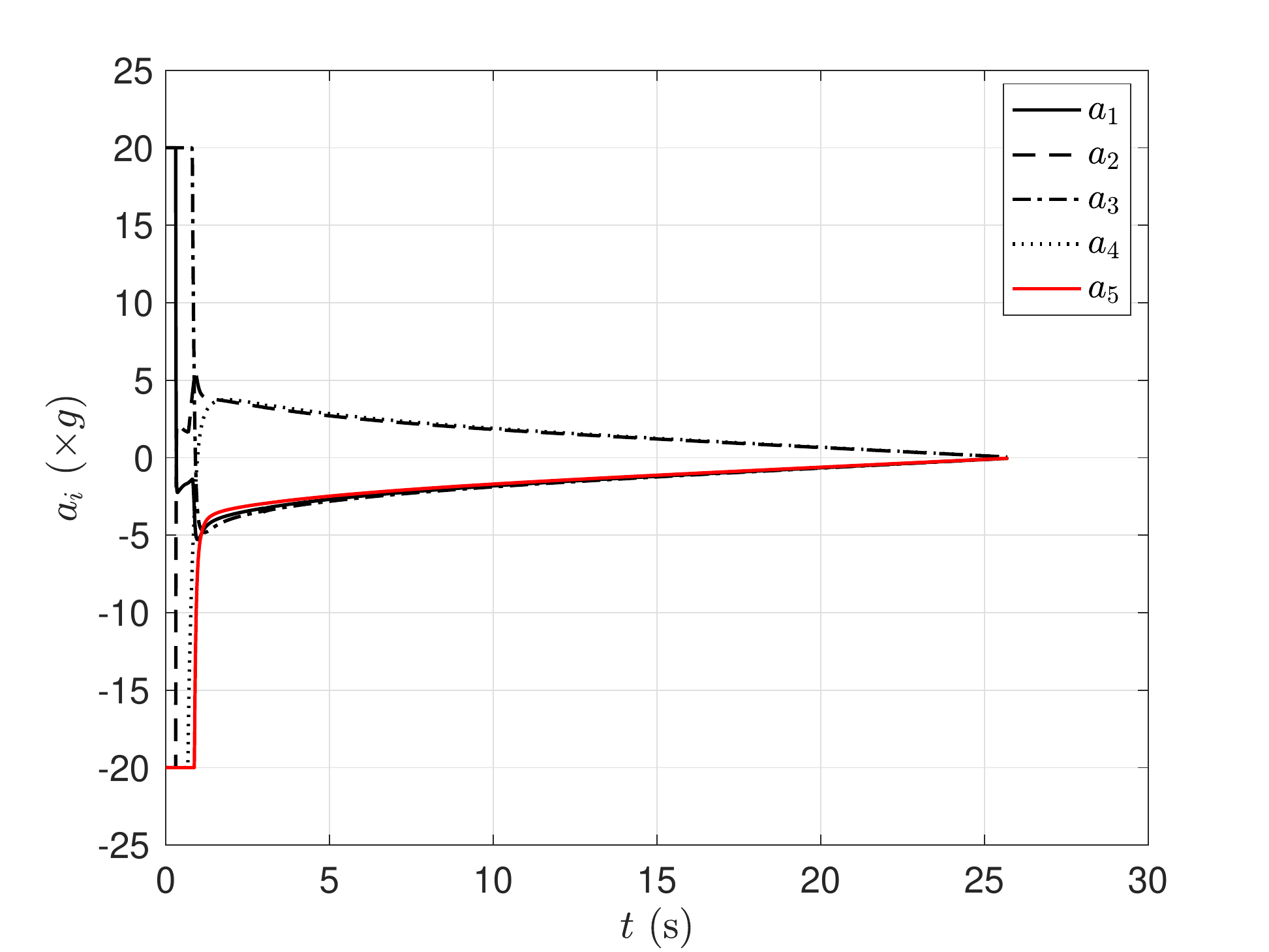}
		\caption{Lateral accelerations.}
		\label{fig:std2am}
	\end{subfigure}
	\hfill
	\begin{subfigure}[t]{0.475\linewidth}
		\centering
		\includegraphics[width=1.1\linewidth]{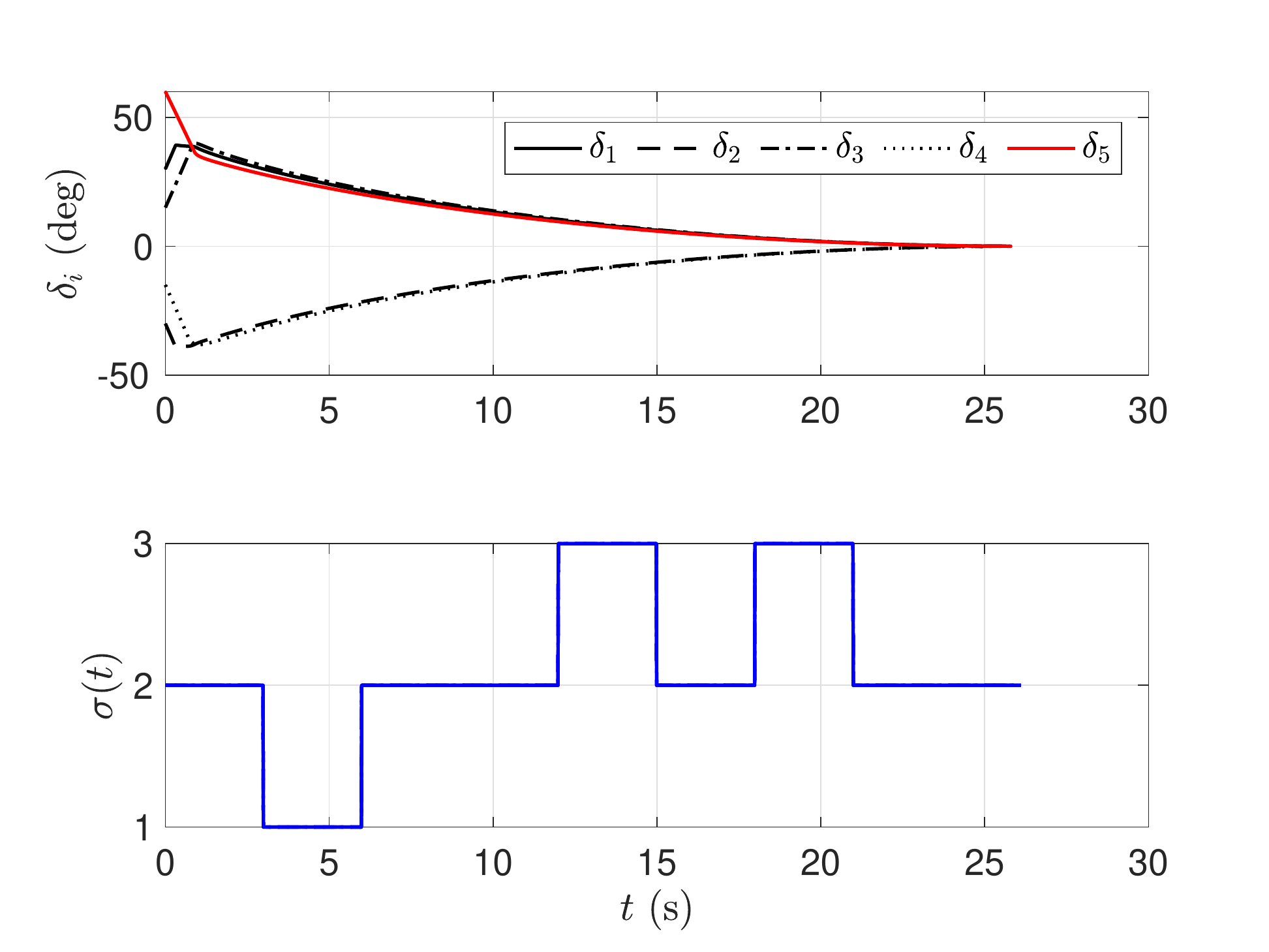}
		\caption{Look angles and the switching signal.}
		\label{fig:std2sigma}
	\end{subfigure}
	\caption{Simultaneous interception of a stationary target (switching topologies).}
	\label{fig:std2}
\end{figure}

In \Crefrange{fig:mand2}{fig:std2}, we show the performance of the proposed cooperative guidance strategy when the interceptors are connected over switching topologies. It is worth noting that under arbitrary switching, the cooperative behavior of the interceptors' swarm remains similar to what has been observed in \Crefrange{fig:mans2}{fig:s2} since the commands \eqref{eq:aidevdyn}, \eqref{eq:aidevcvtdyn} and \eqref{eq:aistdyn2} ensure that the interceptors time-to-go values achieve consensus within $T_s$ irrespective of the manner in which the interceptors communicate.

\subsection{Comparison with Other Cooperative Guidance Strategies}
\begin{figure}[h!]
	\begin{subfigure}[t]{0.32\linewidth}
		\centering
		\includegraphics[width=1.1\linewidth]{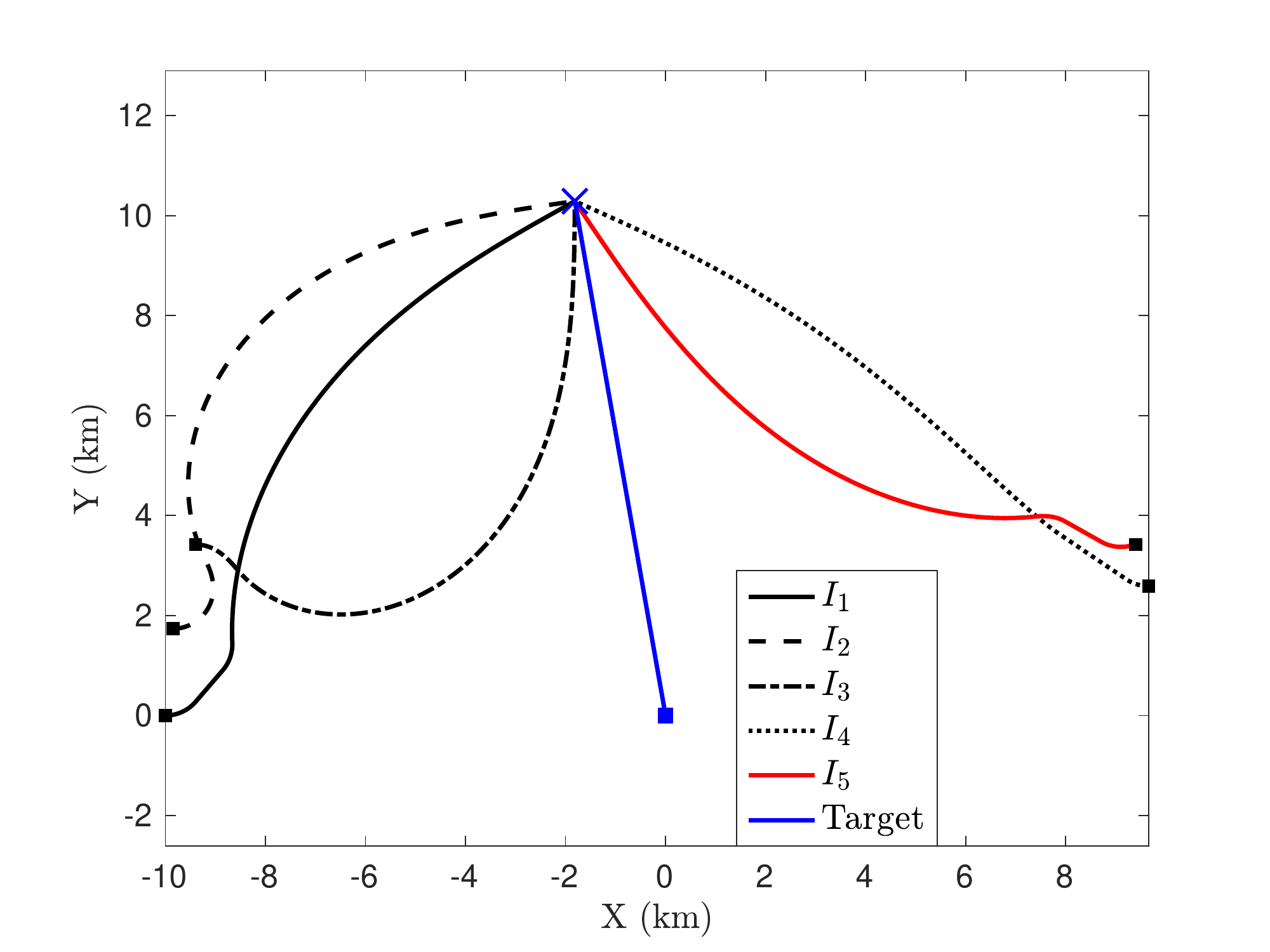}
		\caption{Trajectories.}
		\label{fig:piptrajectory}
	\end{subfigure}
	\begin{subfigure}[t]{0.32\linewidth}
		\centering
		\includegraphics[width=1.1\linewidth]{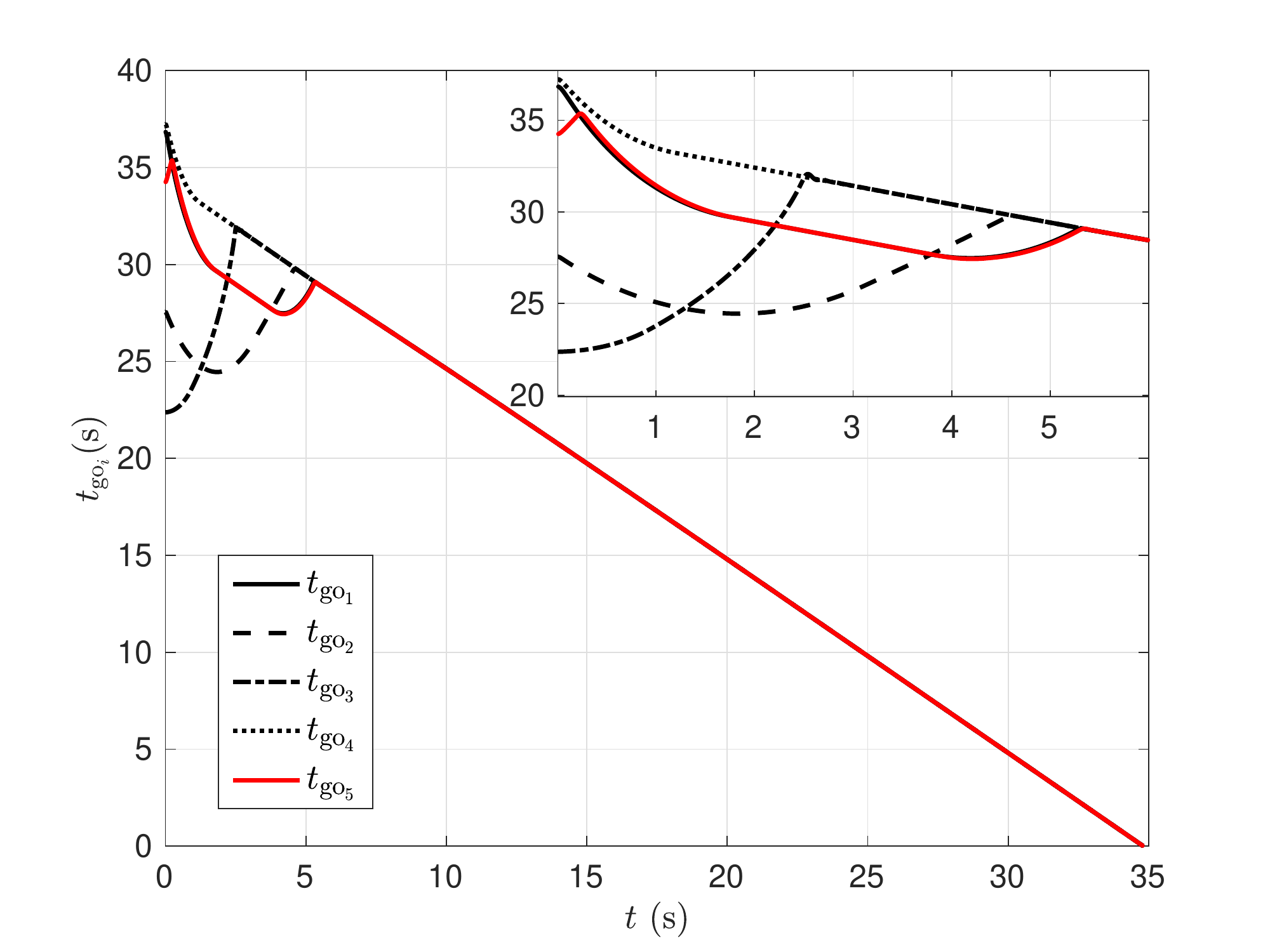}
		\caption{Time-to-go.}
		\label{fig:piptgo}
	\end{subfigure}
	\begin{subfigure}[t]{0.32\linewidth}
		\centering
		\includegraphics[width=1.1\linewidth]{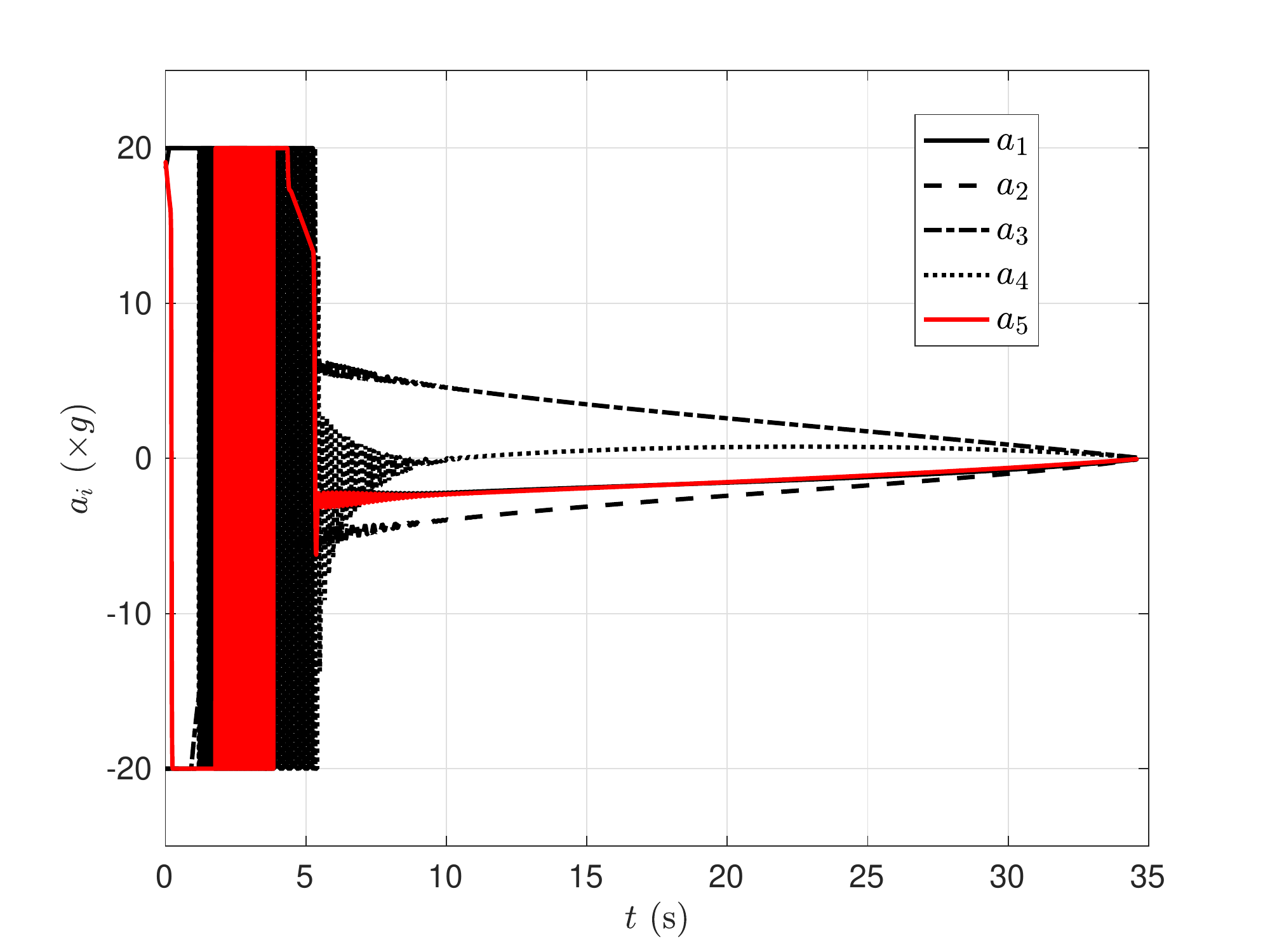}
		\caption{Lateral accelerations.}
		\label{fig:pipsigma}
	\end{subfigure}
	\caption{Simultaneous interception of a constant speed target with an existing guidance strategy \cite{7376242} using the concept of PIP.}
	\label{fig:pip}
\end{figure}

Yet another technique to simultaneously intercept a moving target ($a_\mathrm{T}=0,v_\mathrm{T}\neq 0$) may be designed using the concept of predicted interception point \cite{HOU2019105142,7376242,9000526,HOU2017685,doi:10.2514/1.G001681}. For example, a target moving with constant speed may be treated as a virtual stationary target located at a position that is expected to be the target's actual position after a certain time. This location is commonly known as the predicted interception point whose coordinates $(X_i^\mathrm{PIP},Y_i^\mathrm{PIP})$ are computed for the $i$\textsuperscript{th} interceptor as
\begin{equation}
	X_i^\mathrm{PIP} = X_\mathrm{T} + v_\mathrm{T}t_{\mathrm{go}_i}\cos\gamma_\mathrm{T},~Y_i^\mathrm{PIP} = Y_\mathrm{T} + v_\mathrm{T}t_{\mathrm{go}_i}\sin\gamma_\mathrm{T},
\end{equation} 
where $(X_\mathrm{T},Y_\mathrm{T})$ denotes the instantaneous position of the constant speed target. While this method is simple, it may require updating the predicted interception point for accuracy until consensus in time-to-go is established. In this case, consensus in time-to-go implies consensus in PIP coordinates, and the swarm of interceptors now aim at a fixed PIP whose accuracy is determined by how much precise the time-to-go estimate is. Any erroneous measurement may lead to a miss, and to achieve consensus in time-to-go values early in the engagement may incur heavy lateral acceleration demand. We revisit the scenario considered in \Cref{fig:cvts} using the concept of PIP used in \cite{7376242}, and the results are shown in \Cref{fig:pip}. 

\Cref{fig:pip} portrays the simultaneous interception of a constant speed target using the concept of predicted interception point, that is, the constant speed target is regarded as a virtual stationary target at $(X_i^\mathrm{PIP},Y_i^\mathrm{PIP})$. The interceptors now aim at that point instead of \emph{chasing} the target (see \Cref{fig:piptrajectory}). In this case, the time of consensus in time-to-go is suitably selected using the design parameters as 5 s. We do this in order to make a comparison with the results in \Cref{fig:cvts} wherein $T_s$ was chosen as 5 s. It can be observed from \Cref{fig:piptgo} that interceptors agree upon a common time-to-go value slightly after 5 s. To achieve an agreement this early in the engagement, the lateral acceleration requirement for each interceptor increases significantly in the transient phase (see \Cref{fig:pipsigma}). One may draw similar inferences as in the previous case regarding the lateral acceleration requirement in the terminal guidance phase. However, the actuators used to implement the cooperative guidance may not be able to sustain such large demands if faster consensus in time-to-go is needed. This partially motivates us to resort to deviated pursuit guidance, as shown in the proposed design while vindicating the efficacy of the proposed method.

\subsection{Performance with Autopilot}
\begin{figure}[h!]
	\begin{subfigure}[t]{0.475\linewidth}
		\centering
		\includegraphics[width=1.1\linewidth]{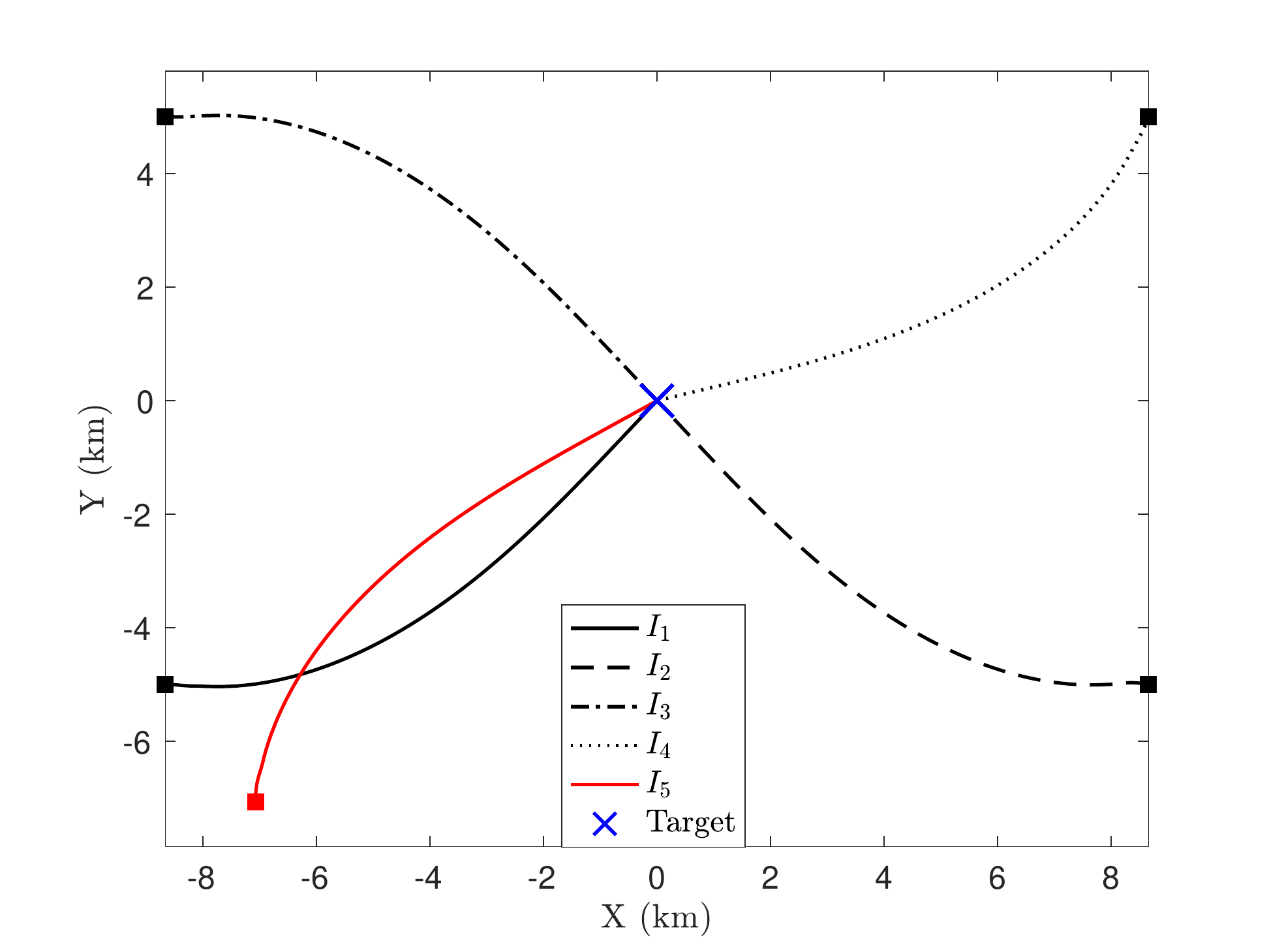}
		\caption{Trajectories.}
		\label{fig:sautopilottrajectory}
	\end{subfigure}
	\hfill
	\begin{subfigure}[t]{0.475\linewidth}
		\centering
		\includegraphics[width=1.1\linewidth]{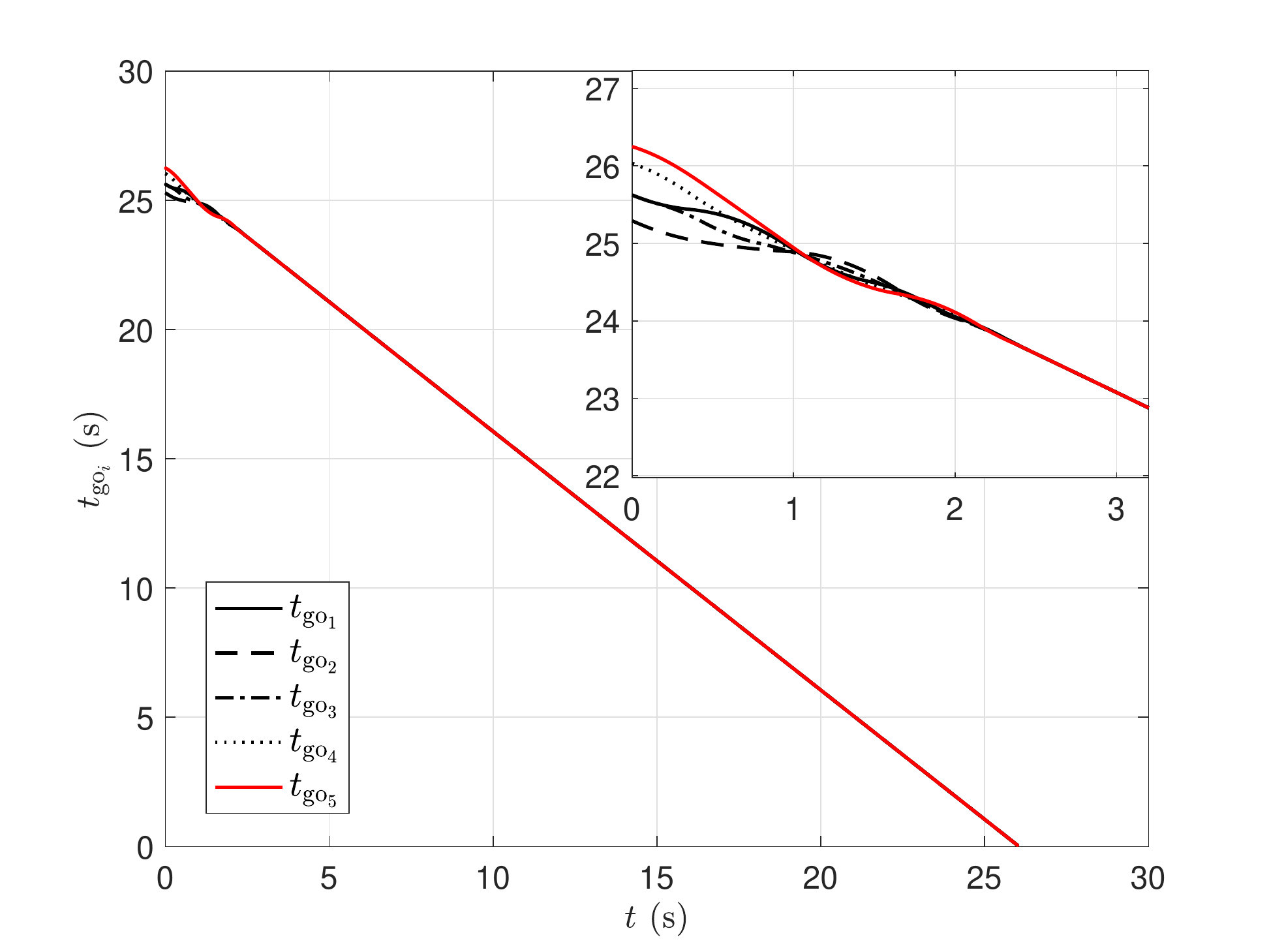}
		\caption{Time-to-go.}
		\label{fig:sautopilottgo}
	\end{subfigure}
	\begin{subfigure}[t]{0.475\linewidth}
		\centering
		\includegraphics[width=1.1\linewidth]{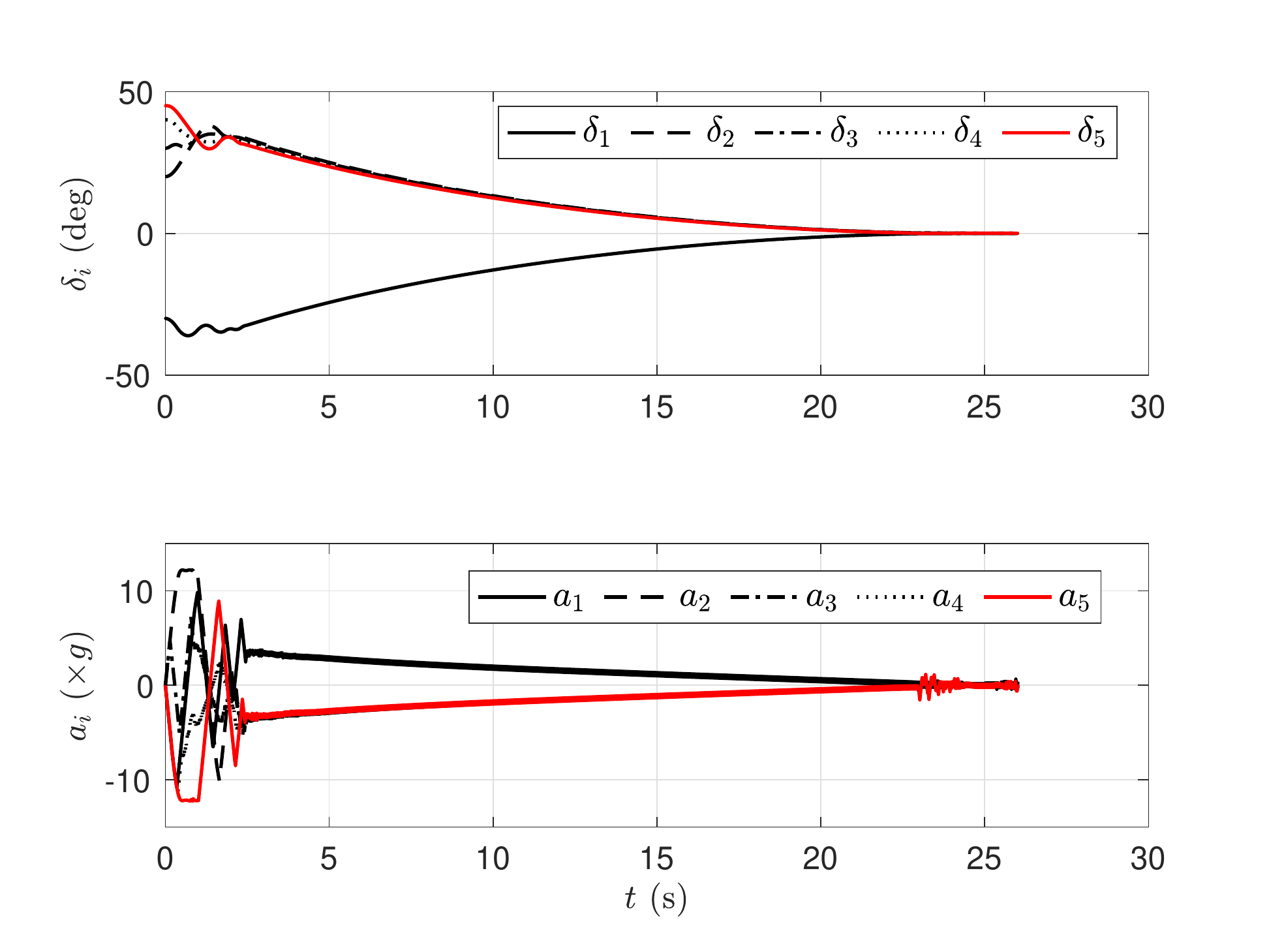}
		\caption{Look angles and lateral accelerations.}
		\label{fig:sautopilotam}
	\end{subfigure}
	\hfill
	\begin{subfigure}[t]{0.475\linewidth}
		\centering
		\includegraphics[width=1.1\linewidth]{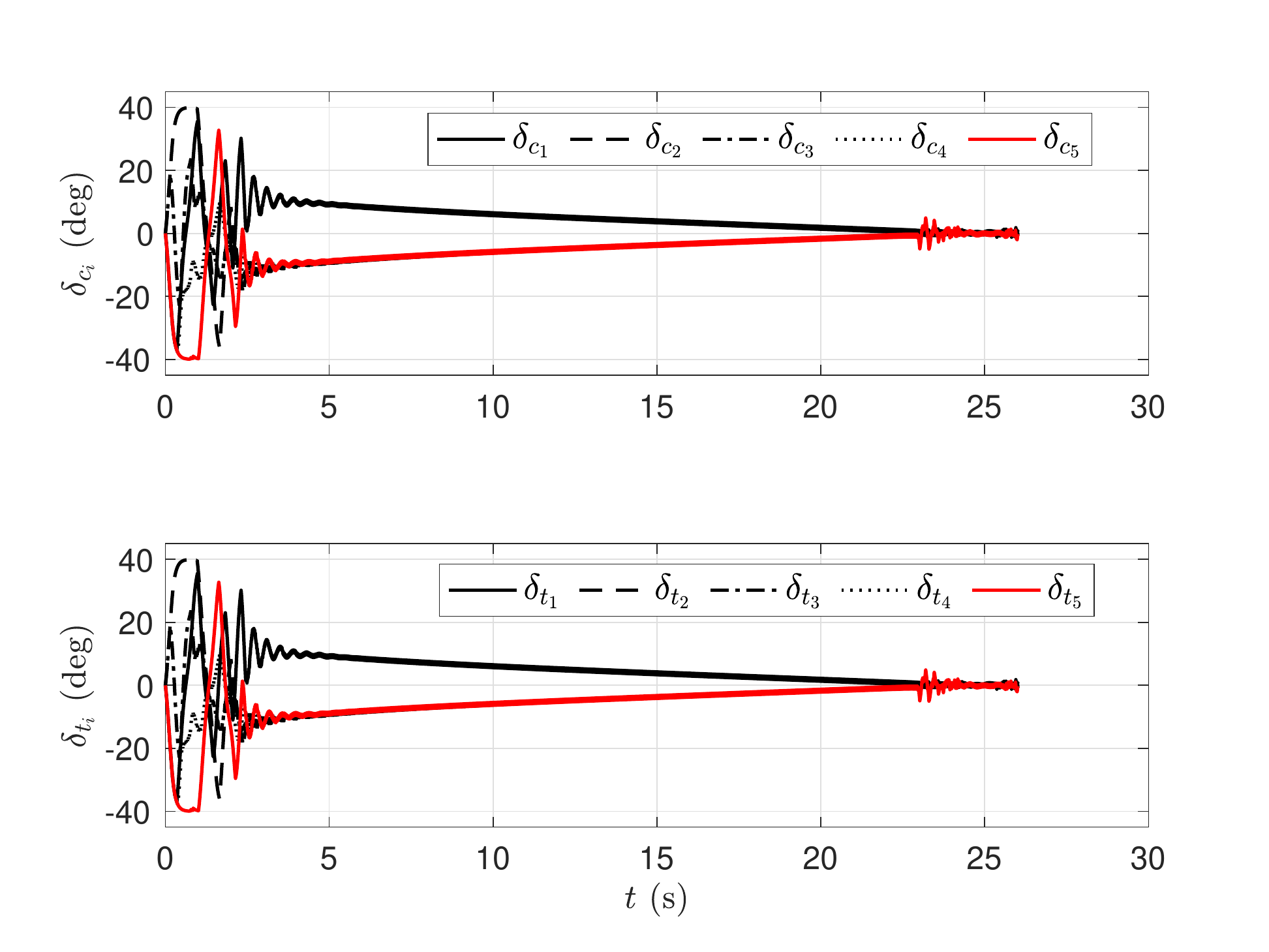}
		\caption{Control surface deflections.}
		\label{fig:sautopilotsigma}
	\end{subfigure}
	\caption{Simultaneous interception of a stationary target by a dual-controlled interceptor.}
	\label{fig:sautopilot}
\end{figure}
In this part, we validate the proposed cooperative scheme on a dual-controlled interceptor model whose separate guidance and control design was presented in \cite{SINHA2021106776}, and the $i$\textsuperscript{th} interceptor's nonlinear dynamics is governed by the equations:
\begin{subequations}\label{eq:missiledynamics}
	\begin{align}
		\dot{\alpha}_i &= q_i-\dfrac{L_i(\alpha_i,\delta_{c_i},\delta_{t_i})}{m_iv_i} = q_i-\dfrac{a_i}{v_i}, \\ 
		\dot{q}_i &= \dfrac{M_i(\alpha_i,q_i,\delta_{c_i},\delta_{t_i})}{J_i},\quad\dot{\tilde{\lambda}}_i = q_i,\\ 
		\dot{\delta}_{c_i} &= \dfrac{\delta_{c_i}^\star - \delta_{c_i}}{\tau_{c_i}},\quad\dot{\delta}_{t_i} = \dfrac{\delta_{t_i}^\star - \delta_{t_i}}{\tau_{t_i}},
	\end{align}
\end{subequations}
where $\alpha_i$, $\tilde{\lambda}_i$, and $q_i$, respectively denote the angle of attack, the pitch angle, and the pitch rate for the $i$\textsuperscript{th} interceptor. The terms $L_i(\cdot)$ and $M_i(\cdot)$ represent the lift and the aerodynamic pitch moment acting on $i$\textsuperscript{th} interceptor, whereas $m_i$ and $J_i$ are its mass and moment of inertia, respectively. The canard and the tail of each interceptor, which are the control surfaces, are described by first-order dynamics with time constants $\tau_{c_i}$ and $\tau_{t_i}$, respectively. Note that consideration of an interceptor's dynamics is more closer to practice than modeling an interceptor's autopilot as first-order lag since aerodynamic parameters are accounted in a dynamic model.

While we haven't considered the dynamics of an interceptor during the derivation of the proposed cooperative commands, we show through \Cref{fig:sautopilot} that our design exhibits satisfactory performance even for the above interceptor model along with the autopilot proposed in \cite{SINHA2021106776} is taken into account. In doing so, the control surface deflection commands generate the necessary lift, and hence the lateral acceleration, to drive the interceptor on the requisite time-constrained collision course. For brevity, we show results for stationary target interception only. For this case, $\theta_i(0) = [30^\circ,150^\circ,-30^\circ,210^\circ,45^\circ]$ while $\gamma_{i}(0) = [0^\circ,170^\circ,0^\circ,250^\circ,90^\circ]$. The predefined-time, $T_s$, is chosen as 3 s. It is observed from the time-to-go plot that even with autopilot dynamics, time-to-go values achieve agreement within 3 s, and a simultaneous target capture is ascertained with reasonable lateral acceleration demands.


\section{Concluding Remarks}\label{sec:conclusions}
In this paper, we proposed predefined-time consensus-based cooperative guidance commands for a swarm of interceptors to capture a  target simultaneously. Using the proposed guidance commands, the interceptors achieved consensus in their time-to-go values at a time that  was independent of their initial values, and can be supplied in the guidance command directly. This prevented the controller gains to be overestimated for a prescribed guidance performance and ensured that the interceptors admit requisite paths that would lead them to the target in the same time. The proposed scheme was designed for various target motions and two approaches to simultaneously intercept a constant speed target were also presented that utilized the concept of predicted interception point and deviated pursuit to facilitate flexible design and implementation depending on the mission. The proposed scheme was also compared with an existing guidance strategy developed against a moving adversary, and further validations on a dual-controlled interceptor model was carried out. In both of these cases, satisfactory performance was observed. The work presented in this paper can be further extended by provisioning for computational constraints, network induced faults and failures, and other resource constraints.

\bibliographystyle{IEEEtran}
\bibliography{references}

\end{document}